\documentstyle[emulateapj,psfig]{article}
\newcommand{\HI}{\ion{H}{1} }
\newcommand{\lya}{Ly$\alpha$ }
\newcommand{\MgII}{\ion{Mg}{2} }
\newcommand{\CIV}{\ion{C}{4} }
\newcommand{\FeII}{\ion{Fe}{2} }
\newcommand{\MgI}{\ion{Mg}{1} }
\newcommand{\w}{$W_0^{\lambda2796}$}
\begin{document}

\title{Discovery of Damped Lyman-Alpha Systems at Redshifts Less Than 1.65
and Results on their Incidence and Cosmological Mass Density$^1$}
\altaffiltext{1}{Based 
on data obtained with the NASA/ESA Hubble
Space Telescope at the Space Telescope Science Institute, which is
operated by AURA, Inc., under NASA contract NAS5-26555.}

\author{Sandhya M. Rao\altaffilmark{2} and David A. Turnshek\altaffilmark{3}
} \affil{Department
of Physics \& Astronomy, University of Pittsburgh, Pittsburgh,
PA 15260, USA}
\altaffiltext{2}{rao@everest.phyast.pitt.edu}
\altaffiltext{3}{turnshek@quasar.phyast.pitt.edu}

\begin{abstract}

We present results from an efficient, non-traditional survey  
to discover damped \lya (DLA) absorption systems with neutral 
hydrogen column densities $N_{HI}\ge2\times10^{20}$ atoms 
cm$^{-2}$ and redshifts $z<1.65$. In the past, identification 
of DLA systems at $z<1.65$  has been difficult due to their
rare incidence and the need for UV spectroscopy to detect 
Ly$\alpha$ absorption at these low redshifts. Our survey relies 
on the fact that all known DLA systems have corresponding 
\MgII absorption. In turn, \MgII absorption systems have been 
well-studied and their incidence at redshifts $0.1<z<2.2$ as a 
function of the \MgII rest equivalent width, \w, is known 
(Steidel \& Sargent 1992). Therefore, by observing the \lya 
line corresponding to identified low-redshift \MgII systems
and determining the fraction of these that are damped, we have
been able to infer the statistical properties of the low-redshift 
DLA population. In an earlier paper (Rao, Turnshek, \& Briggs 
1995), we presented initial results from an archival study with 
data from {\it HST} and {\it IUE}. Now, with new 
data from our {\it HST} GO program, we have more than 
doubled the sample of \MgII systems with available ultraviolet 
spectroscopic data. In total we have uncovered 12 DLA lines in 87 \MgII 
systems with \w $\ge0.3$ \AA. Two more DLA systems were discovered 
serendipitously in our {\it HST} spectra. At the present time 
the total number of confirmed DLA systems at redshifts $z<1.65$ is 23. 

The significant results of the survey are: (1) the DLA absorbers 
are drawn almost exclusively from the population of \MgII absorbers 
which have \w $\ge0.6$ \AA. Moreover, half of all absorption 
systems with  both \MgII\w\ and \FeII$W_0^{\lambda2600}\ge 0.5$ 
\AA\ are DLA systems. (2) The incidence of DLA systems per unit 
redshift, $n_{DLA}$, decreases as a function of decreasing redshift.
The low redshift data are consistent with the larger
incidence of DLA systems seen at high redshift (Wolfe et al. 1995)
and the inferred low incidence for DLA at $z=0$ derived from 21 cm
observations of gas-rich spirals (Rao, Turnshek, \& Briggs 1995). However,
the errors in our determination are large enough that it is not clear
if the decrease per comoving volume begins to be significant at
$z\approx 2$, or possibly does not set in until $z\approx 0.5$.
(3) On the other hand, the cosmological mass density of neutral gas 
in low-redshift DLA absorbers, $\Omega_{DLA}$, is observed to be
comparable to that observed at high redshift. In particular, there 
is no observed trend which would indicate that $\Omega_{DLA}$ at
low redshift is approaching the value at $z=0$, which is a factor of
$\approx 4 - 6.5$ lower than $\Omega_{DLA}$. (4) The low-redshift DLA 
absorbers exhibit a larger fraction of very high column density 
systems in comparison to determinations at both high redshift and 
at $z=0$. In addition, at no redshift is the column density distribution
of DLA absorbers observed to fall off in proportion to $\sim N_{HI}^{-3}$
with increasing column density, a trend that is theoretically predicted 
for disk-like systems. We discuss this and other mounting evidence that
DLA absorption does not arise solely in luminous disks but in a 
mixture of galaxy types.

Although we have doubled the sample of confirmed low-redshift DLA 
systems, we are still confronted with the statistics of small numbers. 
As a result, the errors in the low-redshift determinations of $n_{DLA}$ 
and $\Omega_{DLA}$ are substantial. Therefore, aside from the above 
evolutionary trends, we also discuss associated limitations caused by 
small number statistics and the robustness of our results. In addition, 
we note concerns due to gravitational lensing bias, reliance on the 
\MgII statistics, dust obscuration, and the sensitivity of local \HI\ 
21 cm emission surveys. 

\end{abstract}


\section{Introduction}

QSO absorption-line systems provide one of the most powerful probes
of the gaseous components associated with cosmologically distant
galaxies. Except for the small fraction of absorption-line systems
that are ejected from QSOs, the absorption lines reveal the presence of
intervening gaseous clouds  all the way up to the redshifts of the most
distant QSOs in the Universe.  Thus, independent of galaxy luminosity,
and provided that dust obscuration effects are small,
QSO absorption lines can probe the gas in galaxies up to redshift
$z\approx5$, back to a time when the Universe was less than 10\% of
its present age.  The absorbing clouds are empirically found 
to have neutral hydrogen column densities in the range $10^{12} < N_{HI} <
5\times 10^{21}$ atoms cm$^{-2}$, and they are thought to be associated with
anything from primordial clouds in the intergalactic medium (possibly
the weakest ``Ly$\alpha$ forest'' systems) to low-luminosity galaxies,
interacting systems, low surface brightness (LSB) galaxies, and the evolved
gaseous spheroidal and disk components of the most luminous galaxies.

Yet, despite the wealth of information on QSO absorption lines that
has been gleaned from ground-based studies --- and now from many years
of observation with the {\it Hubble Space Telescope} ({\it HST}) ---
there is still much to be learned about the low-redshift ($z<1.65$),
high column density absorbers referred to as damped Ly$\alpha$ (DLA)
systems. Historically these relatively rare high column density systems
were the subject of intensive searches because, by comparison with the
\lya absorption signature produced by the Galaxy, they were
thought to represent the signature of the most luminous high-redshift
gas-rich disk galaxies (Wolfe et al. 1986, hereafter
WTSC86). However, we now know that this is not the only explanation for
DLA systems and, in fact, the most luminous disk galaxies may not even
dominate the population of DLA galaxies (e.g. Le Brun et al. 1997,
 Rao \& Turnshek 1998).

The DLA systems are classically defined to have neutral hydrogen column
densities $N_{HI}\ge2\times10^{20}$ atoms cm$^{-2}$ in statistical
surveys (WTSC86). There is a practical reason for this definition ---
although damping wings in a \lya absorption-line profile with low
velocity dispersion ($10-20$ km s$^{-1}$) set in at column densities of
a few times $10^{18}$ atoms cm$^{-2}$, they still have relatively small rest
equivalent widths (e.g. $W_0^{\lambda1216} \approx 1$ \AA). Only 
\lya absorption lines with considerably larger rest
equivalent widths, say $W_0^{\lambda1216} \ge 10$ \AA, are easily recognized
in low-resolution spectra.  In turn, low-resolution spectroscopy offers an
efficient way to survey the \lya forest and search for rare DLA
absorption lines.  Thus, a 10 \AA\ rest equivalent width threshold
is  adopted in DLA spectroscopic surveys and this 
corresponds to $N_{HI}\approx2\times10^{20}$ atoms cm$^{-2}$ in the limit of
radiative damping.

In the past, surveys for DLA absorption have been used
to study the distribution of neutral hydrogen at high redshift (WTSC86;
Lanzetta et al. 1991; Wolfe et al. 1995, hereafter WLFC95). These high-redshift
studies indicate that DLA absorbers contain the bulk of the 
observable neutral gas mass in the Universe. However, spectroscopic 
detection of these absorbers using
ground-based telescopes can only be done at redshifts $z \ge 1.65$.
The detection of DLA at lower redshifts, where observations of
Ly$\alpha$ require UV data, has been difficult due to limited
{\it HST} resources. But for $\Lambda=0$ cosmologies and
$q_0$ = 0, $z<1.65$ corresponds to a lookback time of $\approx$ 62\%
of the age of the Universe ($\approx$ 77\% for $q_0$ = 0.5).  Thus, the
determination of the statistics and properties of DLA systems at these
redshifts is crucial for understanding the formation and evolution of
galaxies.  Since the occurrence
of a DLA system is relatively rare and the number of QSOs studied in
the UV has, until recently, been small, progress in this area has been
slow. The {\it HST} QSO Absorption Line Key Project survey found only one
DLA system with redshift $z=1.372$ towards Q0935+417 (Jannuzi
et al.  1998) in a redshift path of $\Delta Z=49$. Moreover, only two DLA 
systems have been confirmed from an {\it IUE} survey (Lanzetta, Wolfe, \& 
Turnshek 1995, hereafter   LWT95) which originally
reported the discovery of 14 low-redshift DLA candidate systems. One of
these is the Key Project system towards Q0935+417 and the other is the
$z=1.010$ system towards EX 0302$-$223 (Pettini \& Bowen 1997; Boiss\'e et
al. 1998). Four of the 14 systems had $W_0^{1216}>10$\AA\ and formed the
final  sample from which they derived the low-redshift DLA statistics; 
the $z=1.010$ system towards EX 0302$-$223 was not one of these
four. Thus, the {\it IUE} data and the statistical results derived 
from them, which were subsequently reported in many studies
(e.g., WLFC95, Storrie-Lombardi et al. 1996a, Storrie-Lombardi, McMahon, 
\& Irwin 1996b, Pei, Fall, \& Hauser 1999), should now be considered unreliable. 

Previously, we used unbiased results that could be extracted from {\it
HST} and {\it IUE} archival spectra to place limits on the properties
of the low-redshift DLA population (Rao, Turnshek, \& Briggs 1995,
hereafter RTB95). In this paper, we discuss our recently completed 
{\it HST}-FOS observing program
to discover low-redshift ($z<1.65$) DLA systems. Coupled with our earlier
archival work, we have now completed an unbiased study of the Ly$\alpha$
absorption line in 87 \ion{Mg}{2} absorption-line systems, 12 of which
we find to be DLA systems which formally meet the classical criterion of $N_{HI}
\ge 2 \times 10^{20}$ atoms cm$^{-2}$. In addition, two new classical 
DLA systems with no existing \MgII information were discovered serendipitously 
in our {\it HST} spectra. Consequently, our
survey has greatly increased the number of known low-redshift DLA systems.
Including the 14 classical DLA systems which were uncovered in this survey,
at the present time there are 23 published low-redshift ($z<1.65$)
DLA systems for which $N_{HI} \ge 2 \times 10^{20}$ atoms cm$^{-2}$ has been
confirmed on the basis of UV spectroscopy of the \lya absorption-line
profile or 21 cm absorption. Although DLA systems with somewhat 
lower column densities can often be detected using {\it HST} spectra, we 
have set our threshold column density for detection to the classical value so
that meaningful comparisons between the statistics of these systems at
low redshift and those found in high-redshift surveys can be made.

In \S2, we describe the survey, some ancillary data which we use,
and the analysis which leads to the identification of the DLA systems.
In \S3, we discuss the statistical properties of the \ion{Mg}{2} sample.
In \S4, we derive the incidence of low-redshift DLA systems, $n_{DLA}(z)$,
and their cosmological mass density, $\Omega_{DLA}(z)$, and compare them
to determinations at high redshift. In \S5, we derive the DLA column
density distribution, $f(N_{HI})$, and consider its evolution. We find
that while the incidence of DLA systems at low redshift shows some indication
of dropping with decreasing redshift to a value consistent with observations
of the \HI\ cross-sections of local spirals, there is no clear evidence that
their cosmological mass density drops from the large values observed
at high redshift to the value inferred at $z=0$ 
(although the error bars remain large).  Moreover,
contrary to previous claims, there is evidence that the DLA column
density distribution flattens at low redshift when   compared to  the
high-redshift DLA column density distribution.  In \S6,  we discuss the
limitations of our results as well as concerns about the possible
effects of obscuration by dust and gravitational lensing.  
We discuss the nature of the DLA absorber population in \S7 and
the conclusions are summarized in \S8.

\section{UV Spectroscopic Survey for DLA Systems}

The details of our spectroscopic survey for DLA systems are presented
below.  First we discuss the unbiased  method of searching for
DLA which allows us to derive the statistical properties of the
DLA absorbers (\S2.1). We then describe the extent to which the survey
relies on archival observations (\S2.2) and new observations (\S2.3);
the role of 21 cm observations in the exclusion and selection of objects
is also described (\S2.4).  The identified DLA systems are presented in
\S2.5 and an object of special interest, which was a DLA candidate but is
now identified as a grouping of \lya absorption lines on a supercluster
size scale, is discussed in \S2.6.

\subsection{Unbiased Selection of \MgII Systems for UV Spectroscopic
Study}

The \ion{Mg}{2} selection method for finding DLA systems with $z<1.65$
was discussed in RTB95. In brief, since all DLA systems discovered so far
exhibit metal-line absorption (Turnshek et al. 1989; Lu et al. 1993; Wolfe
et al. 1993; Lu \& Wolfe 1994), the probability of finding DLA lines
is greatly increased if a search is restricted to QSOs whose spectra
show intervening metal-line absorption. \ion{Mg}{2} absorption lines
can be detected from the ground with optical spectroscopy for redshifts
$0.1<z<2.2$, and their statistical properties are  known (Steidel \&
Sargent 1992, hereafter SS92).  Thus, if the fraction of DLA systems
in an unbiased sample of \ion{Mg}{2} systems can be determined, then
the unbiased incidence of DLA systems at redshifts as low as $z=0.1$
can be constructed from the known incidence of the \ion{Mg}{2} systems.

Our initial \ion{Mg}{2} sample was drawn from a list of over 360
\ion{Mg}{2} systems in the literature with $z<1.65$ (Rao 1994). 
The sample of \ion{Mg}{2} systems used to search for DLA
absorption was constructed after excluding (1) \ion{Mg}{2}
absorption lines with rest equivalent width $W_0^{\lambda2796} < 0.3$ \AA,
(2) systems for which the predicted position of the Ly$\alpha$ line fell
within a broad absorption line (BAL) trough in the spectrum of a 
BAL QSO (Turnshek 1988; Turnshek 1997a,b), (3) previously known \HI\
21 cm absorbers (\S2.4), (4) systems for which the predicted position of the
Ly$\alpha$ line was shortward of a known Lyman series absorption edge (i.e. a
Lyman limit), and (5) \ion{Mg}{2} systems within 1000 km s$^{-1}$ of the
QSO emission redshift. Furthermore, in accordance with the criterion
adopted by SS92, \ion{Mg}{2} systems within 1000 km s$^{-1}$ of each
other were considered to be a single system. This resulted in a sample
of 243 \ion{Mg}{2} systems which could be observed for follow-up study
if unlimited telescope time were available.
These 243 systems define our total possible sample. 

\subsection{Update on {\it HST} and {\it IUE} Archival Data}

The archival study of RTB95 included 43 \ion{Mg}{2}
systems which satisfied the selection criteria above (\S2.1) and which
had adequate UV spectral data (see Table 1 in RTB95) to search for DLA
or DLA candidates. Two DLA lines ($z=0.859$ in PKS 0454+039 and
$z=0.656$ in 3C336 (1622+259)) and four candidates were identified. With
the availability of additional archival data, the archival portion of
the RTB95 survey has now been updated and modified as explained below. 

One of the four RTB95 DLA candidates, the $z=1.0096$ system towards
EX 0302$-$223, has now been confirmed as a DLA system. Two candidates,
the $z=0.1602$ system towards Q0151+045 and the $z=1.2232$ system towards
PG 1247+268, were remeasured using {\it HST}-FOS data and they did
not meet the minimum \HI\ column density criterion. The fourth RTB95 DLA
candidate, the $z=0.5761$ system towards PG 0117$+$213, was observed
with {\it HST}-FOS in spectropolarimetric mode (see figure 3 in Koratkar
et al. 1998 for the recalibrated spectrum). We have determined that
this candidate system is not damped, but discuss it in \S2.6 as an object of
special interest. In addition, although RTB95 discounted the {\it IUE}
spectrum of the $z=1.391$ system in Q0957+561A as showing $N_{HI}
\ge 2 \times 10^{20}$ atoms cm$^{-2}$, we have now remeasured its
spectrum using higher resolution and signal-to-noise ratio
{\it HST}-FOS data and find it to be
classically damped.  Also, we find one of the six new archival systems,
$z=0.633$ in Q1209+107, to be damped. In order to treat all of the
sample data consistently, we have remeasured the \HI\ column densities
and uncertainties for the archival DLA systems by fitting Voigt profiles
to the data, even when the DLA systems in these objects were previously studied
by others (see \S2.5).  These details, including six more
systems that have since appeared in the {\it HST} archives and which we
have added to our observed sample, are given in Table 1.
Finally, we note that in Table 1 of RTB95 it was
erroneously reported that the $z=0.6330$ \ion{Mg}{2} system in Q0420$-$014
was part of our sample, however there is no available UV spectrum of
this object in the required wavelength range. Thus, this system is now
explicitly excluded from the observed sample.  Currently, the total
number of archival \ion{Mg}{2} systems included in our \lya survey is 48.

\begin{deluxetable}{llllcccccl} 
\tablenum{1}
\tablecaption{Update to the RTB95 Archival Sample} 
\tablehead{
\colhead{QSO} & 
\colhead{Coordinate} &
\colhead{$m_V$\tablenotemark{a}} & 
\colhead{$z_{em}$\tablenotemark{a}} 
& \multicolumn{2}{c}{\ion{Mg}{2}} & 
\colhead{Ref.\tablenotemark{b}} &
\colhead{\ion{Mg}{1}\tablenotemark{c}}  
& \colhead{DLA\tablenotemark{c}} 
& \colhead{Source\tablenotemark{d}}\\[.2ex]  \cline{5-6}
\colhead{} & 
\colhead{Designation} & 
\colhead{} &
\colhead{} &
\colhead{$z_{abs}$} & 
\colhead{$W_0^{\lambda2796}$ (\AA)}  & 
\colhead{} & 
\colhead{}
& \colhead{}
& \colhead{}
 }

\startdata 
PG 0117+213 & 0117+213 & 16.1 & 1.491 & 0.5761 & 0.91 & 1 & + & $-$ & Koratkar, PID 6109\\
Q 0151+045 & 0151+045 & 16.9 & 0.404 & 0.1602 & 1.55 & 2 & + & $-$ & Tytler, PID 4396\\ 
EX 0302$-$223 & 0302-223 & 16.4 & 1.409 & 1.0096 & 1.16 & 3 & + & + & Burbidge, PID 6224 \\ 
Q 0333+321 & 0333+321 & 17.5 & 1.259 & 0.9531 & 0.47 & 1 & $-$ &$-$ &  Wills, PID 5441\\
PKS 0454+039 & 0454+039 &  16.5 & 1.343 & 0.8596 & 1.53 & 3 & + & + & Bergeron, PID 5351\\
Q 0957+561A & 0957+561A & 17.0 & 1.414 & 1.3911 & 2.12 & 4 & $+$ & + & Burbidge, PID 5683 \\ 
Q 1209+107 & 1209+107 & 17.8 & 2.193 & 0.3930 & 1.00 & 5 & \nodata &$-$ & Bergeron, PID 5351\\
\nodata & \nodata & \nodata &\nodata  & 0.6295 & 2.92 & 5 & \nodata & + & Bergeron, PID 5351 \\
PG 1247+268 & 1247+268 & 15.6 & 2.043 & 1.2232 & 0.48 & 1 & + & $-$ & Burbidge, PID 5095\\
PG 1248+401 & 1248+401 & 16.1 & 1.032 & 0.7729 & 0.76 & 1 & $-$ &$-$ & Bahcall, PID 5664\\
PKS 1327$-$206 & 1327$-$206 &  17.0 & 1.165 & 0.8500 & 2.11 & 6 & +  &$-$ & Tye, PID 5654\\
Q 1517+239 & 1517+239 & 17.3 & 1.903 & 0.7382 & 0.30 & 7 & $-$  &$-$ & Foltz, PID 5320\\
3C 336 & 1622+239 & 17.5 & 0.927 & 0.6561 & 1.29 & 1 & $-$ & + & Steidel, PID 5304\\
\enddata 
\tablenotetext{a}{\ Apparent magnitude $m$ and emission
redshift $z_{em}$ are from V\'eron-Cetty \& V\'eron (1998).}

\tablenotetext{b}{\ Reference for $z_{abs}$, $W_0^{\lambda2796}$,
and Mg I.}

\tablenotetext{c}{\ '+' indicates the presence of, and '$-$'
indicates the absence of the said absorption line. '...' is entered
where no data are available.}
 
\tablenotetext{d}{\ The Principal Investigator of the {\it HST} archival program and the 
proposal identification number (PID) are listed.}

\tablerefs{(1) SS92;
(2) Bergeron et al. 1988; (3) Petitjean \& Bergeron 1990;
(4) Caulet 1989; (5) Young, Sargent, \& Boksenberg 1982;
(6) Bergeron, D'Odorico, \& Kunth 1987; (7) Sargent, Boksenberg,
\& Steidel 1988}

\end{deluxetable}

\subsection{New {\it HST}-FOS Observations}

In order to extend our original archival study (RTB95) and improve the
statistics, we carried out a large UV spectroscopic survey of low-redshift
\ion{Mg}{2} systems with {\it HST}-FOS in an {\it HST} Cycle 6 program.
The survey was optimized by selecting QSOs that: (1) had more than
one \ion{Mg}{2} system whose \lya line could be observed in an FOS
spectrum but that were physically unassociated, i.e., the difference in
absorption redshifts corresponded to more than 1000 km s$^{-1}$ and 
(2) were bright enough to obtain adequate
signal-to-noise ratios in a single orbit at the expected position of the
\lya line. The 1.0-PAIR aperture was used to obtained all the 
spectra and the FOS grating selection was based
on the expected position of the \lya line. The data generally have
minimum signal-to-noise ratios of S/N $\approx$ 11 for the high 
resolution G190H/RD and G270H/RD gratings and S/N $\approx$ 15 for the
low resolution G160L/RD and G160L/BL gratings. The QSO limiting V magnitude
requirements for obtaining these minimum S/N ratios in 27 minute exposures,
which was typical for our sample, is shown in Figure 1 for the various 
gratings. By the end of Cycle 6,
before FOS was decommissioned, we had used 40 orbits to obtain 38 spectra
of 36 QSOs that included the expected location of the Ly$\alpha$ lines of
60 \ion{Mg}{2} systems (actually, in one case only Ly$\beta$ and higher
order lines were available, but this was sufficient for our purposes).
No data were obtained during 2 orbits due to technical problems with
{\it HST}.

\begin{figure*}
\plotone{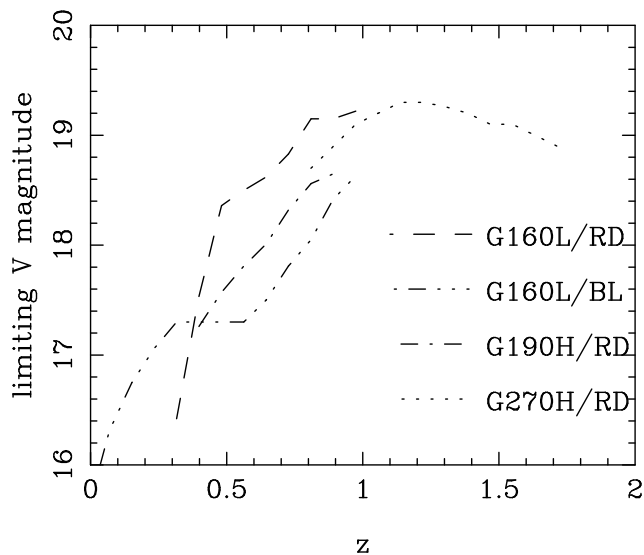}
\caption{Limiting V magnitude of QSOs in the survey set by the criterion
$S/N=11$ for G190H and G270H spectra, and $S/N=15$ for G160L spectra at 
the redshifted wavelength of Ly$\alpha$. For example, the spectrum of a QSO
brighter than V=19.2 observed with the G270H grating will have
$S/N>11$ at the position of a $z=1.4$ Ly$\alpha$ line.}
\end{figure*}

The details of this part of the \ion{Mg}{2} sample and associated new 
{\it HST}-FOS observations are given in Table 2, with entries taken from the 
V\'eron-Cetty \& V\'eron (1998) QSO catalog.
Specifically, column 1 gives the QSO's name;
column 2 gives the QSO's 1950 coordinate designation; 
column 3 gives its V magnitude;
column 4 gives its emission redshift;
column 5 gives the \ion{Mg}{2} absorption redshift;
and column 6 gives the reference from which the absorption
redshift was obtained.  Columns 7, 8, and 9 give details of the
{\it HST}-FOS observation, i.e., grating,  exposure time,
and  observation date, respectively. In cases where two or more
systems were chosen for observation with one grating,
the grating and integration time are entered alongside the lowest
redshift absorption system. Applicable comments are noted in column
10, with an explanation in the footnotes.

Unfortunately, previously unobserved Lyman limit absorption eliminated
information on the \lya absorption line in 21 of the 60 new \ion{Mg}{2}
systems (see column 10 of Table 2). 
In addition, the G160L/RD spectrum of S4 0248+43 was too noisy
near the redshifted ($z=0.3939$) Ly$\alpha$ wavelength ($\approx$1700 \AA)
to determine if the \lya absorption line in this particular \ion{Mg}{2}
system is damped using the existing FOS data alone. However, we
have retained this system in our sample since its damped nature has been
revealed through 21 cm observations (see \S2.4).

Therefore, with the aid of archival and new observations, it is now
possible to determine the nature of the \lya line in a total of 87
\ion{Mg}{2} systems. These 87 systems define our current unbiased
sample, which we will henceforth refer to as the RT sample.
In \S3 we discuss some of the additional properties of this
sample (e.g. other metal lines).  The discussion in the remainder of
this section primarily pertains to the determination of which of these
87 systems have associated DLA lines.

\begin{deluxetable}{lcccccccll}
\footnotesize
\tablewidth{7.0in}
\tablenum{2}
\tablecaption{New {\it HST}-FOS Observations}
\tablehead{  \colhead{QSO\tablenotemark{a}}
                      & \colhead{Coordinate}
                      & \colhead{$m_V$\tablenotemark{a}}
                      & \colhead{$z_{em}$\tablenotemark{a}}
                      & \colhead{$z_{abs}$}
                      & \colhead{Ref.\tablenotemark{b}}
                      & \colhead{FOS }
                      & \colhead{Exposure}
                      & \colhead{Observation} 
                      & \colhead{Notes\tablenotemark{c}} \\ [.2ex]
                      \colhead{} 
                      & \colhead{Designation} 
                      & \colhead{}
                      & \colhead{}
                      & \colhead{\ion{Mg}{2}}
                      & \colhead{}
                      & \colhead{Grating}
                      & \colhead{Time (s)}
                      & \colhead{Date} 
                      & \colhead{} }

\tablecolumns{10}

\startdata
Q 0002$-$422    & 0002$-$422  & 17.2  & 2.758 & 0.8366 & 1 & G270H/RD & 1170 & 2 June 1996 & LL \\*
  &   &   &   & 1.5413 & 2 &   &   &  & \\
PHL 938      & 0058+019  & 17.2  & 1.959 & 0.6128 & 3 & G190H/RD & 1590  & 12 July 1996 & \\
PKS 0119$-$04  & 0119$-$046  & 16.9  & 1.953 & 0.6577 & 4 & G190H/RD & 1590  & 22 July 1996 & \\
B2 0141+33   & 0141+339  & 17.6  & 1.450 & 0.4709 & 5 & G160L/RD & 1580  & 24 Aug 1996 & \\
UM 366       & 0143$-$015  & 17.7  & 3.141 & 1.0383 & 6 & G270H/RD & 1590  & 5 June 1996 & \\*
  &   &   &   & 1.2853 & 6 &   &   &   & \\
UM 675       & 0150$-$202  & 17.4  & 2.147 & 0.7801 & 3 & G190H/RD & 1590  & 12 Sep 1996 & \\
PKS 0229+13  & 0229+131  & 17.7  & 2.065 & 0.3723 & 3 & G160L/BL & 1400  & 29 Aug 1996 & LL \\*
  &   &   &   & 0.4177 & 3 &   &   &   & LL \\
S4 0248+43   & 0248+430  & 17.6  & 1.310 & 0.3939 & 5 & G160L/RD & 1660  & 21 Aug 1996 & poor S/N\\*
  &   &   &   & 0.4515 & 7 &   &  &   & \\
PKS 0421+019 & 0421+019  & 17.0  & 2.055 & 1.3918 &  5 & G270H/RD & 1600  & 20 Aug 1996 & \\*
  &   &   &   & 1.6380 & 5 &   &   &  & \\
PKS 0424$-$13  & 0424$-$131  & 17.5  & 2.166 & 1.0345 & 5 & G270H/RD & 1590  & 13 July 1996 & LL \\*
  &   &   &   & 1.4080 &  5 &   &   &   & \\*
  &   &   &   & 1.5623 & 5 &   &&   & \\
Q 0453$-$423   & 0453$-$423  & 17.1  & 2.661 & 0.7255 & 8 & G190H/RD & 1700  & 22 May 1996 & LL \\
   &   &   &   & 0.9079 & 8 & G270H/RD & 2440  &   & LL \\*
  &   &   &   & 1.1492 & 8 &   &   &   & LL \\*
  &   &   &   & 1.4592 & 8 &   &   &   & LL \\ 
OI 363       & 0738+313  & 16.1  & 0.630 & 0.2213 & 9 & G160L/BL &  1520 & 15 May 1996 & \\
PKS 0823$-$22 & 0823$-$223  & 16.2  & $>$0.910 & 0.9103 & 10 & G270H/RD & 1610  & 26 June 1996 & \\
B2 0827+24   & 0827+243  & 17.3  & 0.941 & 0.5247 & 11 & G160L/RD & 1560  & 28 Oct 1996 & \\
MG 0833+1123  & 0830+115  & 18.0  & 2.979 & 0.8032 & 6 & G190H/RD & 1570  & 15 May 1996 &  LL \\
4C 13.39     & 0843+136  & 17.8  & 1.877 & 0.6064 &  12 & G160L/RD & 1540  & 15 May 1996 & \\
TB 0933+733  & 0933+732  & 17.3  & 2.525 & 1.4789 & 5 & G270H/RD & 1900  & 6 Dec 1996 & \\*
  &   &   &   & 1.4973 & 5 &   &   &   & \\
PKS 0952+17 & 0952+179  & 17.2  & 1.478 & 0.2377 & 5 & G160L/BL & 1430  & 24 Dec 1996 & \\
SBS 0953+549 & 0953+549  & 17.4  & 2.579 & 1.0590 & 5 & G270H/RD & 1790  & 29 Sep 1996 & LL \\*
  &   &   &   & 1.0642 & 5 &   &   &   & LL \\*
  &   &   &   & 1.2624 & 5 &   &   &   & LL \\
MARK 132      & 0958+551  & 16.0  & 1.760 & 0.2413 & 3 & G160L/BL &  1670 & 11 June 1996 & \\
TOL 1035.8$-$27.6  & 1035$-$276  & 19.0  & 2.168 & 0.8230 & 13 & G160L/RD & 1510  & 15 June 1996 & \\
4C 61.20     & 1049+616  & 16.5  & 0.421 & 0.2251 & 9 & G160L/BL & 1660  & 1 June 1996 & \\*
  &   &   &   & 0.3933 & 9 &   &   &   & \\
PKS 1127$-$14  & 1127$-$145  & 16.9  & 1.187 & 0.3130 & 14 & G160L/BL & 1450  & 22 June 1996 & \\
B2 1148+38   & 1148+386  & 17.0  & 1.304 & 0.2130 & 9 & G160L/BL & 1520  & 31 May 1996 & \\*
  &   &   &   & 0.5533 & 5 &   &   &   & \\
UM 485      & 1213$-$002  & 17.0  & 2.691 & 1.5543 & 5 & G270H/RD & 1590  & 2 Dec 1996 & \\ 
TON 1530     & 1222+228  & 16.6  & 2.048 & 0.6681 & 3 & G190H/RD & 1600  & 4 July 1996 & \\
Q 1246$-$057   & 1246$-$057  & 16.7  & 2.224 & 0.6399 & 15 & G190H/RD & 1600  & 12 June 1996 & LL \\
4C 65.15     & 1323+655  & 17.5  & 1.624 & 1.5181 & 16 & G270H/RD & 1820  & 9 Nov 1996 & \\*
  &   &   &   & 1.6101 & 16 &   &   &   & \\
PG 1329+412  & 1329+412  & 17.2  & 1.937 & 1.2820 & 5 & G270H/RD & 1710  & 2 June 1996 & \\*
   &   &   &   & 1.6011 & 5 &   &   & \\
PKS 1354+25  & 1354+258  & 18.0  & 2.006 & 0.8585 & 16 & G270H/RD & 1600  & 17 Nov 1996 & \\*
  &   &   &   & 0.8856 & 16 &   &   &   & \\*
  &   &   &   & 1.4205 & 16 &   &   &   & \\
MC 1511+103  & 1511+103  & 17.7  & 1.546 & 0.4370 & 12 & G160L/RD & 1550  & 25 June 1996 & LL\\
4C 18.43     & 1540+180  & 18.0  & 1.661 & 0.6945 & 16 & G190H/RD & 1570  & 6 July 1996 & LL \\*
  &   &   &   & 0.7294 & 16 &   &   &   & LL \\*
  &   &   &   & 0.7936 & 16 &   &   &   & LL \\
KP 1623.7+26.8B & 1623+269  & 16.0  & 2.521 & 0.3292 & 3 & G160L/BL & 1920  & 7 July 1996 & LL \\*
  &   &   &   & 0.8881 & 3 & G270H/RD & 1630  &   & \\
PG 1715+535  & 1715+535  & 16.5  & 1.940 & 0.3674 & 3 & G160L/BL & 1660  & 8 Nov 1996 & LL \\ 
PKS 1821+10  & 1821+107  & 17.3  & 1.364 & 0.4738 & 5 & G190H/RD & 1590  & 3 June 1996 & LL \\*
  &   &   &   & 0.5702 & 5 &   &   &  & LL \\*
  &   &   &   & 1.2528 & 5 &   &   &  &  \\
3C 395       & 1901+319  & 17.5  & 0.635 & 0.3901 & 17 & G160L/BL & 1460  & 12 Aug 1996 & \\
\enddata

\tablenotetext{a}{\ From V\'eron-Cetty \& V\'eron 1998.}

\tablenotetext{b}{\ The reference for the Mg II absorption line redshift: (1) Sargent et al. 1979;
(2) Lanzetta, Turnshek, \& Wolfe 1987; (3) Sargent, Boksenberg, \& Steidel 1988; (4) Sargent, Young,
\& Boksenberg 1982; (5) SS92; (6) Sargent, Steidel, \& Boksenberg 1989;
(7) Womble et al. 1990; (8) Carswell, Smith, \& Whelan 1977; (9) Boulade et al. 1987; 
(10) Falomo 1990; (11) Ulrich \& Owen 1977; (12) Foltz et al. 1986; (13) Jakobsen, Perryman, \& Cristiani 1988; 
(14) Bergeron \& Boiss\'e 1991; (15) Boksenberg et al. 1978; (16) Barthel, Tytler, \& Thomson 1990;
(17) Aldcroft, Bechtold, \& Elvis 1994.}

\tablenotetext{c}{\ This column indicates that the nature of the Ly$\alpha$ line associated with the 
Mg II absorption system could not be determined either due to an intervening Lyman limit system (LL),
in which case there was no continuum, or because the signal to noise ratio at the expected
position of the Ly$\alpha$ line was very poor.}

\end{deluxetable}

\subsection{Exclusion and Selection of DLA Systems Based on 21 cm Data}

We excluded all previously known 
21 cm absorbers from the RT sample because we had 
{\it a priori} knowledge of the high \HI\ column density of the absorber.
Moreover, {\it HST} UV spectra of the background QSOs were taken by various
researchers for the explicit purpose of studying the damped line 
corresponding to the 21 cm absorption system. The four excluded systems  
(see RTB95) are: $z=0.5420$ in AO 0235+164, $z=0.4369$ in 3C 196 
(0809+483), $z=0.3950$ in PKS 1229$-$021, and $z=0.6922$ in 3C 286 
(1328+307). However, concurrent with this
survey, a survey for \HI\ 21 cm absorption in radio-loud quasars with
known \ion{Mg}{2} systems was being carried out by W. Lane and F. Briggs 
(see Lane et al. 1998a for some initial results). Three of the known
\ion{Mg}{2} systems observed during our {\it HST}-FOS survey have now
been found to be 21 cm absorbers.
 These are the $z=0.3939$ system towards S4 0248+43, the
$z=0.3130$ system towards PKS 1127$-$14, and the $z=0.2213$ system
towards OI 363 (0738+313) (Lane et al. 1998a; Lane et al. 
2000, in preparation). The \lya line in the spectrum of S4 0248+43 was 
the only \lya line which  did not have {\it HST}-FOS data
of sufficient quality to determine if the line was damped.  However,
the availability of the \HI\ 21 cm data, which shows that this line
must be damped (see \S2.5.2 for further details), gives our survey a
certain degree of completeness since, for all observed systems which were
selected in an unbiased manner, we have been able to determine whether
or not they are damped.  In particular, the originally-defined sample
of 243 systems included \ion{Mg}{2} systems for which no information on
the \HI\ column density was published, either from the UV \lya line or
for the 21 cm line. Thus, all information obtained after the sample was
assembled is acceptable, and this does not bias the results. If anything,
the decision to exclude all previously known 21 cm systems might now be
viewed as overly conservative.  
It could be argued that unwarranted exclusion of these systems might cause us
to {\it underestimate} the incidence and mass density of DLA systems
at low redshifts.  In the future, if \HI\ column density information
becomes available for all of the \ion{Mg}{2} systems that satisfy our
selection criteria, all of the previously identified 21 cm absorbers
which were excluded from the sample should be included at that time,
because they would be considered free of selection bias.

Finally, we note that the $z=0.0912$ DLA system towards OI 363
(0738+313), which we detected serendipitously (see \S2.5.5 and Rao \&
Turnshek 1998), has also been detected in 21 cm absorption (Lane et al. 1998b).

\subsection{\ion{Mg}{2} Systems with Identified DLA Absorption}

The QSOs with DLA detections are listed in Table 3 along with the
redshift and rest equivalent width of the \MgII absorption lines, the
redshift and column density of the DLA lines and their uncertainties, and
the source of the Ly$\alpha$ data.
We have determined that 14 DLA lines which meet the classical criterion
for being damped are present in our data set. Twelve of these are found 
in the 87 \MgII systems and two others were found serendipitously. 
Redshifts for the $z=0.3941$ absorption system towards S4 0248$+$43,
the $z=0.0912$ and $z=0.2212$ systems towards OI 363 (0738+313),
and the $z=0.3127$ system towards PKS 1127$-$14, which were derived
from  more accurate \HI\ 21 cm data, are reported without error
(Lane et al. 1998a,b). When deriving redshifts for the remaining DLA
systems, we have not corrected for wavelength zero point
offsets that were reported in FOS data (Rosa, Kerber, \& Keys 1998,
ISR CAL/FOS-149) since we did not find this necessary.  For the DLA
systems without 21 cm redshifts, our derived redshifts are
mostly within $1-2\sigma$ of the \ion{Mg}{2} redshifts. Thus, the
current data are sufficient to identify any DLA lines that correspond
to metal-line absorbers.  Although Voigt profile fits that differed
by 0.0001 in redshift are clearly distinct in fits to DLA lines,  we
have reported redshift uncertainties of 0.001 for data taken with the
G190H and G270H gratings and 0.002 for data taken with G160L gratings,
since the grating resolutions and signal-to-noise ratios do not permit
more accurate wavelength determinations.

\begin{deluxetable}{llccccccl}
\footnotesize
\tablenum{3}
\tablecaption{The DLA Systems Found in the Survey}
\tablehead{ 
\colhead{QSO} & 
\colhead{Coordinate} &
\multicolumn{2}{c}{\ion{Mg}{2}} &
\multicolumn{2}{c}{Ly$\alpha$} &
\colhead{$N_{HI}/10^{20}$} &
\colhead{$\sigma_{N_{HI}}/10^{20}$} &
\colhead{Source\tablenotemark{b}} \\[.2ex] \cline{3-4} \cline{5-6}
\colhead{} &
\colhead{Designation} &
\colhead{$z_{abs}$} &
\colhead{$W^{\lambda2796}_{0}$ (\AA)} &
 \colhead{$z_{abs}$} & 
\colhead{$\sigma_{z_{abs}}$\tablenotemark{a}} &
\colhead{cm$^{-2}$} &
\colhead{cm$^{-2}$} &
\colhead{}
}
\startdata
UM 366 & 0143$-$015 & \nodata & \nodata & 1.613\tablenotemark{c} & 0.001 & 2.0 & 
0.2 &   This Program (Rao, PID 6577)\\
S4 0248+43 & 0248+430 & 0.3939 & 1.86 & 0.3941 & \nodata & 36\tablenotemark{d}
 & 4\tablenotemark{d}  &  This Program (Rao, PID 6577)\\
EX 0302$-$223 & 0302$-$223 & 1.0096 & 1.16 & 1.010 & 0.001 & 2.3 & 0.2 & 
Archive (Burbidge, PID 6224)\\
PKS 0454+039 & 0454+039 & 0.8596 & 1.53 & 0.859 & 0.001 & 4.7 & 0.3 & Archive (Bergeron, PID 5351) \\
OI 363 & 0738+313 & \nodata & \nodata & 0.0912\tablenotemark{c}
 & \nodata & 15 & 2 &  This Program (Rao, PID 6577)\\
\nodata  & \nodata & 0.2213 & 0.52 & 0.2212 & \nodata &  7.9 & 1.4 & This Program (Rao, PID 6577)\\
B2 0827+24 & 0827+243 & 0.5247 & 2.90 & 0.518 & 0.002 & 2.0 & 0.2 & This Program (Rao, PID 6577)\\
TB 0933+733 & 0933+732 & 1.4789 & 0.95 & 1.478 & 0.001 & 42 & 8 & This Program (Rao, PID 6577)\\
PKS 0952+17 & 0952+179 & 0.2377 & 0.63 & 0.239 & 0.002 & 21.0 & 2.5 & This Program (Rao, PID 6577)\\
Q 0957+561A & 0957+561A & 1.3911 & 2.12 & 1.391 & 0.001 & 2.1 & 0.5 & Archive (Burbidge,
 PID 5683)\\
PKS 1127$-$14 & 1127$-$145 & 0.3130 & 2.21 & 0.3127 & \nodata & 51 & 9 & 
This Program (Rao, PID 6577)\\
Q 1209+107 & 1209+107 & 0.6295 & 2.92 & 0.633 & 0.002 & 2.0 & 1.0 & Archive (Bergeron, 
PID 5351)\\
PKS 1354+25 & 1354+258 & 1.4205 & 0.61 & 1.418 & 0.001 & 32 & 2 & This Program (Rao, PID 6577)\\
3C 336 & 1622+239 & 0.6561 & 1.29 & 0.656 & 0.001 & 2.3 & 0.4 & Archive (Steidel, PID
5304)\\ 
\enddata

\tablenotetext{a}{\  No error is reported if $z_{abs}$ is taken from 
21 cm absorption-line data.}

\tablenotetext{b}{\  The Principal Investigator of the {\it HST} program and the proposal 
identification (PID) number are listed.}

\tablenotetext{c}{\ The two serendipitously discovered systems for 
which no \ion{Mg}{2} information is available are also listed.}

\tablenotetext{d}{\ From 21 cm absorption-line data and assuming 
$T_{s}=700$ K (Lane et al. 2000, in preparation).}

\end{deluxetable}

To fit the DLA lines with Voigt profiles, all of the
pipeline-calibrated spectra were first linearly resampled in wavelength
to 4 pixels per resolution element in order to match the quarter-diode
substepping procedure used during the observations. A continuum level was
fitted to each spectrum interactively and the spectrum was normalized
by this continuum. In a few cases, in medium resolution gratings or in
cases where the \HI\ column density was high in a low resolution grating,
we found it necessary to subtract a small correction (usually less than
10\% of the continuum) before normalizing the spectrum in order to force
the center of the damped line to zero flux.  Similar procedures were
adopted by Steidel et al. (1995) and Boiss\'e et al. (1998) to measure
the column density of DLA lines detected in FOS spectra. 
Possible reasons for the non-zero flux at the center of a DLA line in an
FOS spectrum are scattered light within the detector and incorrect dark 
background subtractions during pipeline processing.

A Voigt damping profile, convolved with the appropriate Gaussian line
spread function of the FOS-grating combination, was then fitted to the
data.  The Doppler parameter was set to $b=15$ km s$^{-1}$ for all of
the fits.  Note that for the new {\it HST}-FOS observations we used the
1.0-PAIR aperture which results in the line spread function FWHM being
0.96 FOS diodes, i.e., 3.8 pixels. This corresponds to typical resolutions
(FWHM) of $\approx$ 2.0 \AA\ for the G270H grating,  $\approx$ 1.5 \AA\
for the G190H grating, and $\approx$ 7 \AA\ for the G160L grating.
The exact resolutions in units of \AA\ depend on the wavelength of the
damped line.

The theoretical profiles that best fit the observed damped profiles were
also determined interactively.  In many cases, Ly$\alpha$ forest lines
overlapped with the DLA line and care had to be taken to interactively
exclude them from the fit. In addition, the fact that some of the data
were fairly noisy, with typical signal-to-noise ratios of $\approx 5$
within the damped profile, made fitting a reasonable continuum near the
damped line the most challenging task.  Since continuum placement is
also somewhat subjective, and this usually is the largest contribution
to the uncertainties, conservative estimates for errors in $N_{HI}$ were
determined by shifting the  normalized continuum above and below the
most likely continuum level by an offset measured from the $1\sigma$
(per pixel) error array of the flux spectrum.  The spectrum was then
renormalized to the new levels and Voigt profiles were again fitted to
the DLA lines in the renormalized spectra. The resulting ranges in column
densities derived in this way are reported as the formal uncertainties in
$N_{HI}$. To illustrate the procedure, we show examples of Voigt profile
fits with high and low renormalized continua to the damped profile in
B2 0827+24 (Figures 2a,b).  For this spectrum, the adopted 1$\sigma$
continuum placement error corresponds to 11\% of the continuum; thus, the
``high'' and ``low'' continuum levels were set at 1.11 and 0.89 times
the best fit continuum, respectively. The Voigt profiles that best fit
the data then give $N_{HI}^{high}=2.2\times 10^{20}$ atoms cm$^{-2}$
and $N_{HI}^{low}=1.8\times 10^{20}$ atoms cm$^{-2}$. Thus the column
density for this system is reported as $N_{HI}=(2.0\pm0.2)\times 10^{20}$
atoms cm$^{-2}$ (Table 3, see also \S2.5.7). We have found that the uncertainties
in column densities are generally on the order of 10\%; data noisier than
that for B2 0827+24 usually result in larger uncertainties while
higher signal-to-noise ratio data usually result in lower uncertainties.

Details of the results for each of the 14 identified DLA systems
are given below. In cases where a fit to the DLA profile is made, the resulting
redshift is reported to four decimal places.  However, as noted above,
the accuracy is more realistically $\pm$0.001 for determinations made
with the medium-resolution gratings and $\pm$0.002 for the low-resolution
gratings. \HI\ 21 cm absorption redshifts are adopted when possible.  

\subsubsection{The UM 366 (0143$-$015) $z=1.613$ DLA Line} 

No known \ion{Mg}{2} system was reported in the literature at this
redshift, but we discovered a DLA line at $z=1.613$ serendipitously
in our G270H spectrum. We observed this QSO as part of the survey
because it has two \ion{Mg}{2} absorption-line systems at $z=1.0383$
and $z=1.2853$ (neither is damped). Although no \ion{Mg}{2} information is
available for the $z=1.613$ system, Sargent, Steidel, \& Boksenberg (1989)
did detect \ion{Fe}{2} at this redshift. Eight percent of the continuum
had to be subtracted prior to fitting the profile in order to force the
spectrum to have zero flux at the DLA line center. Figure 3 shows the
normalized spectrum in the vicinity of the DLA line and the best-fit Voigt 
profile with $z=1.6128$ and $N_{HI}=2.0 \times 10^{20}$ atoms cm$^{-2}$ 
overlaid.

\subsubsection{The S4 0248$+$43 $z=0.3941$ DLA Line} 

We obtained an FOS-G160L spectrum of this object to study the \lya lines
of \ion{Mg}{2} systems at $z=0.3939$ and $z=0.4515$. The \lya line of
the $z=0.4515$ system is not damped.  However, the signal-to-noise ratio
in our FOS spectrum at the position of \lya in the $z=0.3939$ system
is very poor, and the nature of this \lya line could not be determined.
The FOS spectrum is shown in Figure 4.  However, as mentioned in \S2.4,
this system was convincingly found to exhibit \HI\ 21 cm absorption in
the radio continuum of the background quasar (Lane et al. 2000, in preparation). 
Three narrow components near $z=0.3941$ are present. They are clustered within 
a 40 km s$^{-1}$ velocity interval and have optical depths between 
15 and 20\%. Assuming that the 21 cm lines are optically thin, their strengths
suggest that the total column density is $N_{HI}=(5.1\pm0.6) \times
10^{18} T_s$ atoms cm$^{-2}$ K$^{-1}$ (Lane et al. 2000, in preparation).  
For a spin temperature of $T_s=700$ K, which is generally found to be the case
(Lane et al. 2000, in preparation), the resulting column density is 
$N_{HI}=(3.6\pm 0.4) \times 10^{21}$ atoms cm$^{-2}$. It is unfortunate 
that the spin temperature had to be assumed for the $N_{HI}$ determination.
 This DLA line should be reobserved in the UV
so that data on the DLA profile can be used to determine $N_{HI}$
independently.

\subsubsection{The EX 0302$-$223 $z=1.010$ DLA Line}

This system was taken from the archives. Originally, it was chosen as a
target for observation without knowledge of the fact that its spectrum
contains a DLA absorption line.  It has been studied by Pettini \&
Bowen (1997) and Boiss\'e et al. (1998).  A portion of the FOS-G270H
normalized spectrum, overlaid with our best-fit DLA profile, is shown
in Figure 5. Defining the level of the continuum
between $2450-2500$ \AA\ was problematic in making the fit
 because of the underlying
\ion{O}{6}/Ly$\beta$ emission line from the QSO. However, the core and
short-wavelength wing of the DLA profile fit the data well. We derive a
redshift of $z=1.0096$ and a column density of $N_{HI}=2.3 \times 10^{20}$
atoms cm$^{-2}$, in agreement with the values derived by Pettini \&
Bowen (1997) and Boiss\'e et al. (1998).

\subsubsection{The PKS 0454+039 $z=0.859$ DLA Line}

This system was also taken from the archives. Again, it was chosen as a
target for observation without knowledge of the fact that its spectrum
contains a DLA absorption line.  It has been studied by Steidel et al. (1995).
The normalized FOS-G190H spectrum in the vicinity of the DLA line, with our best-fit
Voigt profile overlaid, is shown in Figure 6.  We derive $N_{HI}=4.7
\times 10^{20}$ atoms cm$^{-2}$ at redshift $z=0.8591$ with an $\approx$
6\% error in $N_{HI}$.  Steidel et al. (1995) derived  $N_{HI}=5.7 \times
10^{20}$ atoms cm$^{-2}$ with an error of $\approx$ 5\%.  Our smaller
value for $N_{HI}$ results from the exclusion of the 2268 \AA\ absorption
feature (a likely Ly$\alpha$ forest line) from the fit, and agrees with
the value which was derived by Boiss\'e et al. (1998).

\subsubsection{The OI 363 (0738+313) $z=0.0912$ DLA Line}

We obtained an FOS-G160L spectrum of this object to study 
the \lya line in the $z=0.2212$ \ion{Mg}{2} system of OI 363
(0738+313) (see \S2.5.6) and found the DLA system at $z=0.0912$ serendipitously.
Since this is presently the lowest-redshift confirmed classical DLA
system known, we published the result in a separate contribution (Rao
\& Turnshek 1998) and the details of this system are given there. The
normalized spectrum and best-fit Voigt profile with redshift $z=0.0937$
and $N_{HI}=1.5 \times 10^{21}$ atoms cm$^{-2}$ are shown in Figure 7.
This system has also been found to have 21 cm absorption (Lane et
al. 1998b).

\subsubsection{The OI 363 (0738+313) $z=0.2212$ DLA Line}

The predicted position of the \lya line in the $z=0.2212$
system of OI 363 (0738+313) was observed as part of our survey. This
system is also found to be damped. It is the second lowest-redshift
confirmed classical DLA system known. The details of this system 
appeared in Rao \& Turnshek (1998).  The normalized spectrum and best-fit
Voigt profile with redshift $z=0.2224$ and $N_{HI}=7.9 \times 10^{20}$
atoms cm$^{-2}$ are shown in Figure 8.  This system has also been found
to have 21 cm absorption (Lane et al. 1998a).

\subsubsection{The B2 0827$+$24 $z=0.518$ DLA Line} 

We obtained an FOS-G160L spectrum of this object to study the \lya line
of the \ion{Mg}{2} system at $z=0.5247$. Figure 9 shows the
normalized spectrum in the vicinity of the DLA line and the best-fit
Voigt profile with column density $N_{HI}=2.0 \times 10^{20}$ atoms
cm$^{-2}$ and redshift $z=0.5176$. At column densities this low, it is
important to properly take into account the convolution of the Voigt
damping profile with the low-resolution G160L line spread function. For
this spectrum, we did not consider the possibility of subtracting a
fraction of the continuum to force the flux at the center of the DLA
line to zero because, based on the convolution of the damping profile
with the line spread function, some flux at line center is expected.
The discrepancy between our redshift and the Ulrich \& Owen (1977) 
\MgII redshift  is somewhat puzzling. We suspect
that either the zero point wavelength correction for our G160L spectrum
(which we ignored) was atypically large, or there was a wavelength
calibration problem in the original \MgII data.

\subsubsection{The TB 0933$+$733 $z=1.478$ DLA Line} 

According to SS92, two \ion{Mg}{2} systems are present in this object
at $z=1.4789$ and $z=1.4973$. We attempted to investigate \lya in both of
them by obtaining an FOS-G270H spectrum at the expected wavelengths
of Ly$\alpha$. Fortunately, the expected positions of the Ly$\beta$ and
Ly$\gamma$ absorption lines were also covered. The spectrum shows a
partial Lyman limit at 3060 \AA\ ($z\approx2.36$), which sets in just to
the long-wavelength side of the expected positions of the two \lya lines
at 3013 \AA\ ($z=1.4789$) and 3036 \AA\ ($z=1.4973$).  The 
spectrum of this object is shown in Figure 10, with the positions of the first
three Lyman series lines of both systems marked. The system marked
with subscript `1' corresponds to $z=1.4789$, while the one marked with
subscript `2' corresponds to $z=1.4973$. Note that the strong absorption
feature at $\approx$ 2770 \AA\ is a blend of \lya lines and is not damped.

Based on the appearance of the spectrum near \lya at $3000-3025$ \AA, at
least one of the \lya lines in the \ion{Mg}{2} systems at $z=1.4789$ and
$z=1.4973$ appears to be damped, although the interpretation is confused
by the Lyman limit system near $z\approx2.36$.  The determination of
which of these systems is damped is better resolved with the higher
order Lyman lines, particularly Ly$\beta$.  A normalized spectrum in
the vicinity of the damped Ly$\beta$ line, and its best-fit Voigt 
profile with $z=1.4784$ and $N_{HI}=4.2 \times 10^{21}$ atoms cm$^{-2}$ 
overlaid, is shown in Figure 11. We estimate
the uncertainty in the derived column density to be $\approx20$\%;
the high uncertainty reflects the low signal-to-noise ratio near the Ly$\beta$
 line. The same redshift and column density are consistent with the 
 absorption line at the position of Ly$\gamma$.
The Ly$\beta$ line corresponding to the $z=1.4973$ system is not damped.

Another interesting feature in this spectrum is the broad absorption 
depression between $\approx 2520$ \AA\ and $\approx 2600$ \AA.
However, this object is not a BAL QSO (Afanasev et al. 1990; SS92).  
The absorption complex includes both Ly$\beta$
lines at $z=1.4784$ and $z=1.4973$. Assuming that the entire absorption
complex is due to Ly$\beta$, then these wavelengths correspond to
the redshift range $1.457<z<1.535$.  If interpreted as a Hubble flow,
this corresponds to $\approx37h_{65}^{-1}$ Mpc along the line-of-sight
($q_0=0.5$). The extent of the absorption
is similar to what is seen in the spectra of the QSOs TOLOLO 1037-27
and 1038-27 (Jakobsen et al. 1986; Sargent \& Steidel 1987; Jakobsen
\& Perryman 1992), which is speculated to be absorbing material distributed
on a supercluster size scale. However, high-ionization metal-line
absorption (e.g. \ion{C}{4}) is present in the TOLOLO pair, but this
does not appear to be the case here. Unfortunately, given the proximity 
of the DLA line ($z=1.478$) and the Lyman limit edge at $z\approx2.36$,
a corresponding broad
feature at the predicted wavelengths of \lya between $1.457<z<1.535$ 
cannot be searched for to resolve the situation. On the other hand,
if the $2520-2600$ \AA\ feature is primarily due to \lya, and the two
Ly$\beta$ lines just happen to lie within this complex of absorption
lines, then the feature would extend between $1.073<z<1.139$ and it
would have an extent of $\approx 47h_{65}^{-1}$ Mpc ($q_0=0.5$).
 
\subsubsection{The PKS 0952$+$17 $z=0.239$ DLA Line} 

We obtained an FOS-G160L spectrum near the expected position of \lya
in the PKS 0952+17 $z=0.2377$ \ion{Mg}{2} system and find it to
be damped.  The normalized spectrum and our best-fit Voigt profile are
shown in Figure 12.  The fitting procedure yields $N_{HI}=2.1 \times
10^{21}$ atoms cm$^{-2}$ at $z=0.2394$.  Eleven percent of the continuum
level had to be subtracted from the spectrum prior to the fit in order
to force the line center flux to zero.

There are two other absorption lines (most likely \lya) on each side
of the DLA line, which is suggestive of clustering on a wide velocity
scale. If the \ion{Mg}{2} absorber lies within a structure that includes
these systems, this group would be an example of absorbing material on
a size-scale which exceeds a supercluster. Interpreting the velocity
interval as part of the Hubble flow would suggest that it extends
$\approx$ 92$h_{65}^{-1}$ Mpc along the line-of-sight ($q_0=0.5$).
Given the signal-to-noise ratio of the current data and the possibility
of a clustering of lines, we believe that it is important to reobserve
this \lya line with improved signal-to-noise ratio and resolution.
For example, if improved data show that this system is simply a unique
blend of many weak lines which resembles a DLA line (we consider this
possibility unlikely), it would eliminate the third lowest redshift DLA
system from our statistical sample. Moreover, it has a moderately large
column density, and thus plays an important role in our determination
of $\Omega_{DLA}$ at low redshift (\S4.3). Note that this absorber is
not detected in absorption at 21 cm (Lane et al. 2000, in preparation); however, it
would not be unique in this respect since other DLA lines in radio-loud
quasars have gone undetected in 21 cm absorption (e.g. Briggs \& Wolfe 1983).

\subsubsection{The Q0957+561A $z=1.391$ DLA Line}

In RTB95 we used {\it IUE} archival spectra to discount the $z=1.3911$
\ion{Mg}{2} system in this two-component gravitational lens as having
a classical DLA line.  Turnshek \& Bohlin (1993) measured its column
density using a co-addition of many low-resolution {\it IUE} spectra
and obtained $N_{HI}=(7\pm 2.3)\times10^{19}$ atoms cm$^{-2}$. However,
with the higher-resolution {\it HST}-FOS-G270H spectra now available
for each component, we have remeasured the \HI\ column density of this
system by fitting a Voigt profile to the DLA lines in each component
of the lens. The system in component B has $N_{HI}=7\times10^{19}$
atoms cm$^{-2}$ and thus, would not enter into our sample. However,
we derive a column density of $N_{HI}=2.1\times10^{20}$ atoms cm$^{-2}$
for the system in component A, so it does enter into our sample, providing
gravitational lensing does not produce a bias (see below). (Note that if
the system in component A and component B had both met the minimum column
density criterion, this would only have been counted as one system.)
In performing the fits, we had to account for the Ly$\alpha$ emission
line from the QSO since it overlaps the long-wavelength wing of the
DLA line. Part of the normalized spectrum of component A is shown in
Figure 13 along with the best-fit DLA profile.  Prior to normalizing
the spectrum fit, we had to subtract 2.4\% of the continuum level
from the spectrum to force the center of the DLA to have zero flux.
The DLA line and the emission lines, Ly$\alpha$ and \ion{N}{5}, were
fitted simultaneously. The emission lines were fitted with the sum of
four Gaussians added on to a continuum level of unity; the sum was
then attenuated by a Voigt profile at the position of the DLA line.
Each Gaussian profile was specified using three parameters: wavelength,
amplitude, and FWHM. The number of parameters in the fit
is large, and the best-fit parameters for the emission line profiles are
not unique. Therefore, we ran a number of fits to span a reasonable range
of emission-line parameters and determined the best-fit Voigt profile
for each case after convolving the spectrum with a Gaussian whose FWHM
is given by the resolution of the G270H grating. As shown in Figure 14,
a column density $N_{HI}=1.6\times10^{20}$ atoms cm$^{-2}$ was clearly too
low while $N_{HI}=2.6\times10^{20}$ atoms cm$^{-2}$ was too high, with
$N_{HI}=2.1\times10^{20}$ atoms cm$^{-2}$ being the best fit.  Thus, in
Table 3 we report the column density as $N_{HI}=(2.1\pm0.5)\times10^{20}$
atoms cm$^{-2}$ at absorption redshift $z=1.3907$.  Small variations in
the emission line parameters had little affect on the column density
determination. In fact, inclusion of the fourth Gaussian at 2951 \AA,
which was necessary to obtain a better fit to the emission line profile,
had no affect on the column density determination because its relative
amplitude was low and its wavelength was far removed from the DLA
line. Our results are in agreement with the measurements of Zuo et
al. (1997) who derive $N_{HI}=1.9\pm0.3\times10^{20}$ atoms cm$^{-2}$
for component A.

One should be careful when selecting a gravitationally lensed
QSO for inclusion in a DLA absorption-line survey which makes use of a
magnitude limited sample of QSOs. This is because magnification of  
gravitationally lensed QSOs might introduce them into samples that they
would not have otherwise entered.  We discuss this further in \S6.2.2
(see also Smette, Claeskens, \& Surdej 1997).  Our survey is susceptible
to this bias for two reasons: (1) the archival part of the survey included
data taken with the {\it IUE} which could observe only the brightest
QSOs and (2) our {\it HST}-FOS observations were chosen to maximize the
signal-to-noise ratio in one {\it HST} orbit (see Figure 1). Thus, we must
exclude any gravitationally lensed QSO if it is determined that it would
not have been observed with either of these instruments had it not been
lensed. Lensing models show that component A ($V=17$) has been magnified
by a factor of 5.2 (Chae 1999), i.e., the unlensed QSO has a magnitude
of 18.8. However, since Figure 1 shows that a QSO of magnitude 18.8 with 
an absorption system at $z=1.3911$ would still have been included as a 
target for observation with {\it HST}, we have retained this system in our 
sample.

\subsubsection{The PKS 1127$-$14 $z=0.3127$ DLA Line} 

We obtained an FOS-G160L spectrum of this quasar to study the \lya
line of the $z=0.3130$ \ion{Mg}{2} system and found that the \lya line
is damped.  Figure 15 shows the normalized spectrum in the vicinity of the
DLA line and the best-fit Voigt profile with column density $N_{HI}=5.1
\times 10^{21}$ atoms cm$^{-2}$ and redshift $z=0.3127$.  Four percent
of the continuum level was subtracted from the spectrum before the fit
to force the DLA line center to have zero flux. This system is a 21 cm 
absorber (Lane et al. 1998a).

\subsubsection{The Q1209+107 $z=0.633$ DLA Line}

This system was taken from the archives. It was one of the six new
\ion{Mg}{2} systems added to the archives after the initial study
of RTB95.  Originally, it was chosen as a target for observation
without knowledge of the fact that its spectrum contains a DLA
absorption line.  A portion of the normalized FOS-G160L spectrum of
this object, with the best-fit Voigt profile overlaid, is shown in
Figure 16.   The poor signal-to-noise ratio ($\approx
2$) near the position of the DLA line ($\approx$ 1975 \AA) leads to
a rather large (50\%) uncertainty in the measured column density
(Table 3). We derive $N_{HI}=2.0 \times 10^{20}$ atoms cm$^{-2}$
at $z=0.6325$. This result is consistent with that of Boiss\'e et
al. (1998).

\subsubsection{The PKS 1354+25 $z=1.418$ DLA Line}

We obtained an FOS-G270H spectrum of this quasar to study the \lya
line of the $z=1.4205$ \ion{Mg}{2} system and found that the \lya line
is damped.  Figure 17 shows the normalized spectrum in the vicinity of the
DLA line and the best-fit Voigt profile with column density $N_{HI}=3.2
\times 10^{21}$ atoms cm$^{-2}$ and redshift $z=1.4179$. Four percent
of the continuum level was subtracted from the spectrum before the fit
to force the DLA line center to have zero flux. 

\subsubsection{The 3C 336 (1622+239) $z=0.656$ DLA Line}

This system was taken from the archives.  Steidel et al. (1997) studied
the absorption-line systems toward this quasar in considerable detail.
However, originally it was chosen as a target for observation without
knowledge of the fact that its spectrum contains a DLA absorption-line. A
portion of the normalized FOS-G190H spectrum of this object and our best-fit
Voigt profile is shown in Figure 18. We derive $z=0.6555$ and $N_{HI}=2.3
\times 10^{20}$ atoms cm$^{-2}$ for the DLA system. This is consistent
with the result found by Steidel et al. (1997) to within our quoted
error (Table 3).  

\subsection{The PG 0117+213 $z=0.576$ System}

As noted in \S2.2, the recalibrated {\it HST}-FOS spectropolarimetric data
on PG 0117$+$213 has been added to the archival portion of our sample
since it covers the predicted wavelength of \lya in the $z=0.5761$
\ion{Mg}{2} system. The data were kindly provided to us by A.
Koratkar.  The spectrum is shown in Figure 19. We find that the strong
absorption feature at $\approx$ 1920 \AA, which appears as a candidate DLA
line in figure 3 of Koratkar et al. (1998), is more convincingly fitted
with a blend of 9 Gaussian components. A Voigt damping profile with column 
density $N_{HI}=3.0 \times 10^{20}$ atoms cm$^{-2}$ is overplotted to
illustrate that a single damping profile is a poor fit to this complex
of absorption. Therefore, we have discounted
it as a DLA line. In this system, the Ly$\alpha$ absorption line which
corresponds to the published \ion{Mg}{2} absorption redshift lies at
1917 \AA. It is the strongest line in the complex blend of absorption
lines. This blended feature is evidently another example of correlated
structure on a large (supercluster) scale. Assuming a cosmological Hubble
flow, the components at the two extremes of the blend would be separated
by $\approx34h^{-1}_{65}$ Mpc ($q_0=0.5$) along the line-of-sight.

\begin{figure*}
\plotone{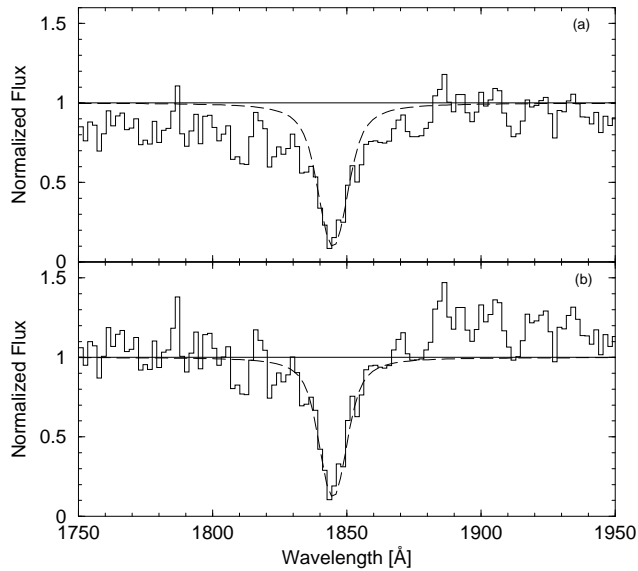}
\caption{This figure illustrates the technique used to estimate uncertainties
in $N_{HI}$. The largest source of uncertainty in measuring $N_{HI}$ comes
from continuum placement in spectra with low signal-to-noise ratios. Best 
fit profiles to the damped line in the spectrum of B2$\;$0827$+$24 are shown here,
but with continua set at levels equal to $1\sigma$ above (panel {\it a}) and below
(panel {\it b}) the most likely continuum level near the position of the damped 
line (Figure 9 shows the fit with the most likely continuum level). For this
spectrum, $1\sigma=11$\ of the most likely continuum, and the column
densities derived from the fits are $N_{HI}^{high}=2.2\times 10^{20}$ atoms 
cm$^{-2}$ from panel ($a$) and $N_{HI}^{low}=1.8\times 10^{20}$ atoms cm$^{-2}$ 
from panel ($b$).}
\end{figure*}

\begin{figure*}
\plotone{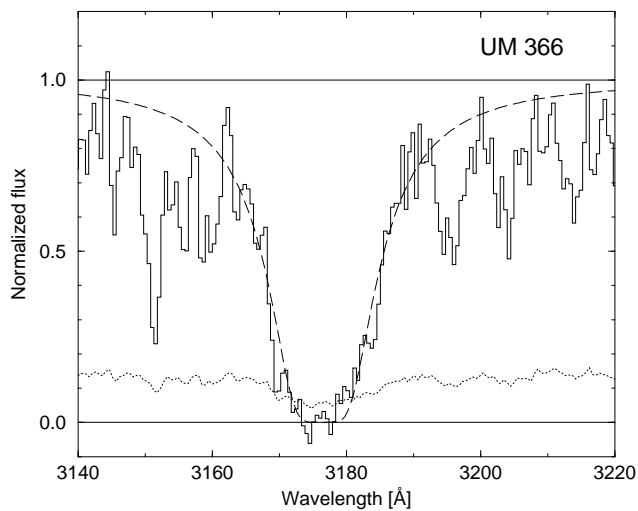}
\caption{Voigt profile fit to the DLA line in the normalized spectrum of UM 366
(0143$-$015) with $z=1.6128$ and $N_{HI}=2.0 \times 10^{20}$ atoms cm$^{-2}$.
The 1$\sigma$ error spectrum is shown as the dotted line.}
\end{figure*}

\begin{figure*}
\plotone{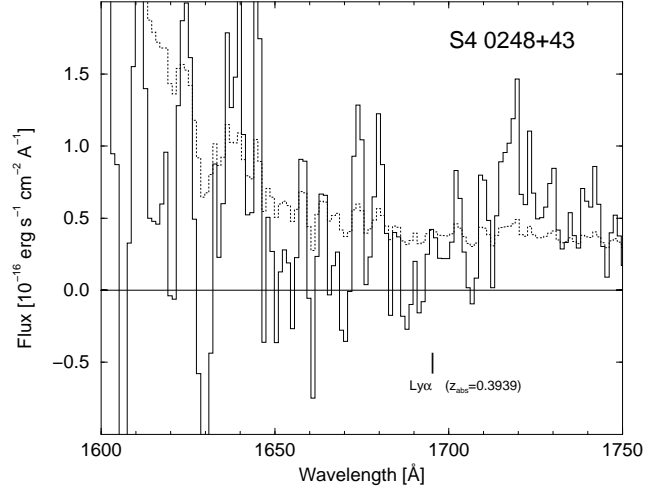}
\caption{A section of the unusable FOS-G160L spectrum of S4 0248$+$43 is shown
with the position of the Ly$\alpha$ line corresponding to the $z_{abs}=0.3939$
Mg II system marked. The $1\sigma$ error spectrum is shown as the dotted
line.}
\end{figure*}

\begin{figure*}
\plotone{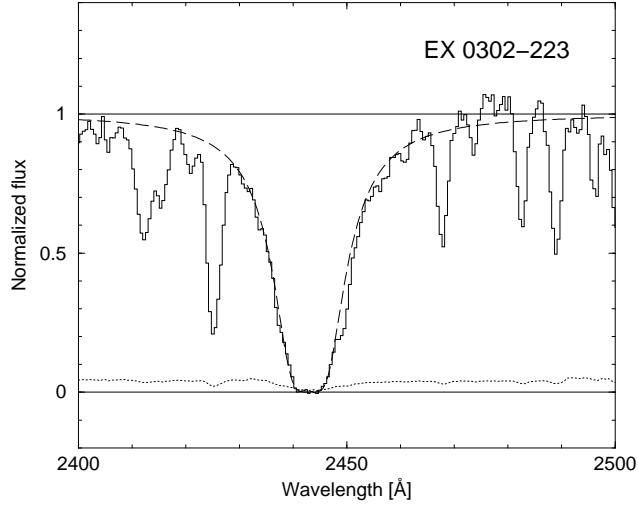}
\caption{Voigt profile fit to the DLA line in the normalized spectrum of 
EX 0302$-$223
with $z=1.0096$ and $N_{HI}=2.3 \times 10^{20}$ atoms cm$^{-2}$.
The 1$\sigma$ error spectrum is shown as the dotted line.}
\end{figure*}

\begin{figure*}
\plotone{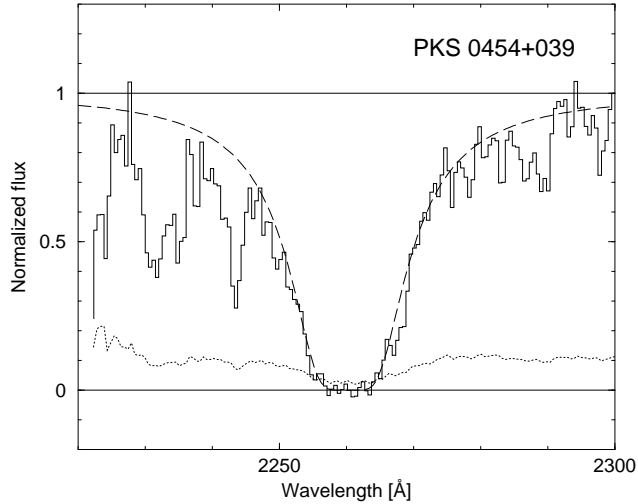}
\caption{Voigt profile fit to the DLA line in the normalized spectrum of  
PKS 0454$+$039 with $z=0.8591$ and $N_{HI}=4.7 \times 10^{20}$ atoms cm$^{-2}$.
The 1$\sigma$ error spectrum is shown as the dotted line.}
\end{figure*}

\begin{figure*}
\plotone{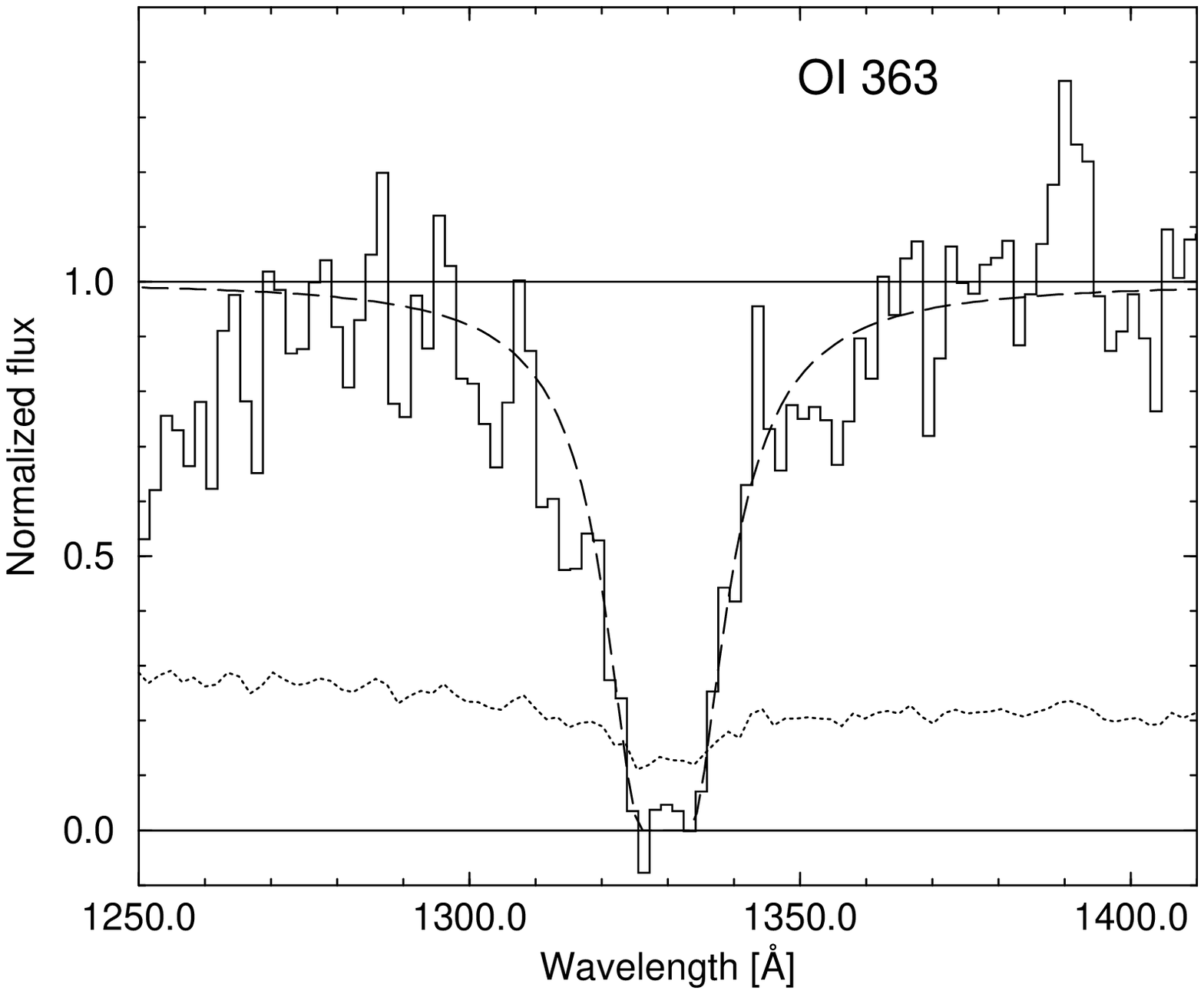}
\caption{Voigt profile fit to the DLA line in the normalized spectrum of OI 363 
(0738$+$313) with $z=0.0937$ and $N_{HI}=1.5 \times 10^{21}$ atoms cm$^{-2}$.
The 1$\sigma$ error spectrum is shown as the dotted line.}
\end{figure*}

\begin{figure*}
\plotone{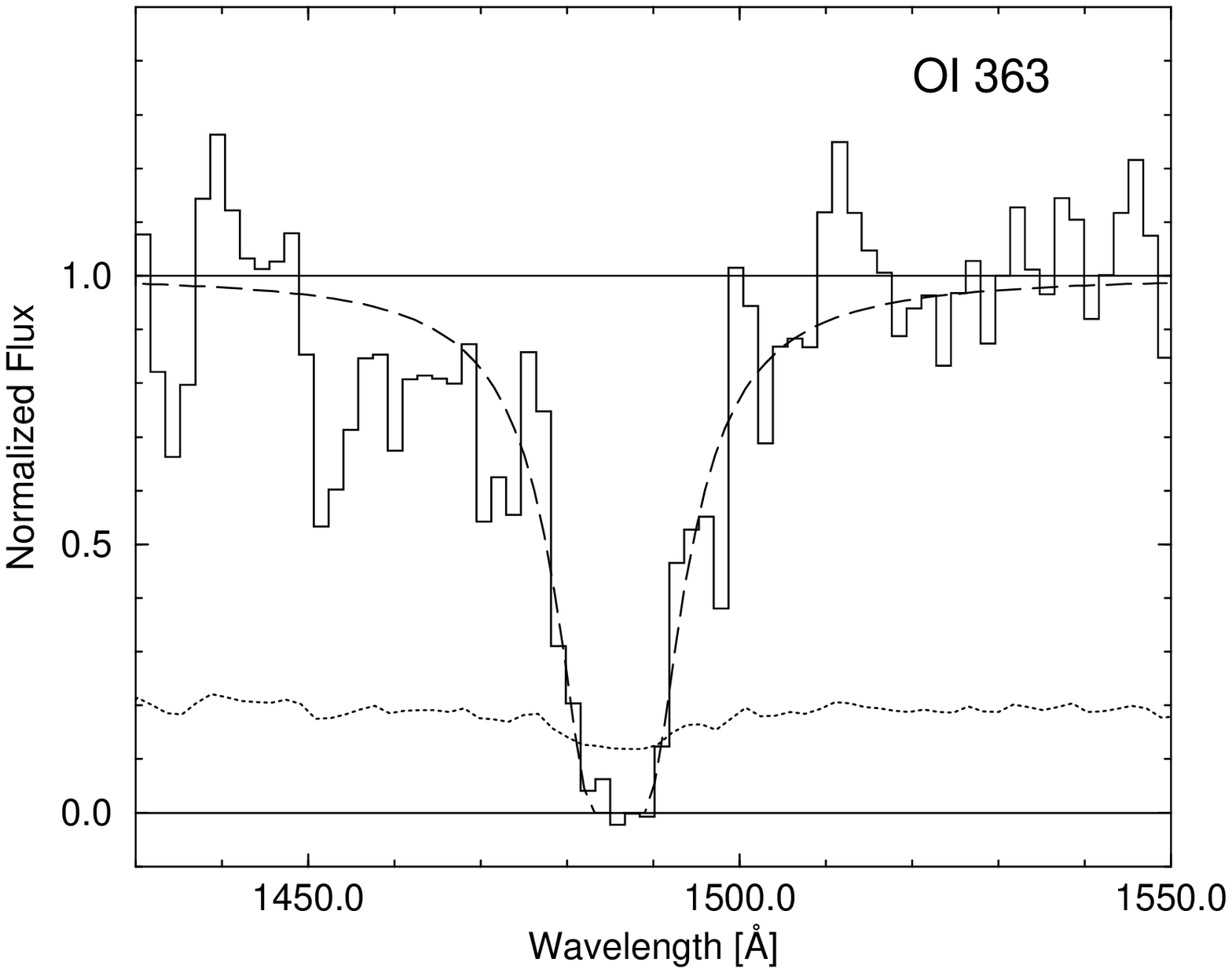}
\caption{Voigt profile fit to the DLA line in the normalized spectrum of OI 363 
(0738$+$313) with $z=0.2224$ and $N_{HI}=7.9 \times 10^{20}$ atoms cm$^{-2}$.
The 1$\sigma$ error spectrum is shown as the dotted line.}
\end{figure*}

\begin{figure*}
\plotone{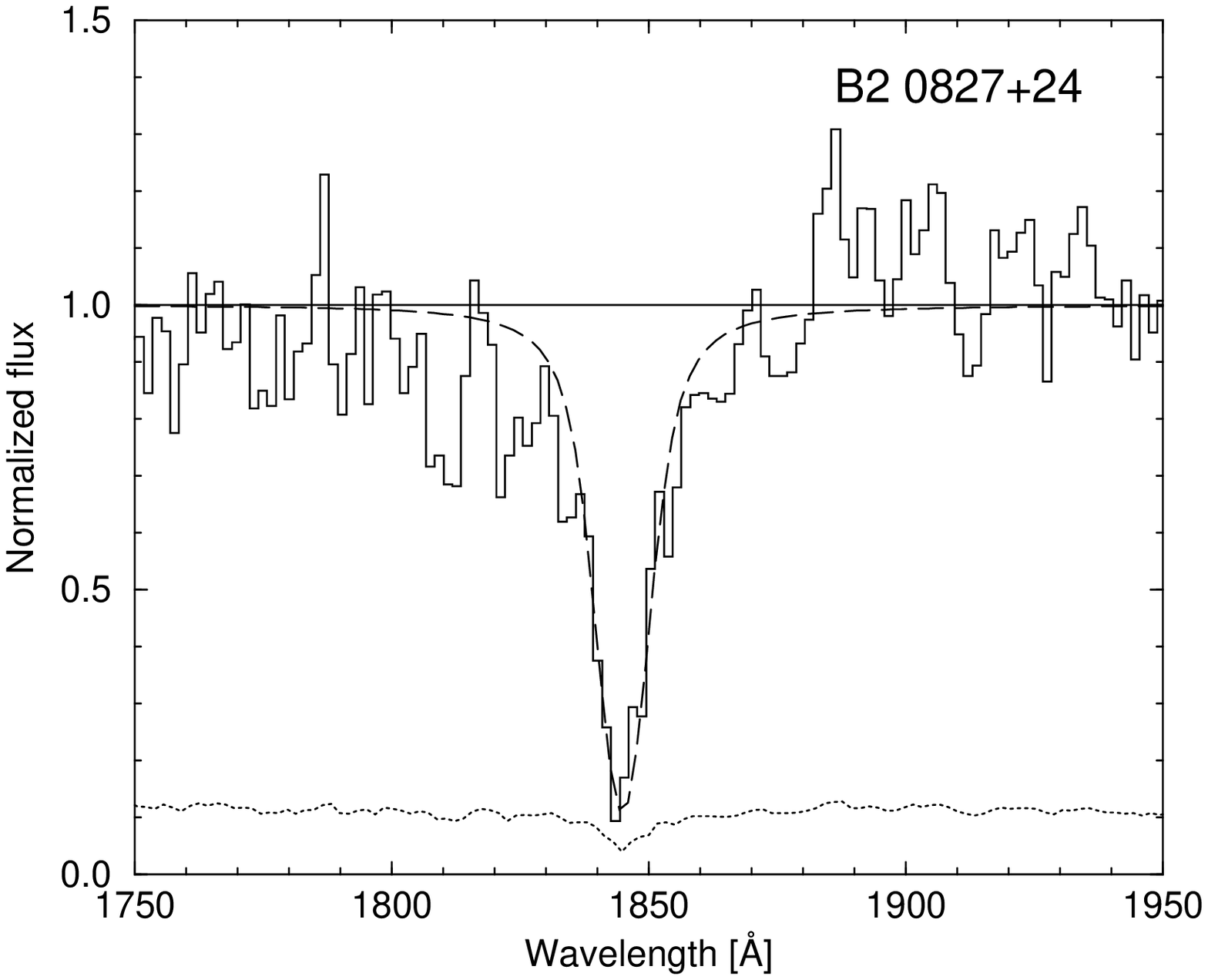}
\caption{Voigt profile fit to the DLA line in the normalized spectrum of 
B2 0827$+$24 with $z=0.5176$ and $N_{HI}=2.0 \times 10^{20}$ atoms cm$^{-2}$.
The 1$\sigma$ error spectrum is shown as the dotted line.}
\end{figure*}

\newpage

\begin{figure*}
\epsscale{1.5}
\plotone{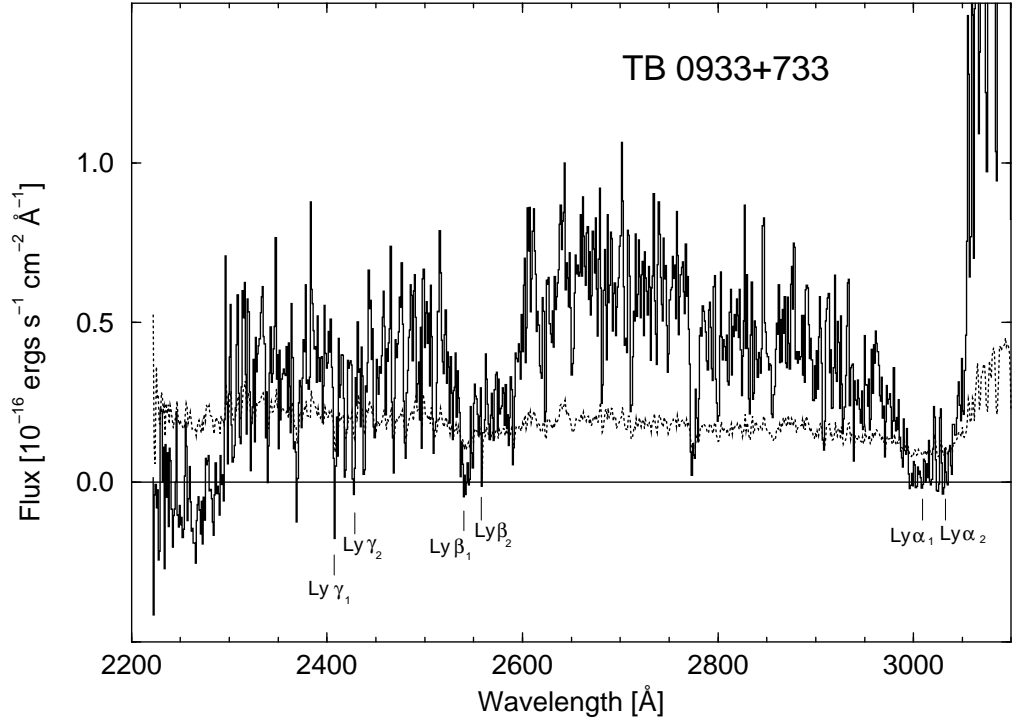}
\caption{The $HST$-FOS G270H spectrum of TB 0933$+$733. 
The 1$\sigma$ error spectrum is shown as the dotted line. Lyman limit systems
are seen at 2300 \AA\ ($z=1.522$) and at 3060 \AA\ ($z=2.255$). The positions
of the Lyman series lines of the two Mg II absorption systems at (1) $z=1.4787$
and (2) $z=1.4973$ are marked. The absorption line at 2770 \AA\ is a blend. 
Absorption in the form of a broad 
trough extending between 2520 \AA\ and 2600 \AA\ can also
be seen; it suggests correlated structures on supercluster size scales (see
text).}
\end{figure*}

\begin{figure*}
\epsscale{1.0}
\plotone{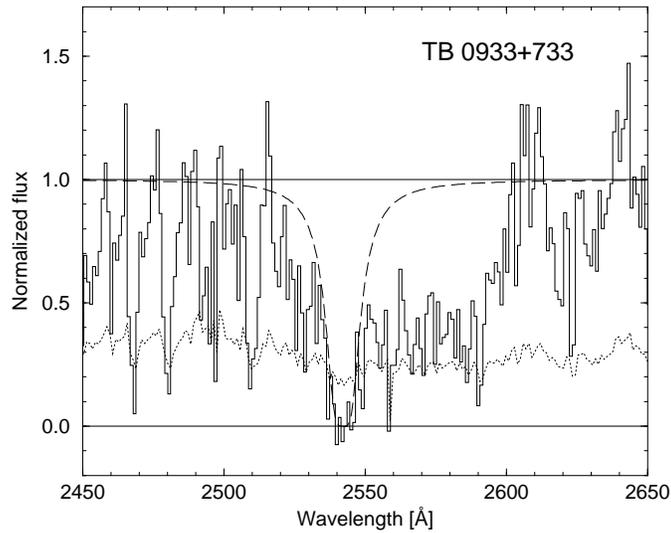}
\caption{Voigt profile fit to the damped Ly$\beta$
line in the normalized spectrum of TB 0933$+$733
with $z=1.4784$ and $N_{HI}=4.2 \times 10^{21}$ atoms cm$^{-2}$.
The 1$\sigma$ error spectrum is shown as the dotted line.}
\end{figure*}

\begin{figure*}
\plotone{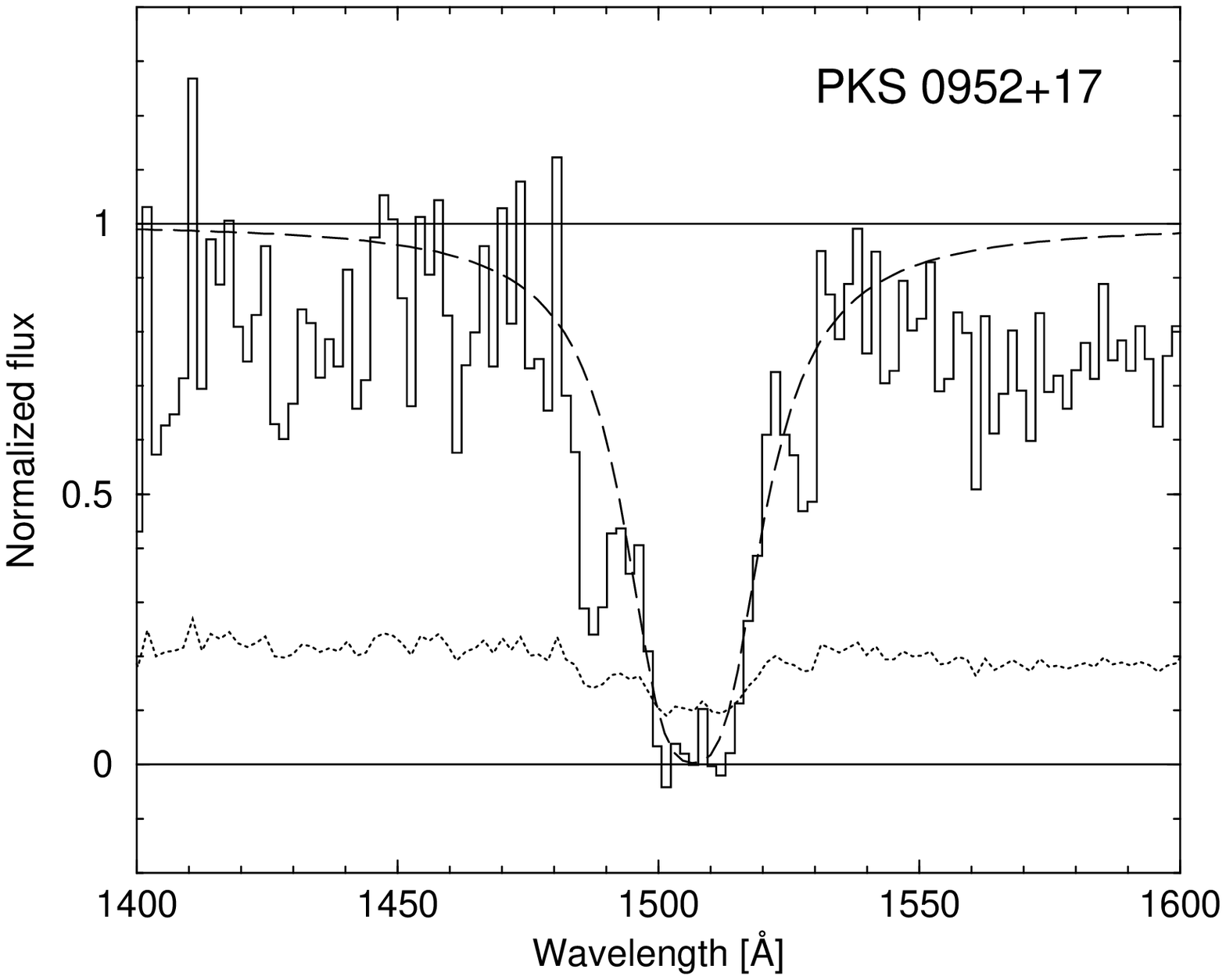}
\caption{Voigt profile fit to the DLA line in the normalized spectrum of  
PKS 0952$+$17 with $z=0.2394$ and $N_{HI}=2.1 \times 10^{21}$ atoms cm$^{-2}$.
The 1$\sigma$ error spectrum is shown as the dotted line.}
\end{figure*}

\begin{figure*}
\plotone{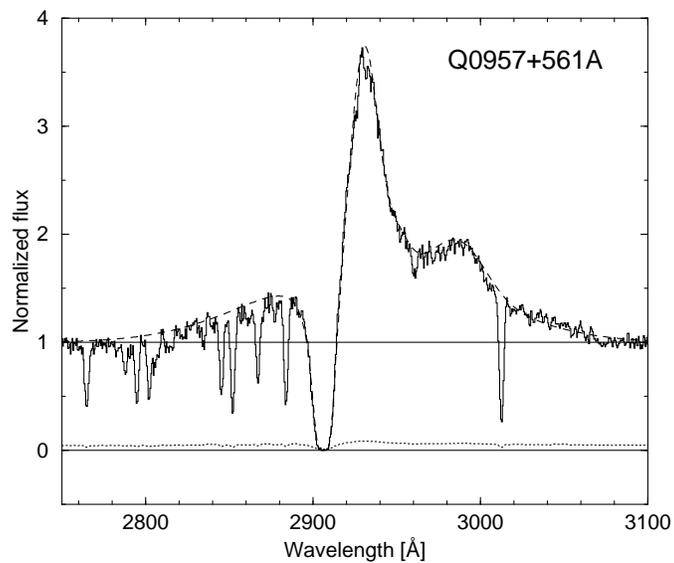}
\caption{Composite profile fit to the emission lines Ly$\alpha$ and N V,
and the DLA absorption line in the spectrum of Q0957+561A. The emission lines
are best fit by a sum of 4 Gaussians added on to a continuum level of unity.
The best fit Voigt profile  has $z=1.3907$ and $N_{HI}=2.1 \times 10^{20}$ 
atoms cm$^{-2}$. The 1$\sigma$ error spectrum is shown as the dotted line.}
\end{figure*}

\begin{figure*}
\plotone{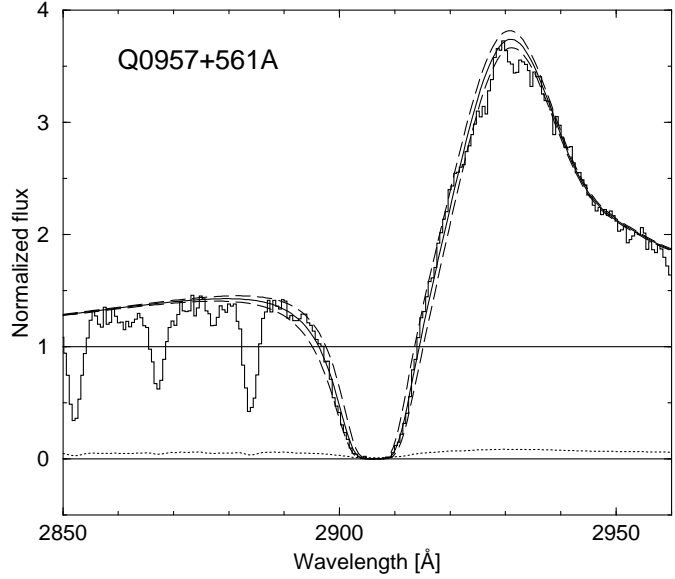}
\caption{The DLA absorption line in the spectrum of Q0957$+$561A with the 
best fit composite profile shown as the solid line 
($N_{HI}=2.1\times 10^{20}$ atoms cm$^{-2}$). The dashed lines show that a
column density of
$N_{HI}=1.6\times 10^{20}$ atoms cm$^{-2}$ (upper dashed line) is 
clearly too low and $N_{HI}=2.6\times 10^{20}$ atoms cm$^{-2}$ (lower dashed
line) is clearly too high. We report the column density of this system
as $N_{HI}=(2.1 \pm 0.5) \times 10^{20}$ atoms cm$^{-2}$.
The 1$\sigma$ error spectrum is shown as the dotted line.}
\end{figure*}

\begin{figure*}
\plotone{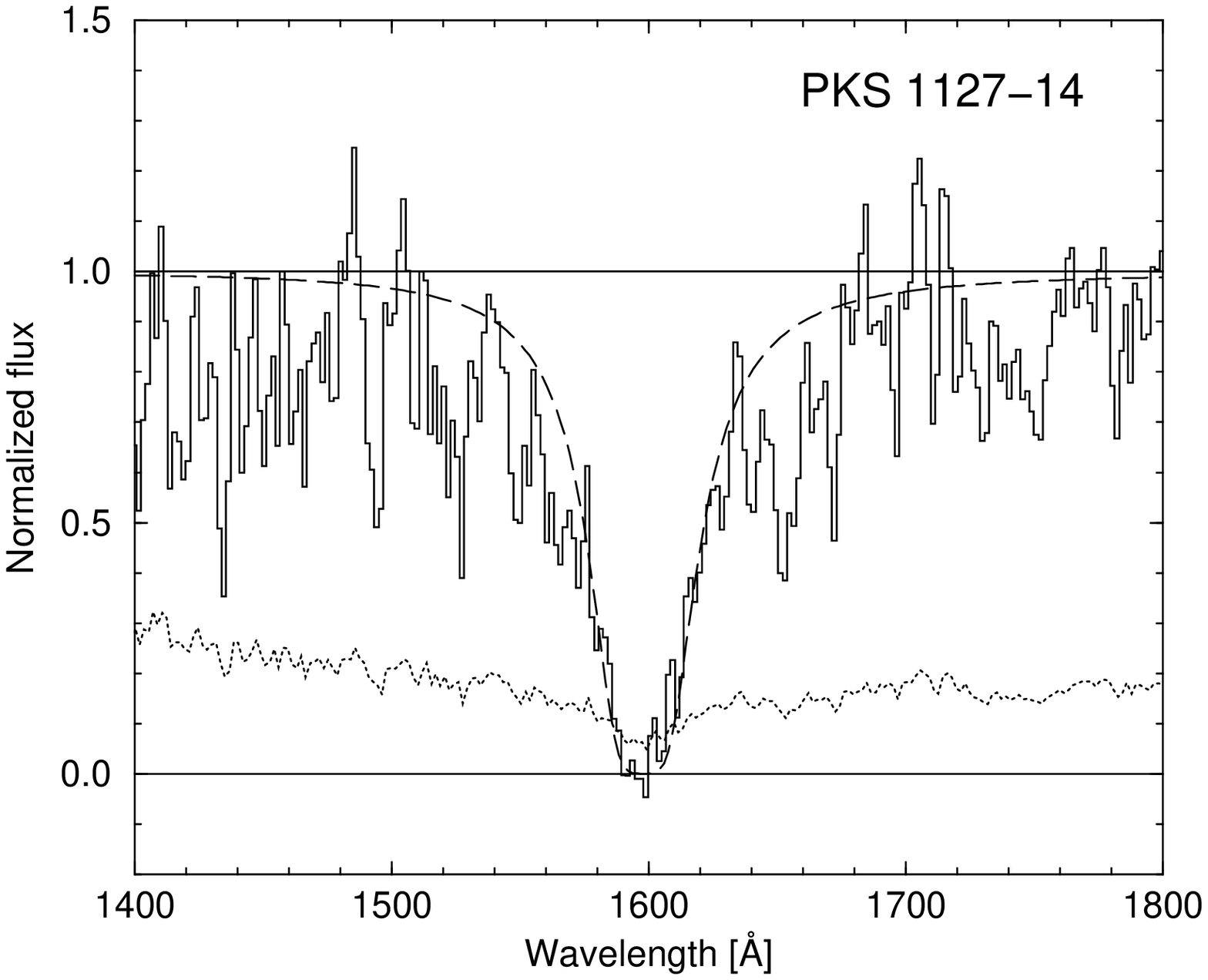}
\caption{Voigt profile fit to the DLA line in the normalized spectrum of  
PKS 1127$-$14 with $z=0.3127$ and $N_{HI}=5.1 \times 10^{21}$ atoms cm$^{-2}$.
The 1$\sigma$ error spectrum is shown as the dotted line.}
\end{figure*}

\begin{figure*}
\plotone{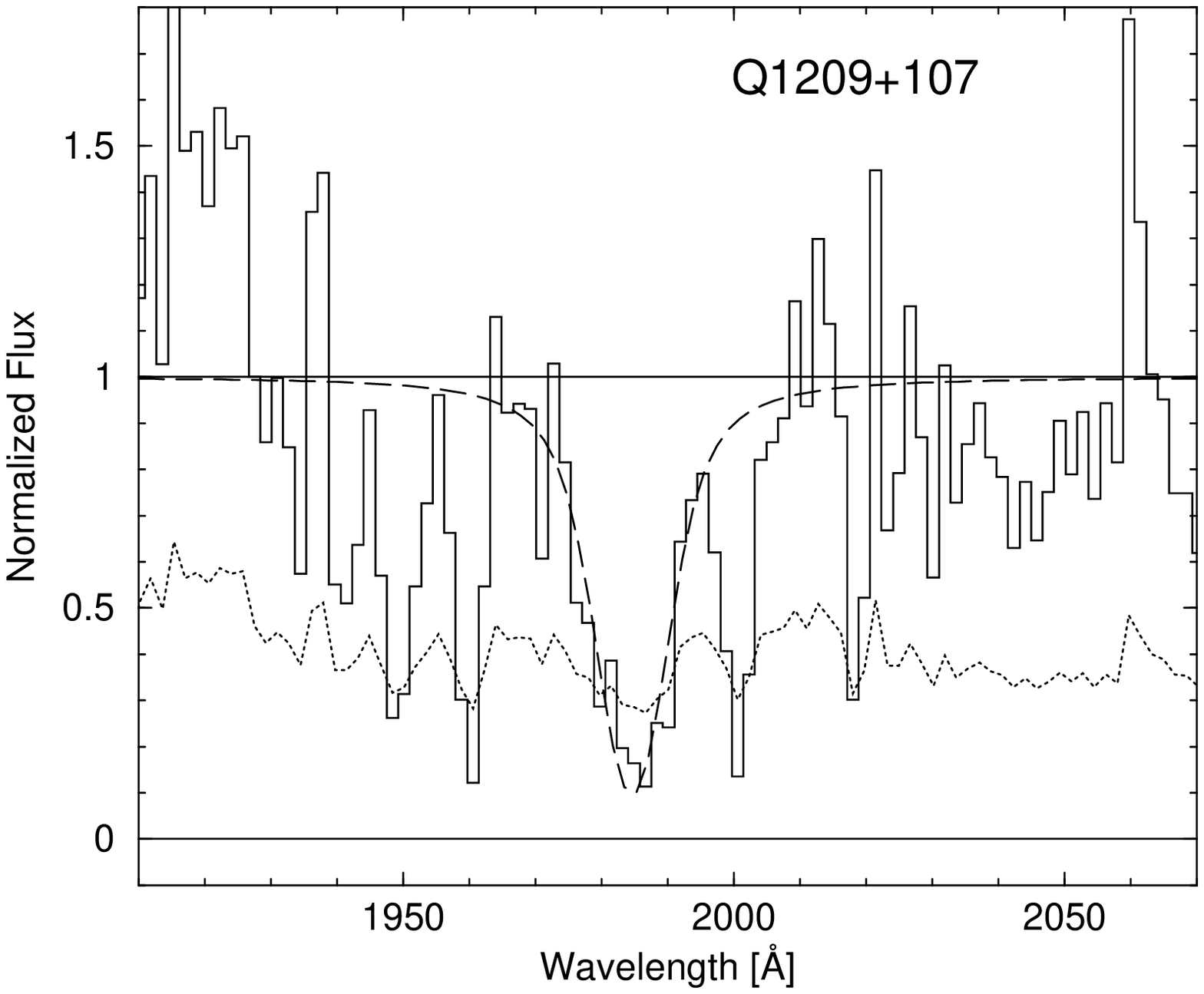}
\caption{Voigt profile fit to the DLA line in the normalized spectrum of  
Q1209+107 with $z=0.6325$ and $N_{HI}=2.0\times 10^{20}$ atoms cm$^{-2}$.
The 1$\sigma$ error spectrum is shown as the dotted line.}
\end{figure*}

\begin{figure*}
\plotone{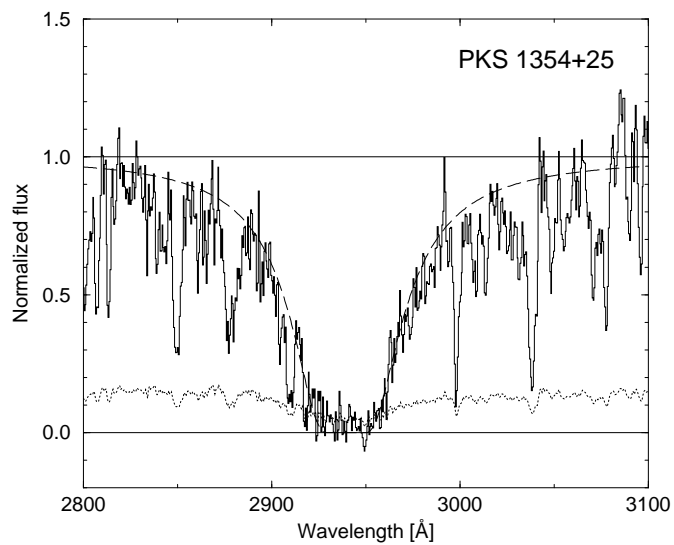}
\caption{Voigt profile fit to the DLA line in the normalized spectrum of  
PKS 1354+25 with $z=1.4179$ and $N_{HI}=3.2 \times 10^{21}$ atoms cm$^{-2}$.
The 1$\sigma$ error spectrum is shown as the dotted line.}
\end{figure*}

\begin{figure*}
\plotone{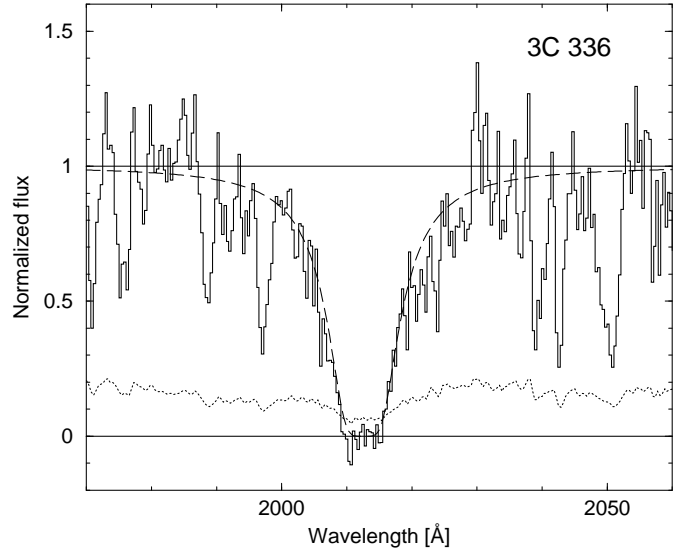}
\caption{Voigt profile fit to the DLA line in the normalized spectrum of  
3C 336 (1622+239) with $z=0.6555$ and $N_{HI}=2.3\times 10^{20}$ atoms cm$^{-2}$.
The 1$\sigma$ error spectrum is shown as the dotted line.}
\end{figure*}

\begin{figure*}   
\plotone{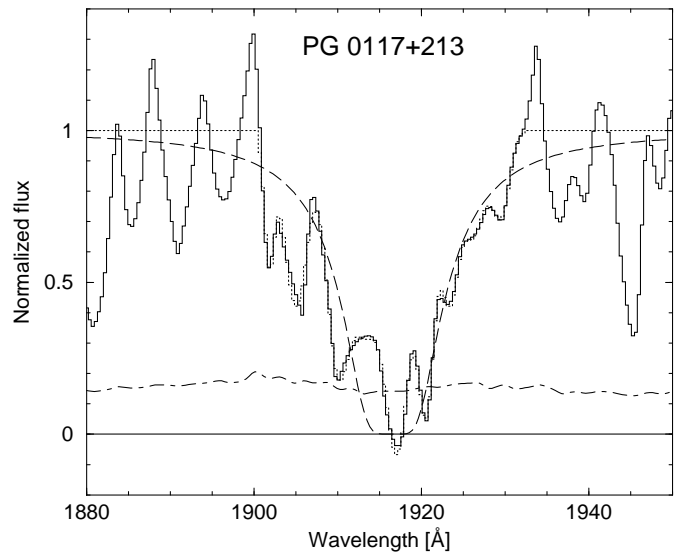}
\caption{The strong absorption feature in the $HST$-FOS G190H spectrum of 
PG 0117$+$213 is fit with a blend of 9 Gaussian components (dotted line).
A single Voigt damping profile with $N_{HI} = 3.0 \times 10^{20}$ atoms
cm$^{-2}$ is overplotted as the dashed line and is a poor fit to the data. 
The deepest component is at the redshift of the Mg II system and has rest 
equivalent width $W_0^{\lambda1216}=2.7$ \AA.
The 1$\sigma$ error spectrum is shown as the dot-dashed line.}
\end{figure*}

\clearpage

\section{Statistical Properties of the \MgII Sample}

Table 4 lists the results on the detection of DLA in the 87 \MgII
systems which make up our low-redshift sample.  We call this sample the
RT sample to distinguish it from the SS92 \MgII sample.  Column 1 of
Table 4 gives the QSO's 1950 coordinate designation; 
column 2 gives the absorption redshift;
columns $3-6$ give the available rest equivalent widths (or upper limits)
of the \MgII$\lambda2796$, \MgII$\lambda2803$, 
\MgI$\lambda2852$, and \FeII$\lambda2600$ absorption lines, 
respectively; column 7 gives the references for the absorption-line data;
and column 8 gives the column density of the classical DLA absorption 
line if present in the system.

In RTB95, we discussed the rationale behind using a sample of \ion{Mg}{2}
systems with a boot-strapping method to determine the statistical
properties of low-redshift DLA systems. Inherent in this method is
the reliance on the empirically determined statistical properties of
\ion{Mg}{2} systems by SS92 and the assumption that the \ion{Mg}{2}
samples of SS92 and RT represent the same parent population of absorbers
for $W_0^{\lambda2796} \ge 0.3$ \AA. Therefore, to justify the results
which we  present in \S4, we discuss the statistical properties
of the SS92 and RT samples and show that they are equivalent (\S3.1).
We also discuss the properties of classical DLA absorbers in the RT
sample and show that they are primarily drawn from the population of \MgII
absorbers with $W_0^{\lambda2796} \ge 0.6$ \AA\ (also \S3.1).  We consider
the possible presence of trends between the existence of classical DLA
absorption and: (1) the \MgII absorption-line doublet ratio (\S3.1.1),
(2) \FeII$\lambda2600$ absorption (\S3.1.2), (3) and \ion{Mg}{1} $\lambda2852$
absorption (\S3.1.3).  Finally, we derive the effective  
redshift path covered in our survey (\S3.2).

\subsection{The \MgII$\lambda2796$ Rest Equivalent Width Distribution}

Figure 20 shows the \MgII$\lambda2796$ rest equivalent width
(\w) distributions for the SS92 sample ($N_{SS92}$),
the RT sample ($N_{RT}$), and the DLA subsample of the RT sample
($N_{RT-DLA}$).  The SS92 \MgII data have been scaled to the RT
\MgII data. It is clear that the SS92 and RT \w\ distributions
are very similar; this is demonstrated more rigorously below.
It is also striking that while 43 of the 87 RT systems have $0.3 \le
W_0^{\lambda2796} < 0.6$ \AA, only one of the 12 DLA systems falls in
this interval.  The remaining 11 DLA systems have $W_0^{\lambda2796}
\ge 0.6$ \AA. Therefore, the probability of finding a DLA system in a
\ion{Mg}{2} sample with $W_0^{\lambda2796}\ge0.3$ \AA\ is 
12/87 $=$ 13.8\%, but the probability becomes  11/44 $=$
25.0\% if the \w\ threshold is increased to \w$\ge 0.6$ \AA.  

The normalized cumulative \w\ distribution functions for the three
above samples (SS92, RT, and RT-DLA) are shown in Figures 21 and 22
for the $W_0^{\lambda2796}\ge0.3$ \AA\ and $W_0^{\lambda2796}\ge0.6$
\AA\ thresholds, respectively. The maximum deviations between these
distributions have been used in a Kolmogorov-Smirnov (KS) test to
determine the probability that any two of the samples are drawn from
the same parent population.  The results of the KS tests are given in
Table 5.  The SS92 and RT samples are completely consistent with being
drawn from the same parent population of absorbers for \w $\ge 0.3$
\AA; and the samples are nearly identical for \w $\ge 0.6$ \AA. This
validates our assumption that the statistical properties of the SS92
\MgII sample are applicable to the RT \MgII sample.  On the other hand,
the KS statistic shows that there is only a 2\% probability
that the RT-DLA subsample is drawn from the same parent population as
the $W_0^{\lambda2796}\ge0.3$ \AA\ RT sample, however the probability
increases to 36\% for the $W_0^{\lambda2796}\ge0.6$ \AA\
RT sample.  Aside from this   threshold
correlation (the tendency for $N_{HI} \ge 2 \times 10^{20}$ atoms
cm$^{-2}$ systems to be present when \w $\ge 0.6$ \AA), Figure
23 shows that there is no further trend between $N_{HI}$ and \w\
above these thresholds.  Nevertheless, we suspect that this threshold
correlation is revealing something fundamental about the absorbers that
give rise to classical DLA absorption-line systems.

Ionization equilibrium studies have shown that QSO absorbers must be
optically thick at the Lyman limit ($N_{HI} > 3 \times 10^{17}$ atoms
cm$^{-2}$) for appreciable amounts of Mg$^{+}$ to be present (Bergeron \&
Stasi\'nska 1986).  This is because Mg$^{+}$ will only survive the presence
of a strong background radiation field if it is embedded in an optically
thick \HI\ cloud which will shield it from the ionizing radiation. This
prediction has been verified since it has also been empirically shown that
\MgII and Lyman limit systems arise in the same population of absorbers
(SS92 and references therein). Thus, it is understood that $N_{HI} >
3 \times 10^{17}$ atoms cm$^{-2}$ is generally necessary for the detection 
of a \MgII system\footnote{However, this may not be true for the population 
of weak \MgII absorbers with \w$<0.3$\AA\ (Churchill et al. 1999a).}.  
Our above empirical analysis suggests that, in addition,
\w $\ge 0.6$ \AA\ is almost always necessary (but not sufficient)
to have $N_{HI} \ge 2 \times 10^{20}$ atoms cm$^{-2}$. {\it Why should
this be the case?} Before we discuss this (\S3.1.4), we wish to note the 
empirical results from several other comparisons which we have considered 
in an attempt to understand the nature of the DLA population.

\subsubsection{The \ion{Mg}{2} Doublet Ratio and DLA Absorption}

Figure 24 is a plot of the \MgII doublet ratio,
\w/$W_0^{\lambda2803}$, versus \w\ for the RT 
sample. The DLA systems from the RT sample are marked with open squares.
We have also plotted measurements from the literature
for the four 21 cm absorbers that we   
excluded from our unbiased \MgII sample (\S2.4). Although we have not
included these previously known 21 cm absorbers
 in the statistical analysis of low-redshift DLA 
systems, their properties may be relevant for assessing the nature of
the low-redshift DLA population.  The horizontal dashed line in the figure
has \w/$W_0^{\lambda2803} = 1.5$; all of the DLA systems
have \MgII doublet ratios  that fall below this line indicating that 
their \MgII lines lie somewhat close to the saturated part of the curve 
of growth. However, the \MgII doublet ratio exceeds 1.5 about 30\% of
 the time in non-DLA systems.
 
\subsubsection{\MgII$\lambda2796$ versus \FeII$\lambda2600$ and 
DLA Absorption}

Bergeron \& Stasi\'nska (1986) used photoionization models to show
that \MgII systems with strong \FeII absorption have \HI column densities 
in excess of a few times $10^{19}$ atoms cm$^{-2}$.
Le Brun et al. (1997), in their search for DLA galaxies, used the criterion 
\w/$W_0^{\lambda2600}$ $\approx 1$ as a predictor of DLA 
in an absorption-line system. 

\begin{figure*}
\plotone{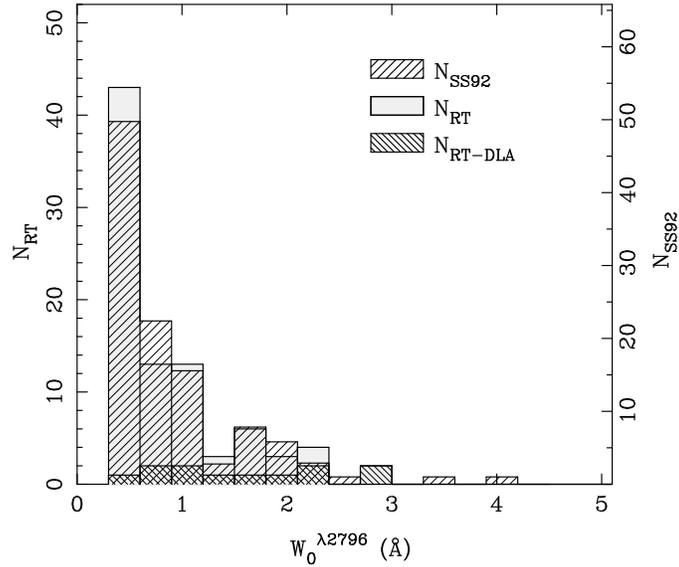}
\caption{The Mg II $\lambda2796$ rest equivalent width
(\w) distributions for the SS92 sample ($N_{SS92}$),
the RT sample ($N_{RT}$), and the DLA subsample of the RT sample
($N_{RT-DLA}$).  The SS92 Mg II data have been scaled to the RT
Mg II data. The figure shows that the SS92 and RT distributions are
very similar. Also note that while only 1 of the 43 RT systems with 
$0.3 \le W_0^{\lambda2796} < 0.6$ \AA\ is a DLA system, 11 of the 44 
RT systems with $W_0^{\lambda2796} \ge 0.6$ \AA\ are DLA systems.}
\end{figure*}

\begin{figure*}
\plotone{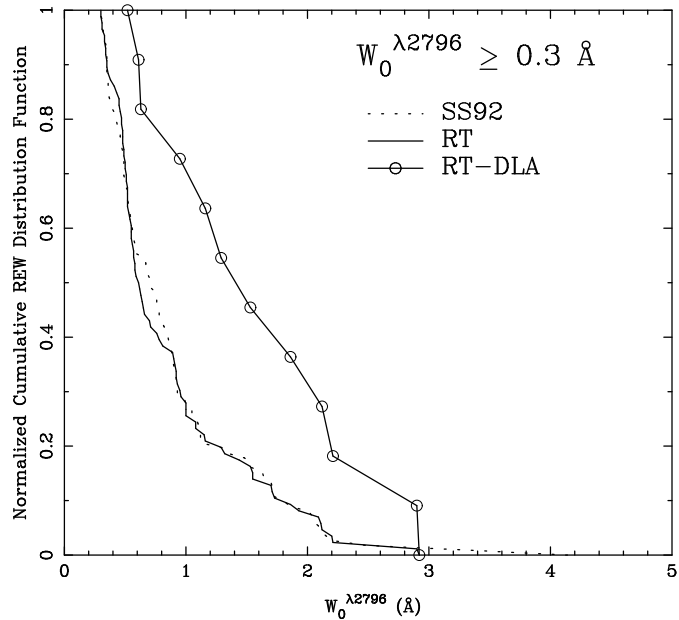}
\caption{The normalized cumulative Mg II rest equivalent width (\w)
distributions for the SS92, RT, and RT-DLA samples with 
$W_0^{\lambda2796} \ge 0.3$ \AA. The figure shows
that the RT and SS92 distributions are nearly identical.  The maximum
deviations between any two of these distributions are used in
a K-S test to determine the probability that they are drawn from
the same parent population. The results are given in Table 5.}
\end{figure*}

\begin{figure*}
\plotone{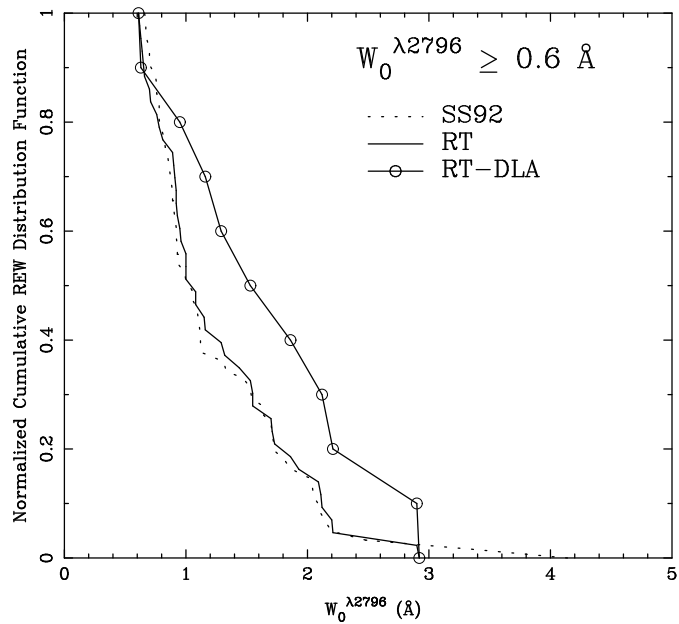}
\caption{Same as Figure 21, but for $W_0^{\lambda2796} \ge 0.6$ \AA.}
\end{figure*}

\begin{figure*}
\plotone{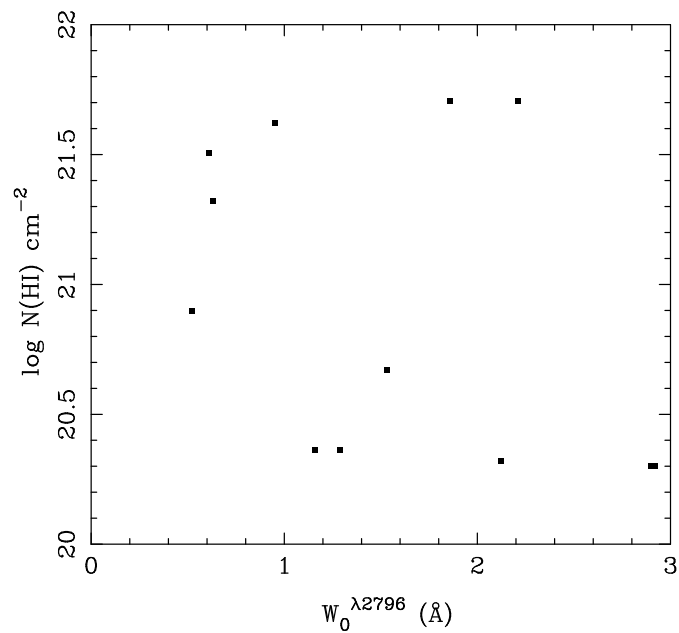}
\caption{A plot of the logarithm of the H I\ column density versus 
Mg II $\lambda2796$ rest equivalent width for the 12 DLA systems
in the RT sample showing that no trend exists between these two
quantities.}
\end{figure*}

\clearpage
  
\begin{deluxetable}{lccccccc}
\footnotesize
\tablenum{4}
\tablecaption{The RT \ion{Mg}{2} Sample\tablenotemark{a}}
\tablehead{ 
\colhead{QSO} & 
\colhead{$z_{\rm abs}$} &
\colhead{$W^{\lambda2796}_{0}$ (\AA)} &
\colhead{$W^{\lambda2803}_{0}$ (\AA)} &
\colhead{$W^{\lambda2852}_{0}$ (\AA)} &
\colhead{$W^{\lambda2600}_{0}$ (\AA)} &
\colhead{References} &
\colhead{$N_{HI}/10^{20}$}  \\[.2ex]
\colhead{} &
\colhead{\ion{Mg}{2}} &
\colhead{\ion{Mg}{2}} &
\colhead{\ion{Mg}{2}} &
\colhead{\ion{Mg}{1}} &
\colhead{\ion{Fe}{2}} &
\colhead{} &
\colhead{atoms cm$^{-2}$} 
}
\startdata
0002$-$422 & 1.5413 & 0.48 & 0.32 & $<$0.3 & $<$0.1 & 1 & \nodata  \\
0002+051   & 0.8516 & 0.92 & 0.84 & 0.16 & \nodata & 2 & \nodata   \\
0058+019   & 0.6128 & 1.70 & 1.51 & 0.35 & 1.39 & 3 & \nodata   \\
0117+213   & 0.5761 & 0.91 & 0.93 & 0.23 & 0.88 & 2 & \nodata   \\
\nodata    & 1.0473 & 0.47 & 0.26 & $<$0.2 & $<$0.1 & 2 & \nodata \\
0119$-$046 & 0.6577 & 0.30 & 0.22 & $<$0.1 & $<$0.1 & 4 & \nodata   \\
0141+339   & 0.4709 & 0.78 & 0.65 & $<$0.3 & $<$0.7 & 2 & \nodata  \\
0143$-$015 & 1.0383 & 0.64 & 0.53 & $<$0.1 & $<$0.1 & 5 & \nodata   \\*
\nodata    & 1.2853 & 0.56 & 0.33 & $<$0.1 & $<$0.1 & 5 & \nodata \\
0150$-$202 & 0.7800 & 0.36 & 0.21 & \nodata & $<$0.3 & 3 & \nodata   \\
0151+045   & 0.1602 & 1.55 & 1.55 & $<$0.5 & $<$0.5 & 6 & \nodata   \\
0215+015   & 1.3447 & 1.93 & 1.64 & 0.32 & 1.36 & 7  & \nodata  \\
0248+430   & 0.3939 & 1.86 & 1.42 & 0.70 & 1.03 & 2 & 51  \\*
\nodata    & 0.4515 & 0.34 & 0.29 & $<$0.5 & $<$0.1 & 8 & \nodata   \\
0302$-$223 & 1.0096 & 1.16 & 0.96 & 0.18 & 0.63 & 9 & 2.3   \\
0333+321   & 0.9531 & 0.47 & 0.33 & $<$0.3 & $<$0.2 & 2 & \nodata  \\
0421+019   & 1.3918 & 0.35 & 0.31 & $<$0.3 & $<$0.1 & 2 & \nodata  \\*
\nodata    & 1.6380 & 0.35 & 0.28 & $<$0.2 & $<$0.4 & 2 & \nodata  \\  
0424$-$131 & 1.4080 & 0.55 & 0.35 & $<$0.3 & 0.44 & 2 & \nodata  \\*
\nodata    & 1.5623 & 0.38 & 0.39 & $<$0.2 & $<$0.2 & 2 & \nodata  \\
0454$-$220 & 0.4744 & 1.44 & 1.31 & 0.35 & 0.97 & 10 & \nodata   \\*
\nodata    & 0.4834 & 0.45 & 0.27 & 0.09 & 0.16 & 10 & \nodata   \\
0454+039   & 0.8596 & 1.53 & 1.40 & 0.37 & 1.11 & 2 & 4.7  \\*
\nodata    & 1.1538 & 0.57 & 0.36 & $<$0.2 & $<$0.1 & 2 & \nodata  \\
0710+119   & 0.4629 & 0.62 & 0.29 & 0.24 & $<$0.4 & 11 & \nodata   \\
0735+178   & 0.4240 & 1.32 & 1.03 & 0.18 & 0.87 & 12 & \nodata   \\
0738+313   & 0.2213 & 0.52 & 0.50 & $<$0.2 & $<$0.6 & 13 & 7.9   \\
0742+318   & 0.1920 & 0.33 & 0.23 & $<$0.2 & \nodata & 13 & \nodata   \\
0823$-$223 & 0.9103 & 1.15 & 0.68 & $<$0.4 & \nodata & 14 & \nodata   \\
0827+243   & 0.5247 & 2.90 & 2.20 & \nodata & 1.90 & 15,16 & 2.0   \\
0843+136   & 0.6064 & 1.08 & 0.67 & $<$0.1 & $<$0.7 & 17 & \nodata   \\
0933+732   & 1.4789 & 0.95 & 1.15 & $<$0.3 & 0.76 & 2 & 42   \\*
\nodata    & 1.4973 & 1.71 & 1.98 & 0.67 & 1.15 & 2 & \nodata  \\
0952+179   & 0.2377 & 0.63 & 0.79 & $<$0.4 & \nodata & 2 & 21  \\
0957+561A  & 1.3911 & 2.12 & 1.81 & 0.19 & 1.67 & 18,19,20 & 2.1   \\
0958+551   & 0.2413 & 0.55 & 0.38 & $<$0.2 & \nodata & 3 & \nodata   \\
1035$-$276 & 0.8242 & 1.08 & 0.87 & $<$0.6 & 0.52 & 21 & \nodata   \\
1038+064   & 0.4416 & 0.66 & 0.57 & $<$0.2 & $<$0.2 & 2 & \nodata   \\
1040+123   & 0.6591 & 0.58 & 0.42 & $<$0.2 & $<$0.2 & 11 & \nodata  \\  
1049+616   & 0.2255 & 0.51 & 0.56 & $<$0.1 & $<$0.4 & 13 & \nodata   \\*
\nodata    & 0.3937 & 0.34 & 0.29 & \nodata & $<$0.1 & 13 & \nodata  \\
1100$-$264 & 1.1872 & 0.51 & 0.28 & $<$0.2 & $<$0.2 & 9 & \nodata   \\*
\nodata    & 1.2028 & 0.54 & 0.43 & 0.27 & $<$0.2 & 9,22 & \nodata   \\
1115+080   & 1.0431 & 0.31 & 0.18 & $<$0.2 & $<$0.1 & 2 & \nodata   \\
1127$-$145 & 0.3130 & 2.21 & 1.90 & 1.14 & 1.14 & 23 & 51   \\ 
1137+660   & 0.1164 & 0.50 & 0.53 & $<$0.2 & $<$0.2 & 24 & \nodata   \\
1148+386   & 0.2130 & 0.96 & 0.37 & $<$0.3 & \nodata &  13 & \nodata   \\*
\nodata    & 0.5533 & 0.92 & 0.99 & $<$0.2 & $<$0.2 & 2 & \nodata   \\
1206+459   & 0.9277 & 1.00 & 0.79 & $<$0.2 & $<$0.1 & 2 & \nodata  \\
1209+107   & 0.3930 & 1.00 & 0.54 & $<$0.2 & $<$0.4 & 25 & \nodata   \\*
\nodata    & 0.6295 & 2.92 & 2.05 & $<$0.4 & 1.50 & 25 & 2.0  \\
1213$-$002 & 1.5543 & 2.09 & 1.65 & $<$0.2 & 1.11 & 2 & \nodata   \\
1222+228   & 0.6681 & 0.43 & 0.41 & $<$0.1 & $<$0.1 & 3 & \nodata   \\
1229$-$021 & 0.7571 & 0.52 & 0.48 & $<$0.1 & \nodata & 11 & \nodata   \\
1241+176   & 0.5506 & 0.48 & 0.37 & $<$0.2 & $<$0.2 & 2 & \nodata   \\
1246$-$057 & 1.2015 & 0.90 & 0.75 & $<$0.4 & 0.40 & 2 & \nodata  \\*
\nodata    & 1.6453 & 0.52 & 0.58 & $<$0.3 & 0.38 & 2 & \nodata   \\ \tablebreak
1247+267   & 1.2232 & 0.48 & 0.38 & 0.11 & \nodata & 2 & \nodata   \\
1248+401   & 0.7729 & 0.76 & 0.50 & $<$0.4 & 0.32 & 2 & \nodata   \\
1254+047   & 0.5193 & 0.46 & 0.38 & $<$0.2 & 0.40 & 2 & \nodata  \\\ 
1317+277   & 0.2887 & 0.33 & 0.31 & $<$0.2 & \nodata & 2 & \nodata  \\*
\nodata    & 0.6596 & 0.49 & 0.33 & $<$0.2 & 0.17 & 2 & \nodata   \\
1323+655   & 1.5181 & 0.57 & 0.53 & $<$0.1 & \nodata & 26 & \nodata \\*
\nodata    & 1.6101 & 2.20 & 1.85 & 0.16 & 0.88 & 26 & \nodata   \\
1327$-$206 & 0.8526 & 2.11 & 1.48 & $<$0.4 & 0.76 & 27,28  & \nodata  \\
1329+412   & 1.2820 & 0.49 & 0.31 & $<$0.3 & \nodata & 2  & \nodata  \\*
\nodata    & 1.6011 & 0.70 & 0.35 & $<$0.2 & $<$0.2 & 2  & \nodata  \\ 
1331+170   & 1.3284 & 1.73 & 1.14 & $<$0.1 & 0.32 & 2 & \nodata   \\
1338+416   & 0.6213 & 0.31 & 0.17 & $<$0.3 & $<$0.1 & 2 & \nodata   \\
1354+195   & 0.4563 & 0.89 & 0.82 & 0.16 & 0.32 & 2 & \nodata \\
1354+258   & 0.8585 & 1.00 & 0.86 & $<$0.1 & $<$0.2 & 26 & \nodata   \\*
\nodata    & 0.8856 & 0.81\tablenotemark{b} & 0.57\tablenotemark{b}
 & $<$0.2 & $<$0.2 & 26 & \nodata   \\*
\nodata    & 1.4205 & 0.61 & 0.50 & 0.20 & 0.55 & 26 & 32   \\
1421+330   & 1.1725 & 0.53 & 0.40 & $<$0.1 & $<$0.1 & 2 & \nodata   \\
1517+239   & 0.7382 & 0.30 & 0.34 & \nodata & $<$0.1 & 3 & \nodata   \\
1622+239   & 0.6561 & 1.29 & $<$1.69 & $<$0.4 & 1.13 & 2 & 2.3   \\*
\nodata    & 0.8914 & 1.55 & 1.27 & $<$0.3 & 1.32 & 2 & \nodata   \\
1623+269   & 0.8881 & 0.93 & 0.75 & 0.14 & 0.21 & 3 & \nodata   \\
1634+706   & 0.9903 & 0.58 & 0.42 & $<$0.2 & $<$0.1 & 2 & \nodata   \\
1704+608   & 0.1634 & 0.59 & 0.52 & $<$0.5 & $<$0.2 & 13 & \nodata   \\*
\nodata    & 0.2220 & 0.55 & 0.33 & $<$0.2 & \nodata & 13 & \nodata  \\  
1821+107   & 1.2528 & 0.71 & 0.48 & 0.09 & $<$0.3 & 2 & \nodata   \\
1901+319   & 0.3901 & 0.45 & 0.15 & $<$0.1 & \nodata & 11 & \nodata   \\ 
2128$-$123 & 0.4299 & 0.41 & 0.37 & 0.10 & 0.27 & 9  & \nodata  \\
2145+067   & 0.7906 & 0.52 & 0.41 & $<$0.2 & $<$0.1 & 2 & \nodata   \\
2223$-$052 & 0.8472 & 0.65 & 0.42 & $<$0.2 & $<$0.4 & 29,30 & \nodata   \\
2326$-$477 & 1.2610 & 0.50 & 0.38 & $<$0.1 & $<$0.1 & 9 & \nodata \\ 
\enddata
\tablenotetext{a}{\ 
Upper limits are either given by the authors or estimated by us from
the published spectra. If the line is part of a blend, then the rest
equivalent width entered is an upper limit. The upper limit is a $1\sigma$ 
upper limit if estimated by us from the published spectrum. In some cases, 
the absorption line ($\lambda2600$ or $\lambda2852$) was clearly visible in 
a spectrum but was not identified by the authors since the rest equivalent 
width of the line did not meet their detection criterion, usually $5\sigma$. 
In these cases, we estimated the equivalent width of the line from the 
spectrum and tabulated the measurement as an upper limit. The trends that 
we have established in Figures 24, 25, and 26 are evident from the 
measurements of rest equivalent widths by the authors of the published
spectra; our upper limit measurements only serve to provide a certain
degree of completeness to our sample and, in fact, do not indicate or
contribute to any trends by themselves.}

\tablenotetext{b}{\ 
The rest equivalent widths quoted in reference (26) are in error.}

\tablerefs{(1) Lanzetta, Turnshek, \& Wolfe 1987,
(2) SS92,
(3) Sargent, Boksenberg, \& Steidel 1988,
(4) Sargent, Young, \& Boksenberg 1982,
(5) Sargent, Steidel, \& Boksenberg 1989,
(6) Bergeron et al. 1988, 
(7) Bergeron \& D'Odorico 1986,
(8) Womble et al. 1990,
(9) Petitjean \& Bergeron 1990,
(10) Tytler et al. 1987,
(11) Aldcroft, Bechtold, \& Elvis 1994,
(12) Boksenberg, Carswell, \& Sargent 1979, 
(13) Boiss\'e et al. 1992,
(14) Falomo 1990,
(15) Wills 1978,
(16) Ulrich \& Owen 1977, 
(17) Foltz et al. 1986,
(18) Weymann et al. 1979, 
(19) Caulet 1989, 
(20) Wills \& Wills 1980, 
(21) Dinshaw \& Impey 1996, 
(22) Boiss\'e \& Bergeron 1985, 
(23) Bergeron \& Boiss\'e 1991,
(24) Bahcall et al. 1993, 
(25) Young, Sargent, \& Boksenberg 1982, 
(26) Barthel,  Tytler, \& Thomson  1990, 
(27) Kunth \& Bergeron 1984, 
(28) Bergeron, D'Odorico, \& Kunth 1987, 
(29) Le Brun et al. 1993, 
(30) Miller \& French 1978.
}
\end{deluxetable}
\clearpage

\begin{deluxetable}{ccc}
\tablenum{5}
\tablewidth{4.0in}
\tablecaption{K-S Test Results\tablenotemark{a}} 
\tablehead{
\colhead{Samples} &
\colhead{$W_0\ge0.3$ \AA} & 
\colhead{$W_0\ge0.6$ \AA} 
}

\startdata
RT \& SS92 &  0.65 & 0.83 \\
SS92 \& RT-DLA & 0.01 & 0.15 \\
RT \& RT-DLA & 0.02 & 0.36 \\
\enddata
\tablenotetext{a}{\ Probability that the two samples are drawn from
the same parent population.}
\end{deluxetable}

\begin{figure*}
\plotone{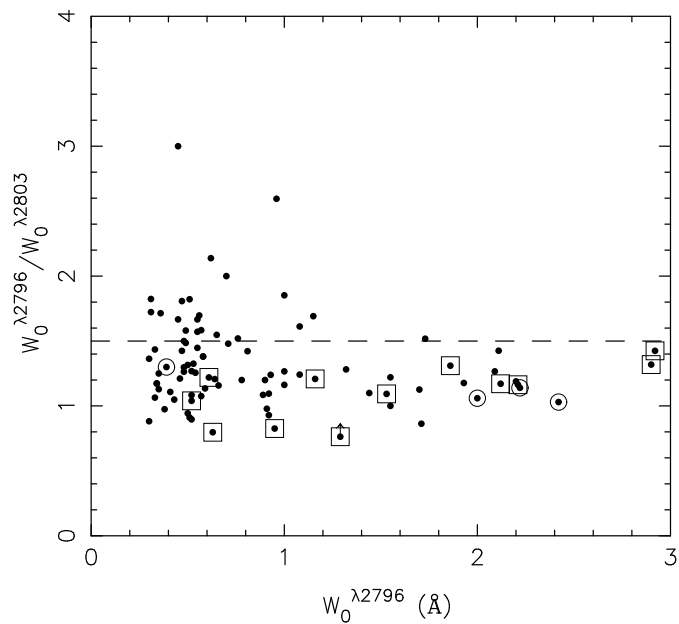}
\caption{A plot of the Mg II doublet ratio, \w/$W_0^{\lambda2803}$, 
versus \w\ for the RT sample (see Table 4). The DLA systems from the 
RT sample are marked with open squares.
The open circles represent previously known 21 cm absorbers that were
excluded from our unbiased Mg II sample. The horizontal dashed line 
has \w/$W_0^{\lambda2803} = 1.5$. The upward pointing arrow indicates
an upper limit to the measured value of $W_0^{\lambda2803}$.}
\end{figure*}

\begin{figure*}
\plotone{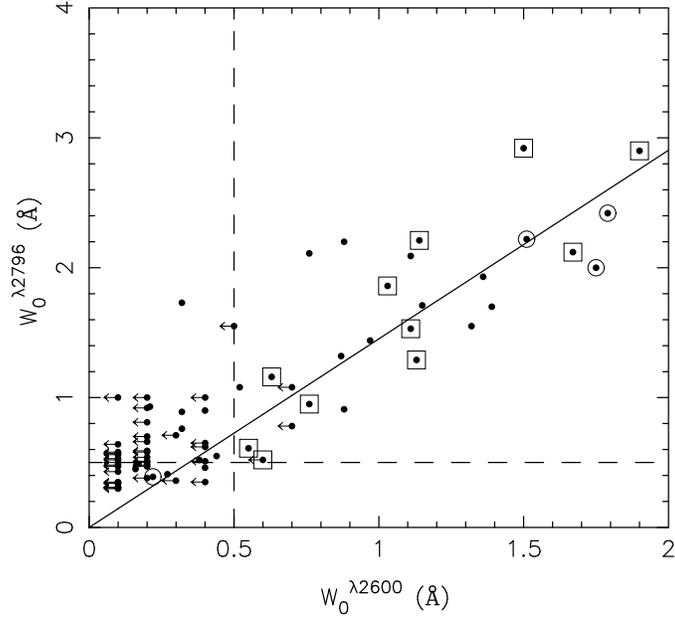}
\caption{A plot of Mg II \w\ versus Fe II $W_0^{\lambda2600}$. 
The DLA systems from the RT sample are marked with open squares.
The open circles represent previously known 21 cm absorbers that were
excluded from our unbiased Mg II sample.  Left pointing arrows indicate
upper limits to the measured value of $W_0^{\lambda2600}$ (see Table 4).
The horizontal and vertical dashed lines identify the region
for which \w$>0.5$ \AA\ and $W_0^{\lambda2600}>0.5$ \AA. 
All the non-DLA systems in this regime have H I column densities 
$N_{HI}>10^{19}$ atoms cm$^{-2}$. The solid line has a slope of 1.45.}
\end{figure*}

\begin{figure*}
\plotone{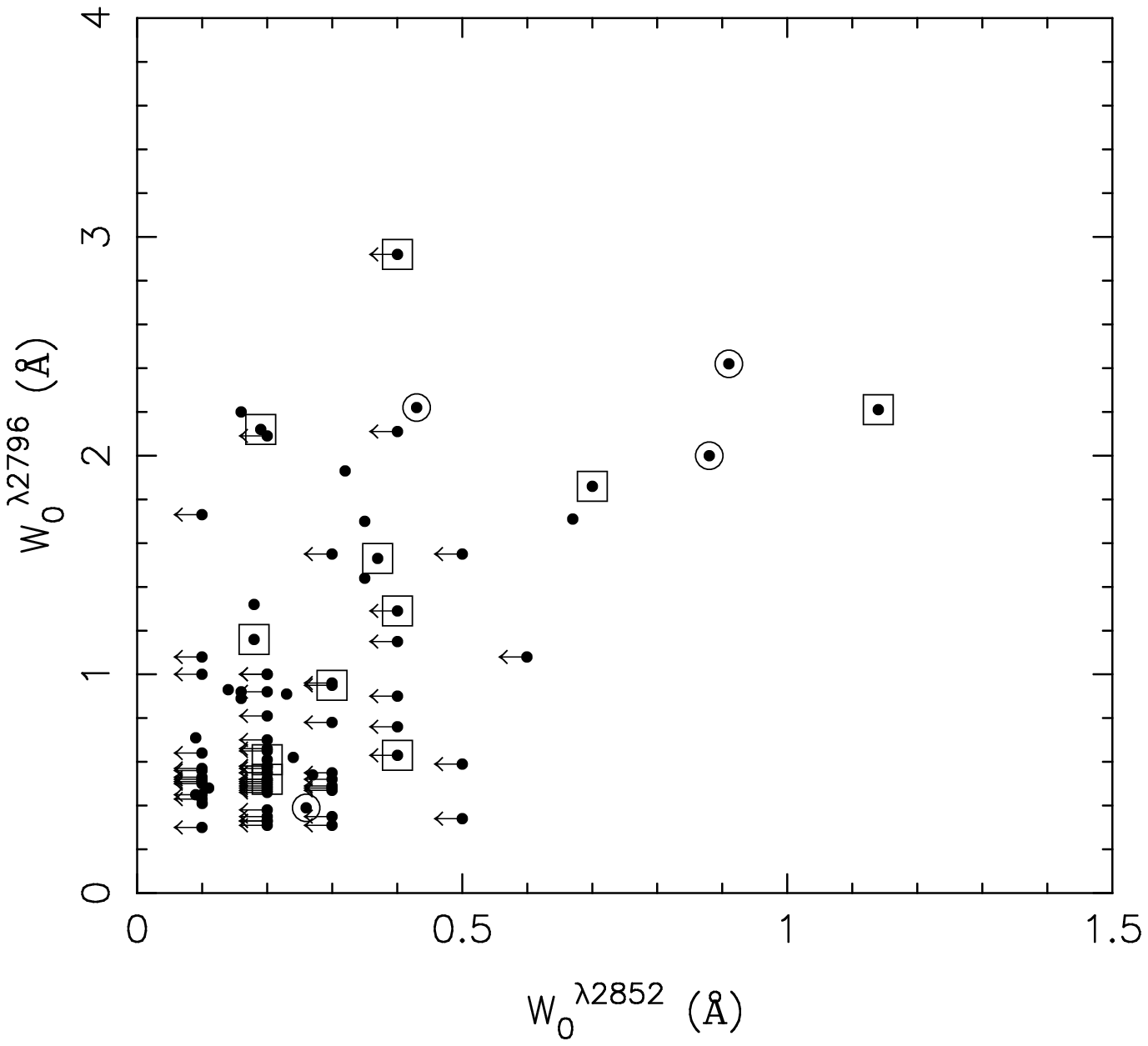}
\caption{A plot of Mg II \w\ versus Mg I $W_0^{\lambda2852}$. 
The DLA systems from the RT sample are marked with open squares.
The open circles represent previously known 21 cm absorbers that were
excluded from our unbiased Mg II sample.  Left pointing arrows indicate
upper limits to the measured value of $W_0^{\lambda2852}$ (see Table 4).}
\end{figure*}

We have used the \MgII systems
in the RT sample with $W_0^{\lambda2600}$ measurements to further test this.
Figure 25 shows \w\ versus $W_0^{\lambda2600}$ for systems for which 
\FeII equivalent width measurements are available in the literature, or
for which we were able to derive upper limits from published spectra 
(see Table 4). As in Figure 24, the DLA systems in the RT sample are
marked with squares and the 21 cm absorbers which are not part of the RT 
sample are marked with circles.
The horizontal and vertical dashed lines identify the region in the 
\w\ versus $W_0^{\lambda2600}$  plane for which \w$>0.5$ \AA\ and
$W_0^{\lambda2600}>0.5$ \AA. Reliable equivalent widths are available in the
literature for 24 out of 27 of these systems, where detections of 
\FeII$\lambda2600$ usually exceed the  $5\sigma$ level of significance. 
Several results of interest emerge from this plot. First, there is a 
linear correlation between \w\ and $W_0^{\lambda2600}$. If we force a
least squares fit to the 24 data points to pass through the origin, we obtain 
\w$=1.45W_0^{\lambda2600}$.  The correlation between \MgII and \FeII 
in high-redshift,
predominantly low-excitation, absorption-line systems was noted by 
Bergeron \& Stasi\'nska (1986). They found that a strong \MgII system
with \w $\ge 1$ \AA\ was invariably accompanied by strong \FeII absorption.
Second, excluding upper limits, 10 of the 21 systems in the RT sample 
($\approx$ 48\%) with \w$>0.5$ \AA\ and
$W_0^{\lambda2600}>0.5$ \AA\ are DLA systems. Third, the non-DLA systems
in this regime all have \HI column densities $N_{HI}>10^{19}$ atoms cm$^{-2}$.
Fourth, with the exception of the known $z=0.692$ 21 cm absorber 
towards 3C 286 which has unusually low metal-line equivalent widths
(\w$=0.39$ \AA\ and $W_0^{\lambda2600}=0.22$ \AA, Cohen et al. 1994), 
all of the DLA systems
lie in the  upper-right region of the plot where \w$>0.5$ \AA\ and 
$W_0^{\lambda2600}>0.5$ \AA. In the future, these results can be used to 
further improve the selection criteria for searching for DLA. {\it In particular, 
there is  approximately a 50\% probability of finding a classical
DLA system in metal-line systems with \w$>0.5$} \AA\ {\it and 
$W_0^{\lambda2600}>0.5$} \AA.

\subsubsection{\MgII$\lambda2796$ versus \ion{Mg}{1} $\lambda2852$ and 
DLA Absorption}

Figure 26 shows \w\ versus $W_0^{\lambda2852}$. It can be
 seen that the DLA systems do not lie in a special region in the
\MgII$\lambda2796$ versus \ion{Mg}{1} $\lambda2852$ rest equivalent
width plane. Evidently the presence of DLA does not  guarantee the
presence of detectable \ion{Mg}{1} $\lambda2852$ in medium resolution
spectroscopic surveys. However, although the statistics are small, 
all systems with $W_0^{\lambda2852} \ge 0.7$ \AA\ have DLA.

\subsubsection{Discussion}

The \MgII absorption lines under study in these samples usually have
Mg$^+$ column densities which place them on or near the saturated regime
on a curve-of-growth plot. The DLA systems  generally appear to exhibit 
somewhat greater saturation. In either case, \w\ is a poor indicator of \MgII
column density, and a better indicator of the number of clouds along
the sight-line and their kinematic spread in velocity (e.g. Petitjean \&
Bergeron 1990, SS92, Churchill 1997).  Thus, our results indicate that
the low-redshift DLA systems predominantly arise along sight-lines
in which the number of clouds and their spread in velocity is
systematically larger in comparison to the $0.3\le$\w$<0.6$ \AA\
systems.  The velocity spread, i.e., the difference between the minimum
velocity and maximum velocity components, must be $> 64$ km s$^{-1}$,
since a \MgII$\lambda$2796 absorption line which is completely black
over the line profile must be spread over a $64$ km s$^{-1}$ interval
to produce a line with $W_0^{\lambda2796}=0.6$ \AA. Empirically, since
we find that the \MgII lines are not necessarily completely black, the
velocity spread must typically exceed this value.  However, observations
in the \HI\ 21 cm absorption line of DLA systems usually show that the
components which give rise to the Voigt damping profile
are spread over a much smaller velocity interval, which
indicates that only a small fraction of the clouds contributing to
\w\ give rise to the DLA line.

Other interesting correlations which pertain
to high column density systems (but not necessarily DLA systems)
have been identified by Churchill et al. (1999b) using high-resolution
(6 km s$^{-1}$) Keck observations of \MgII systems.  (High-resolution
data are not generally available for our systems.)  In their figure 2b
Churchill et al. (1999b) plot \w\ versus $\omega_v^{\lambda2796}$,
where $\omega_v^{\lambda2796}$ is a velocity weighted optical depth
which is sensitive to high-velocity ``outlier'' clouds with small rest
equivalent widths.  The systems with $N_{HI}\ge 10^{19}$ cm$^{-2}$ in this
plot (i.e. neutral-hydrogen-rich systems), which include all of the systems 
in Figure 25 with \w$>0.5$ \AA\ and $W_0^{\lambda2600}>0.5$ \AA\ and
of which the DLA systems
form a subset, are isolated and occupy the top-central region of this
plane, i.e., they have \w $> 0.9$ \AA\ and intermediate values of
$\omega_v^{\lambda2796}$.  Figure 2a of Churchill et al. (1999b)  
shows that these neutral-hydrogen-rich systems lie in an isolated region
of the \MgII\w\ versus \CIV$W_0^{\lambda1548}$ plane, corresponding to smaller
$W_0^{\lambda1548}$/\w\ ratios. Lastly, figure 3b from Churchill et al. 
(1999b) shows that for the 
few cases in which the absorbing galaxies have been identified, the impact 
parameters are among the smallest observed.
These are evidently positive correlations for neutral-hydrogen-rich
systems that may also hold for classically damped systems.

If we consider these pieces of information together, the picture
that emerges suggests that the metal-line absorption components 
associated with a DLA system, which generally correspond to 
components other than those which give rise to the DLA line itself, 
are systematically different in several ways:
(1) There is a relatively large number of absorbing components spread
over a substantial velocity interval (to produce \w $\ge 0.6$ \AA),
but for a fixed \w\ the absorbing components making up the associated
metal-line system will be more closely spaced in comparison to cases
where DLA is not present. This is suggestive of a 
clustering of clouds in the vicinity of the cloud that gives rise to the 
DLA line itself. (2) The \CIV$W_0^{\lambda1548}$/\MgII\w\ metal-line
ratio is relatively small and the \FeII$W_0^{\lambda2600}$/\MgII\w\ metal-line 
ratio is relatively high, signifying a lower ionization phase of the DLA 
absorbing region. (3) The transverse separation (impact parameter)
from the sight-line for any identified DLA absorbing galaxy is
relatively small in comparison to systems which do not have a DLA line.
This qualitative description of the kinematic and ionization structure
of DLA systems and the associated absorbing galaxies appears to be roughly
consistent with the observations, and should be incorporated into
any successful model describing the nature of DLA absorbers.

\subsection{The Effective Redshift Path of the RT Survey, $\Delta Z_{RT}$}

In conventional surveys, like the SS92 \MgII survey, searches for
absorption lines are carried out by observing the continuum regions of
an unbiased sample of QSOs down to some minimum equivalent width threshold over
the total redshift path of the survey. We use the parameter $\Delta Z(W_{min})$
to represent the total redshift path covered down to absorption-line 
rest equivalent width, $W_{min}$,  at mean redshift, $<z>$.
The $\Delta Z(W_{min})$ parameter is computed by adding all of the individual
redshift intervals covered by all survey spectra with data good
enough to detect $W_{min}$ at the $4-5\sigma$ level of significance.
This can also be expressed as the integral of the redshift path density,
$g(z,W_{min})$, over redshift,
\begin{equation} 
\Delta Z(W_{min}) = \int^{\infty}_{0}g(z,W_{min})dz.  
\end{equation}
Descriptions of these standard techniques have been given by WTSC86,
Lanzetta, Turnshek, \& Wolfe (1987), SS92, and others.  However, the RT
survey is not a conventional survey that samples redshifts paths along
random sight lines. We search for DLA lines at specific redshifts along
predetermined sight lines. Nevertheless, we do have indirect
information which allows us to infer an {\it effective} total redshift
path for \MgII $\lambda2796$ covered in the RT survey, 
$\Delta Z_{RT}(W_{min}^{\lambda2796})$, at mean redshift, $<z>_{RT}$.
Note that this is also the redshift path for the detection of DLA lines.

To determine $\Delta Z_{RT}(W_{min}^{\lambda2796})$, we need to determine
the effective redshift path density of the RT survey,
$g_{RT}(z,W_{min}^{\lambda2796})$. In turn, $g_{RT}(z,W_{min}^{\lambda2796})$
can be written as
\begin{equation}
g_{RT}(z,W_{min}^{\lambda2796}) = \frac{m_{RT}(z,W_{min}^{\lambda2796})}
{n(z,W_{min}^{\lambda2796})},
\end{equation}
where $m_{RT}(z,W_{min}^{\lambda2796})$ is the redshift distribution of 
\MgII systems in the RT sample and 
$n(z,W_{min}^{\lambda2796})$ is the number of \MgII systems per unit 
redshift as a function of redshift. The 
$m_{RT}(z,W_{min}^{\lambda2796})$ distributions for $W_{min}^{\lambda2796}$
= 0.3 \AA\ and 0.6 \AA\ are shown in Figure 27.  The redshift
evolution of the number density of \MgII systems can be parameterized by
\begin{equation}
n(z,W_{min}^{\lambda2796})=n_0(W_{min}^{\lambda2796})
(1+z)^{\gamma(W_{min}^{\lambda2796})},
\end{equation}
where $n_0(W_{min}^{\lambda2796})$ and $\gamma(W_{min}^{\lambda2796})$
are obtained from Table 4 of SS92. These values are
$n(z=1.12, W_{min}^{\lambda2796}=0.3) = 0.97$ 
[which implies $n_0(W_{min}^{\lambda2796}=0.3) = 0.54$], 
$n(z=1.17, W_{min}^{\lambda2796}=0.6) = 0.52$ 
[which implies $n_0(W_{min}^{\lambda2796}=0.6)= 0.24$], 
$\gamma(W_{min}^{\lambda2796}=0.3) = 0.78$, and
$\gamma(W_{min}^{\lambda2796}=0.6) = 1.02$.   
The results of equation (2) for $W_{min}^{\lambda2796}=0.3$ \AA\ and 0.6 \AA\
 are shown in Figure 28.  For comparison, SS92 show in their figure 4 the
number of sight-lines for these same values of $W_{min}^{\lambda2796}$;
this is equivalent to redshift path density.

\begin{figure*}
\plotone{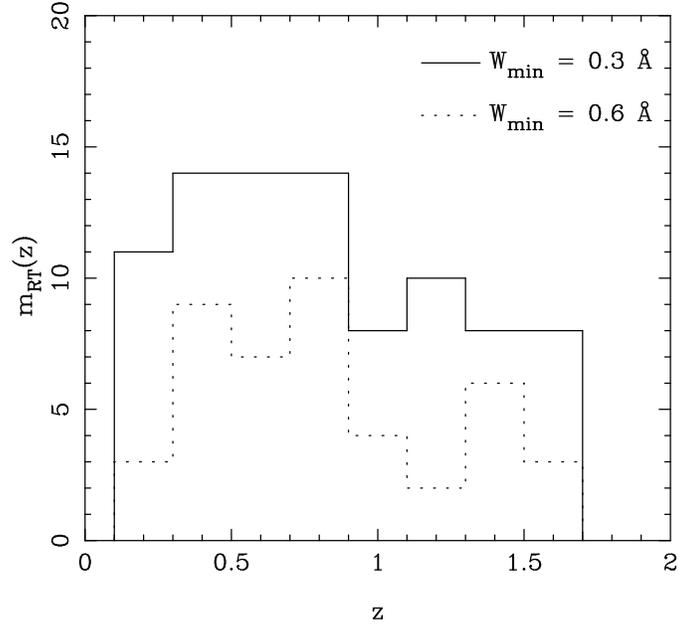}
\caption{The observed redshift distribution of Mg II systems in the 
RT sample, $m_{RT}(z,W_{min}^{\lambda2796})$,  
for $W_{min}^{\lambda2796}$ = 0.3 \AA\ and 0.6 \AA.}
\end{figure*}

\begin{figure*}
\plotone{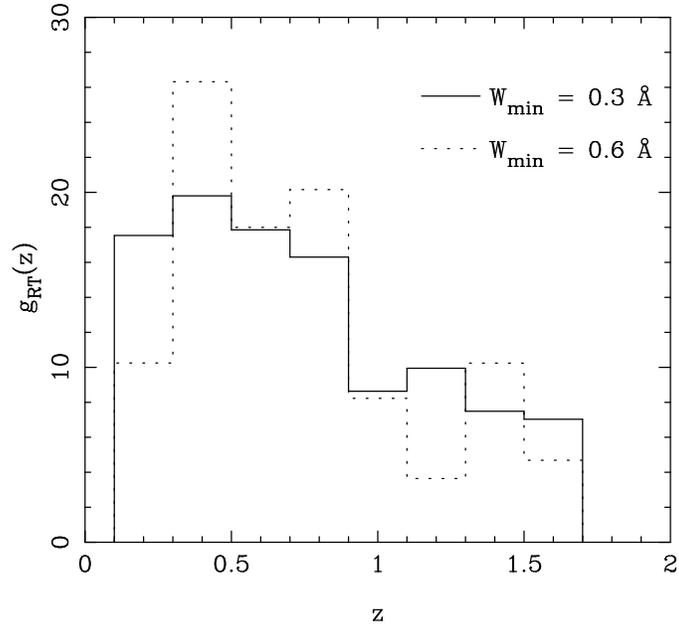}
\caption{The observed effective redshift path density of the RT survey,
$g_{RT}(z,W_{min}^{\lambda2796})$, for $W_{min}^{\lambda2796}$ = 
0.3 \AA\ and 0.6 \AA.}
\end{figure*}

Using equation (1), the effective redshift path for the RT survey is
found to be $\Delta Z_{RT}(W_{min}=0.3) = 104.6$ with mean redshift
$<z>_{RT}=0.83$ and 87 systems, and 
$\Delta Z_{RT}(W_{min}=0.6) = 103.7$
with the same mean redshift $<z>_{RT}=0.83$ and 44 systems.  
To confirm these results we have verified that the detected number of \MgII 
absorption-line systems is given by
$${\cal N}_{MgII}(W_{min}^{\lambda2796}) = 
\int^{\infty}_{0}m_{RT}(z,W_{min}^{\lambda2796})dz$$
\begin{equation}
= \int^{\infty}_{0}g_{RT}(z,W_{min}^{\lambda2796})n(z,W_{min}^{\lambda2796})dz,
\end{equation}
where ${\cal N}_{MgII}(W_{min}^{\lambda2796}=0.3)=87$ and
${\cal N}_{MgII}(W_{min}^{\lambda2796}=0.6)=44$.  For comparison, 
we give the number of systems, mean redshifts, and redshift paths of the
RT and SS92 samples in Table 6. 
The fact that SS92 were able to study $\approx$ 28\%
more systems than in the RT survey over a redshift path length that
was only $\approx$ 9\% larger is understandable since the SS92 survey
had $<z>_{SS92}=1.12$, while the RT survey has $<z>_{RT}=0.83$. After correcting
for the non-equivalent redshift paths for different $W_{min}^{\lambda2796}$
thresholds, we find that (per unit redshift) $\approx$
55\% of the \MgII systems in SS92 have \w $\ge 0.6$
\AA, while this fraction is $\approx$ 50\% for the RT survey.  This is  
consistent with expectations since the SS92 and RT samples are derived
from the same parent population of absorbers (\S3.1).

\begin{deluxetable}{cccccccc}
\tablenum{6}
\tablewidth{5.0in}
\tablecaption{Number of Systems, Mean Redshift, and 
Redshift Path for the RT and SS92 Samples}
\tablehead{ 
\colhead{} &
\multicolumn{3}{c}{RT} &
\colhead{} &
\multicolumn{3}{c}{SS92} \\[.2ex] \cline{2-4} \cline{6-8}
\colhead{$W_{min}$ (\AA)} & 
\colhead{$N_{MgII}$} &
\colhead{$<z>$} &
\colhead{$\Delta Z$} &
\colhead{} &
\colhead{$N_{MgII}$} &
\colhead{$<z>$} &
\colhead{$\Delta Z$}
}
 
\startdata
0.3 & 87 & 0.83 & 104.6 &  & 111 & 1.12 & 114.2 \\
0.6 & 44 & 0.83 & 103.7 &  & 67 & 1.17 & 129 \\
\enddata
\end{deluxetable}

These determinations should be useful in attempts to compare the
effectiveness of our survey strategy, and our results, to conventional
surveys. At the present time, the observed incidence of DLA in the 
various surveys appears to be acceptably consistent given the small 
number statistics (Jannuzi et al. 1998).
However, to make a reliable prediction of the number of systems
which will be detected in a survey one should explicitly make use of
equation (4) instead of simply multiplying the total redshift path, $\Delta Z$,
at a mean redshift $<z>$ by the expected number of systems per unit redshift,
$n(<z>)$.

\section{The Low Redshift DLA Statistics}

In this section, we first discuss the steps taken to correct for any 
Malmquist bias in the RT survey (\S4.1). We then derive results on the
incidence of low-redshift DLA systems, $n_{DLA}(z)$ (\S4.2), and their
corresponding cosmological neutral gas mass density, $\Omega_{DLA}(z)$
(\S4.3). We use the $n_{DLA}(z)$
and $\Omega_{DLA}(z)$ notation deliberately to refer 
only to classical DLA systems having neutral hydrogen column densities
 $N_{HI} \ge 2 \times 10^{20}$ atoms cm$^{-2}$. In
particular, $\Omega_{DLA}$ explicitly excludes hydrogen in   molecular
and ionized forms. It also excludes neutral hydrogen columns that may
have been missed due to selection effects, such as obscuration of
background QSOs by dust.  Since \MgII lines are present in these
  systems, the gas is clearly enriched with metals. However, there is
obviously insufficient dust to cause them to be overlooked.
The presence of heavy elements in the neutral components of DLA systems
does not make a significant contribution to $\Omega_{DLA}(z)$, however the
presence of helium does. Therefore, helium from big bang nucleosynthesis
is explicitly included in the calculation of $\Omega_{DLA}(z)$. We 
assume a cosmology with $\Lambda=0$ and use $H_0=65$ km s$^{-1}$
Mpc$^{-1}$ ($h_{65}=H_0/65$) for all calculations. Where applicable, we
report results assuming $q_0=0$ and $q_0=0.5$.

\subsection{Correction for Malmquist Bias}

A Malmquist bias can be introduced into a statistical study when a
truncation or censoring effect influences sample selection.  This is a
potential problem in absorption-line surveys since sample selection is
usually based on a minimum rest equivalent width or, for cases in which
 Voigt damping profiles are fitted to the data,
the \HI\ column density. This effect is significant if the column density
is a steep-enough function near the threshold column density.
For example, a Malmquist bias would be introduced if we identified and
used only those systems with fitted Ly$\alpha$ profiles that indicated
$N_{HI} \ge 2\times10^{20}$ atoms cm$^{-2}$ because the measurements
have associated errors and in reality some of the systems are likely to
have $N_{HI} < 2\times10^{20}$ atoms cm$^{-2}$. At the same time, there
might be many more systems formally measured to lie below the \HI\ column
density threshold if errors were not taken into account, but in reality
some of these are likely to lie above the threshold.

Thus, our initial intent was to form a DLA sample which included those
\MgII systems with measured \HI\ column densities as small as $2\sigma$
less than $2\times10^{20}$ atoms cm$^{-2}$. However, we have not identified
and measured any systems which are slightly below the threshold.  This is likely
to be an artifact of small number statistics\footnote{Based on the column
density distribution we derive in \S5, the expected number of systems 
with column density $1.6\times10^{20} \le N_{HI} \le 2\times10^{20}$
atoms cm$^{-2}$ is $\sim2$. The probability of finding none is 13.5\%.}.  
Consequently, in order to approximately correct our sample for the effect 
of the Malmquist bias, we have simply removed half of the systems with 
$N_{HI} \approx 2\times10^{20}$ atoms cm$^{-2}$, reasoning that about half 
of them are likely to have actual \HI\ column densities slightly below the 
threshold as suggested by the measurement errors. The details are given below.

Specifically, our study has yielded 87 \MgII systems for which we have 
information on the nature of the \lya line.  Twelve of these systems were
found to have DLA with column densities $N_{HI}\ge2\times10^{20}$
cm$^{-2}$ and two more were found serendipitously (\S2.5). 
However, as can be seen from Table 3, four systems have
measurement errors that suggest that their true \HI\ column
densities may lie below the threshold value. These are 
the $N_{HI}=2.0\pm0.2\times10^{20}$ atoms cm$^{-2}$ $z=0.518$ 
system toward B2 0827+24,
the $N_{HI}=2.1\pm0.5\times10^{20}$ atoms cm$^{-2}$ $z=1.391$ 
system toward Q0957+561A,
the $N_{HI}=2.0\pm1.0\times10^{20}$ atoms cm$^{-2}$ $z=0.633$ 
system toward Q1209+107, and
the $N_{HI}=2.3\pm0.4\times10^{20}$ atoms cm$^{-2}$ $z=0.656$ 
system toward 3C 336 (1622+239).
As noted above, we found no systems which lie slightly below the \HI\
column density threshold.  In order to correct for the Malmquist bias we
have therefore excluded half of these four systems by giving them weight
0.5 when forming the DLA statistical sample for the purpose of determining
$n_{DLA}(z)$ and $\Omega_{DLA}(z)$. The correction will have the net effect of 
lowering our $n_{DLA}(z)$ determination by 20\% (\S4.2), but in fact it has
a minor effect on the determination of $\Omega_{DLA}(z)$ since the 
contributions to $\Omega_{DLA}(z)$ come from the highest column density systems. 
After correcting for Malmquist bias, we effectively have 10 DLA
systems among 87 \MgII systems with \w $\ge 0.3$ \AA\ and 9 DLA systems
among 44 \MgII systems with \w $\ge 0.6$ \AA.

It should be noted that, in principle, a Malmquist bias must also affect
the \MgII samples themselves since they are defined using threshold values for
\w. Since our particular \MgII sample was compiled from various sources,
 it is difficult to assess the degree to which the Malmquist bias
affects the sample selection for \w $\ge 0.3$ \AA.  However, 
with the exception of one system, our DLA detections
arise in the \w $\ge 0.6$ \AA\ subsample (\S3.1), so we believe that 
the effects
of truncation at some \w\ threshold are understood. Moreover, when
we considered effects due to \w\ selection in RTB95, we found that
when measurement errors were well-determined, the resulting statistical results 
remained unchanged for samples with different threshold \w.  Therefore,  
 we will ignore the effect of any Malmquist bias in the \MgII samples.

\subsection{$n_{DLA}(z)$}

In \S3.1 we showed that the low-redshift DLA absorbers are almost
exclusively drawn from the  population of absorbers with \w
$\ge0.6$ \AA.  Therefore, we will use the \w $\ge 0.6$ \AA\
\MgII sample to derive the statistical properties of DLA systems at
low redshift. We start with the redshift evolution of the number
density distribution of \MgII systems with \w $\ge 0.6$ \AA,
$n_{MgII}(z,W_{min}^{\lambda2796}=0.6)$.  SS92 find that
\begin{equation}
n_{MgII}(z,W_{min}^{\lambda2796}=0.6)=(0.24\pm0.10)(1+z)^{1.02\pm0.53}.
\end{equation}
$n_{DLA}(z)$ can then be determined by noting the fraction of DLA
absorbers among the \MgII absorber population. Given the small number of
DLA systems, we can reasonably split the redshift range, $0.11<z<1.65$,
into at most two intervals. Recall that $z=0.11$ is the low redshift
limit of optical spectroscopic surveys for \MgII  and $z=1.65$ is
approximately the upper redshift limit of UV surveys for Ly$\alpha$.
The mean redshift of the 44 \MgII systems with \w $\ge0.6$ \AA\ is
$<z>=0.83$.  We observed 21 \MgII systems in the interval $0.11<z<0.83$
with $<z>=0.49$. Employing the half-weighting method to correct for
Malmquist bias (\S4.1), three DLA systems in this redshift interval are
assigned weight 1.0 and three others have weight 0.5. We observed 23
\MgII systems in the interval $0.83<z<1.65$ with $<z>=1.15$.  Four of
these are DLA systems with an assigned weight of 1.0, and one other 
has weight 0.5. Thus, using equation (5) we find,
\begin{eqnarray}
n_{DLA}(z=0.49) & = & 
(\frac{3^{+2.92}_{-1.66}}{21} + 0.5\frac{3^{+2.92}_{-1.66}}{21})\times \nonumber \\
 & & (0.24\pm0.10)(1+0.49)^{1.02\pm0.53} \nonumber \\
& = & 0.08^{+0.06}_{-0.04}
\end{eqnarray}
and
\begin{eqnarray}
n_{DLA}(z=1.15) & = & 
(\frac{4^{+3.16}_{-1.91}}{23}+0.5\frac{1^{+2.30}_{-0.83}}{23})\times \nonumber \\
& & (0.24\pm0.10)(1+1.15)^{1.02\pm0.53} \nonumber \\
& = & 0.10^{+0.10}_{-0.08}
\end{eqnarray}
where errors derived from small-number Poisson statistics are reported
(Gehrels 1986). These  results are shown in Figure 29.

Figure 29 shows the overall results on $n_{DLA}(z)$ in
the redshift interval $0<z<4.7$.  The logarithm of $n_{DLA}(z)$ is plotted
against $\log(1+z)$. The corresponding linear redshift scale is also shown.
The points at $z>1.65$ are taken directly from  WLFC95
and Storrie-Lombardi \& Wolfe (2000,
in preparation, kindly communicated to us by L. Storrie-Lombardi). The
highest-redshift data include  recent {\it confirmed} DLA detections. The
$z=0$ point is derived from the observed \HI\ distribution in local
spiral galaxies (RTB95) and represents the probability of
intercepting \HI\ columns with $N_{HI}\ge2\times10^{20}$ atoms cm$^{-2}$.

Due to small number statistics, the error bars shown in Figure 29 are
significant, exceeding approximately 60\% in the low-redshift bins and
approximately 20\% in the moderate to high redshift bins (excluding the
highest redshift bins). Nevertheless, it is appropriate to address the
issue of whether the new low-redshift result on $n_{DLA}(z)$ can be {\it
directly} compared to the higher redshift results as done in Figure 29.
This is because, even after setting aside the issue of possible biases 
(e.g. biases due to gravitational lensing or dust obscuration), the 
completeness of the higher redshift surveys should be considered.
These surveys include the ones described in WTSC86, Lanzetta et al. (1991),
 LWT95, and WLFC95.  WLFC95 explain that their results
are derived from the statistical analysis of 62 DLA systems, but that
``the complete statistical sample is likely to exceed 62 DLA systems,
with additional systems coming from the 102 untested candidates with
(Ly$\alpha$ absorption rest equivalent widths) $5 \le W \le 10$ \AA.''
They further explain that ``the 102 untested candidates are likely to
be confirmed (to be classical DLA absorption systems) at a rate lower
than one in four.''  Thus, the problem with {\it directly} comparing the
low-redshift results on $n_{DLA}(z)$ to the high-redshift results (Figure
29), is the absence of a correction for the high-redshift incompleteness.
One way to account for high-redshift incompleteness is to perform a
statistical analysis which uses the considerably more complete subset
of DLA systems which have $N_{HI} \ge 4\times10^{20}$ atoms cm$^{-2}$
(e.g. as done in Lanzetta et al. 1991). However, since our low-redshift
survey does not suffer from this problem, the method we adopt
here is to correct for the high-redshift incompleteness. The details
of this correction are discussed below and the corrected $n_{DLA}(z)$
results are shown in Figure 30.

\begin{figure*}
\plotone{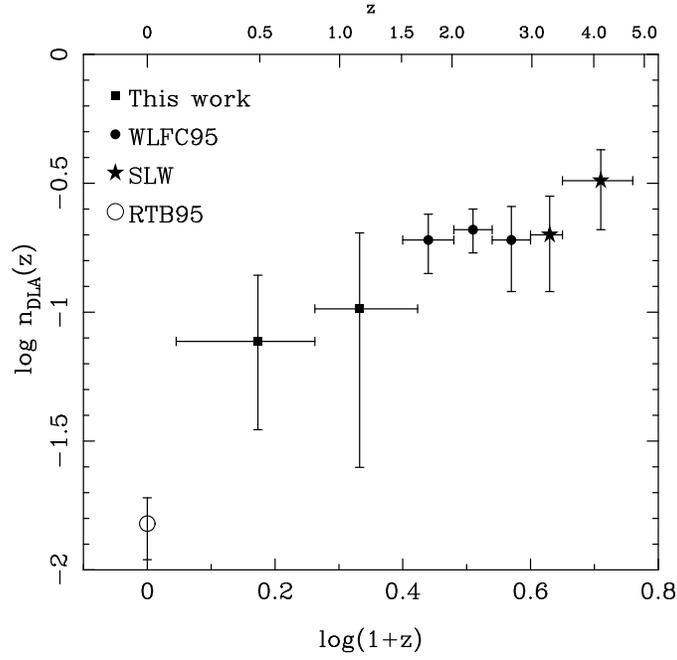}
\caption{A plot of the logarithm of $n_{DLA}(z)$, the number of DLA systems per 
unit redshift along a line of sight, as a function of the logarithm of $(1+z)$. 
The corresponding linear redshift scale is also shown. The new low-redshift
results are shown as solid squares. The solid circles are from WLFC95 and
the solid stars are from Storrie-Lombardi \& Wolfe (in preparation). These
data include recent {\it confirmed} DLA detections. The
$z=0$ point is derived from the observed H I distribution in local
spiral galaxies (RTB95) and represents the probability of
intercepting H I columns with $N_{HI}\ge2\times10^{20}$ atoms cm$^{-2}$.
Vertical bars correspond to $1\sigma$ errors and horizontal bars indicate
bin sizes.}
\end{figure*}

\begin{figure*}
\plotone{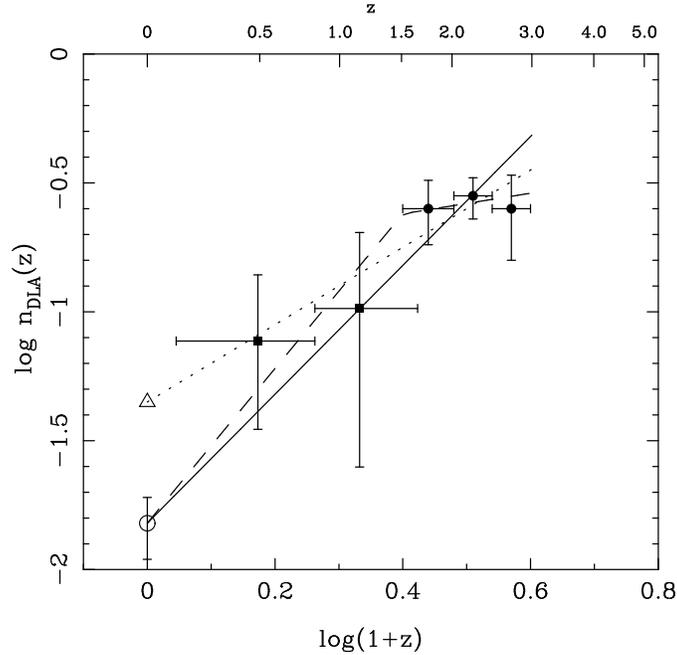}
\caption{Same as Figure 29 but with a Malmquist bias 
correction factor applied to 
the high-redshift data of WLFC95 (see text). Results for $z>3$ are not 
included since the high-redshift study of Storrie-Lombardi \& Wolfe 
is in preparation, and the correction factors, if required, cannot be assessed. 
Three different forms of the power-law fit are shown (see equation (8)).
The solid line has $\gamma=2.5$ and is forced to pass through
the $z=0$ data point, the dotted line has $\gamma=1.5$ and does not 
include the $z=0$ data point. Extrapolation of this power-law to $z=0$
(the open triangle) results in a value that is $\approx3$ times larger 
than the observed incidence at $z=0$. The dashed line has $\gamma=0.4$ for
$z>1.5$ and $\gamma=3.0$ for $z<1.5$.}
\end{figure*}

First, we note that eight of the 102 untested candidates referred to
by WLFC95 come from the low-redshift IUE sample of  LWT95. In addition,
four of the 62 DLA systems in the WLFC95 statistical sample have redshift
$z<1.65$. This suggests that there are 58 confirmed high-redshift DLA
systems in the WLFC95 statistical sample plus another 94 ``untested'' candidates
as of 1995. Our current count of untested high-redshift candidates
shows that 89 remain.  
To assess how the untested candidates may affect the high-redshift
$n_{DLA}(z)$ results, we note the following: (1) in the WLFC95 $1.5 < z \le
2.0$ bin there are 18 confirmed DLA systems and 33 untested candidates
(in reality this bin can only extend down to $z \approx 1.65$), (2) in
the WLFC95 $2.0 < z \le 2.5$ bin there are 25 confirmed DLA systems and 39
untested candidates, (3) in the WLFC95 $2.5 < z \le 3.0$ bin there are 11
confirmed DLA systems and 12 untested candidates, and (4) in the WLFC95
$3.0 < z \le 3.5$ bin there is one confirmed DLA system and 5 untested
candidates.  If the effective success rate for untested candidates (i.e.
minus the correction for ``confirmed'' DLA systems that turn out to lie
slightly below the $N_{HI}=2\times10^{20}$ atoms cm$^{-2}$ threshold)
is taken to be 20\%, this indicates that the high-redshift $n_{DLA}(z)$
results in WLFC95 should be increased significantly relative to the
reported $n_{DLA}(z)$ errors (see Figure 29). Evidence for such
a high effective success rate is readily apparent in figure 2 of Lanzetta
et al. (1991). Thus, we find correction factors of $1.37\pm0.14$, 
$1.31\pm0.11$, and $1.22\pm0.14$ for bins $(1)-(3)$, respectively. 
The errors on the
correction factors suggest that differences between the correction factors 
for different bins are not significant. Thus, an overall correction factor
of $1.31\pm0.07$ appears to apply to the high redshift data of WLFC95. The
new corrected results on $n_{DLA}(z)$ in the redshift interval $0\le z\le 3$ 
are shown in Figure 30 and we adopt them for the remainder of our analysis.
Results for $z>3$ are not considered since the high-redshift study of
Storrie-Lombardi \& Wolfe is in preparation and the correction factor, if 
required, cannot be assessed. This new result is still
qualitatively consistent with a linear decline in the incidence of DLA
systems with decreasing redshift. 

While the statistics are poor, trends can nevertheless be extracted from the 
data. The decline in $n_{DLA}(z)$ can be expressed using the parameterization
that is normally adopted to describe the statistics of QSO absorption-line
systems, 
\begin{equation} 
n_{DLA}(z) = n_0(1+z)^{\gamma}.
\end{equation} 
Therefore, several forms of this power-law parameterization are shown on Figure 30.  
First, a single power-law is overplotted to include all of the data, including 
the $z=0$ point (i.e. the open circle, RTB95). The power-law has index $\gamma = 2.5$,
but a single power law may not fit all of the data satisfactorily.  Second is a 
single $\gamma=1.5$ power law overplotted on all of the data except the 
$z=0$ point.  This very adequately describes
the incidence of DLA systems in QSO absorption-line surveys, but the
extrapolation over-predicts the inferred incidence of DLA systems at $z=0$
by a factor of $\approx$ 3. Third is a broken power-law overplotted on all the
data, including the $z=0$ point. For convenience, we have assumed that
the break occurs at redshift $z=1.5$, consistent with the determination
of the position of the break from the {\it HST} QSO Absorption-Line Key Project
data which applies to the lower column density Ly$\alpha$ forest lines
(Weymann et al. 1998).  However, at $z<1.5$ the Key Project
determined $\gamma \approx 0.16$ for the
Ly$\alpha$ forest but we find that a much steeper decline, with 
$\gamma \approx 3.0$, applies for the DLA systems if the $z=0$ point is included.
At $z>1.5$ Bechtold (1994) found $\gamma \approx 1.85$ for the Ly$\alpha$ 
forest but we find $\gamma\approx 0.4$ for the DLA systems. Thus, the data 
indicate that the broken power-law indices for the
DLA systems are significantly different from the indices determined
 for the Ly$\alpha$ forest. For the DLA systems, the incidence is nearly flat at 
moderate-to-high redshift, but then must decline rapidly at low redshift to
fit the $z=0$ point, while the Ly$\alpha$ forest shows a more rapid decline 
at high redshift but then flattens at low redshift.
However,  the Key Project data do show evidence for a steepening
in the incidence of Ly$\alpha$ forest lines with increasing rest equivalent 
width. The incidence of DLA systems at $z<1.5$ is qualitatively 
consistent with this trend, but it is not consistent with this trend at 
$z>1.5$. Moreover, any break in the DLA power-law fit may actually occur
at some redshift other than 1.5, but this cannot be determined from
 the present data.

{\it What do these results imply?}
Recall that for $\Lambda=0$ cosmologies, the comoving number density of
absorbers is consistent with no intrinsic evolution if $\gamma=0.5$ for
$q_0=0.5$ or if $\gamma=1.0$ for $q_0=0$. Thus, if we neglect the $z=0$
point, the present data on $n_{DLA}(z)$ with $\gamma=1.5$ shows some evidence 
for real intrinsic evolution, but the statistics are not good enough to 
definitively rule out no evolution
in a $q_0=0$ universe. However, if a no-evolution trend is
valid down to $z=0$, it implies that the \HI\ 21 cm emission surveys are
missing $\approx$ 70\% of gas-rich systems with $N_{HI}\ge2\times10^{20}$ 
atoms cm$^{-2}$ locally. We consider this possibility unlikely since  
\HI\ 21 cm emission surveys  have
not revealed a significant population of gas-rich objects that are not
already included in optical surveys (Briggs 1997; Zwaan et al. 1997). These
\HI\ surveys are sensitive down to $N_{HI}>10^{18}$ atoms cm$^{-2}$ and are
capable of detecting masses $M_{HI} > 2\times10^7 h_{65}^{-2}$ M$_{\odot}$.
Alternatively, the need for a large value for the power-law index, e.g.,
$\gamma \approx 2.5$ or the broken power law with $\gamma=0.4$ and $\gamma=3$
($z>1.5$ and $z<1.5$, respectively), arises solely because of the inclusion of 
the $z=0$ data point. Both this steep single power-law or the broken
power-law imply that the DLA absorbers are undergoing
significant evolution. At $z\approx2$ the incidence of DLA
systems is a factor $\approx$ 17 times larger than inferred for $z=0$. 
This is $\approx 6$ times larger than what is predicted in a $q_0=0$
no-evolution universe ($\approx 10$ times larger if $q_0=0.5$).  Since
$n_{DLA}(z)$ is proportional to the number of absorbers per comoving
volume times their \HI\ cross-section, the decline in incidence suggests
that at $z < 2$, during the last 67\% ($q_0=0$) of the
age of the Universe (81\% for $q_0=0.5$), there was either a decrease
in the effective \HI\ cross-sections of DLA absorbers, a decrease in
the number of DLA absorbers per comoving volume, or a combination of
both effects.  A decrease in cross-section may correspond to a collapse
phase for DLA absorbers, while a decrease in the number of DLA absorbers
may be the result of mergers among them.

\subsection{$\Omega_{DLA}(z)$}

The cosmological mass density of neutral gas in DLA systems, $\Omega_{DLA}(z)$, 
is determined using $n_{DLA}(z)$ and the mean column density of DLA 
systems as a function of redshift, $\langle N_{HI}(z)\rangle$, as follows 
\begin{equation}
\Omega_{DLA}(z)=\frac{\mu 8 \pi G m_{H}}{3cH_0}
\frac{n_{DLA}(z) \langle N_{HI}(z)\rangle}{(1+z)}
\end{equation}
for $q_0=0$ and 
\begin{equation}
\Omega_{DLA}(z)=\frac{\mu 8 \pi G m_{H}}{3cH_0}
\frac{n_{DLA}(z) \langle N_{HI}(z)\rangle}{(1+z)^{1/2}}
\end{equation}
for $q_0=0.5$. Here $\mu=1.3$ corrects for a neutral gas composition of
75\% H and 25\% He by mass, $G$ is the gravitational constant, and
$m_H$ is the mass of the hydrogen atom. A Hubble constant of 
$H_0=65$ km s$^{-1}$ Mpc$^{-1}$ is used for all calculations. 
To determine $\Omega_{DLA}(z)$ at
low redshift we use the DLA systems and associated column densities given
in Table 3.  As before, we split the redshift interval $0.11<z<1.65$
into two redshift bins. Three DLA systems with weight 1.0 and
three DLA systems with weight 0.5 are in the lower redshift bin;
four DLA systems with weight 1.0 and one with weight 0.5 are in the
higher redshift bin.  For $q_0=0$, we find
$$\Omega_{DLA}(z=0.49) = \left(1.82^{+1.48}_{-1.00}\right)\times 10^{-3}$$ and
$$\Omega_{DLA}(z=1.15) = \left(1.45^{+1.45}_{-1.17}\right)\times 10^{-3}.$$
For $q_0=0.5$, we find
$$\Omega_{DLA}(z=0.49) = \left(2.23^{+1.55}_{-1.22}\right)\times 10^{-3}$$ and
$$\Omega_{DLA}(z=1.15) = \left(2.13^{+2.13}_{-1.71}\right)\times 10^{-3}.$$

\begin{figure*}
\centerline{\psfig{file=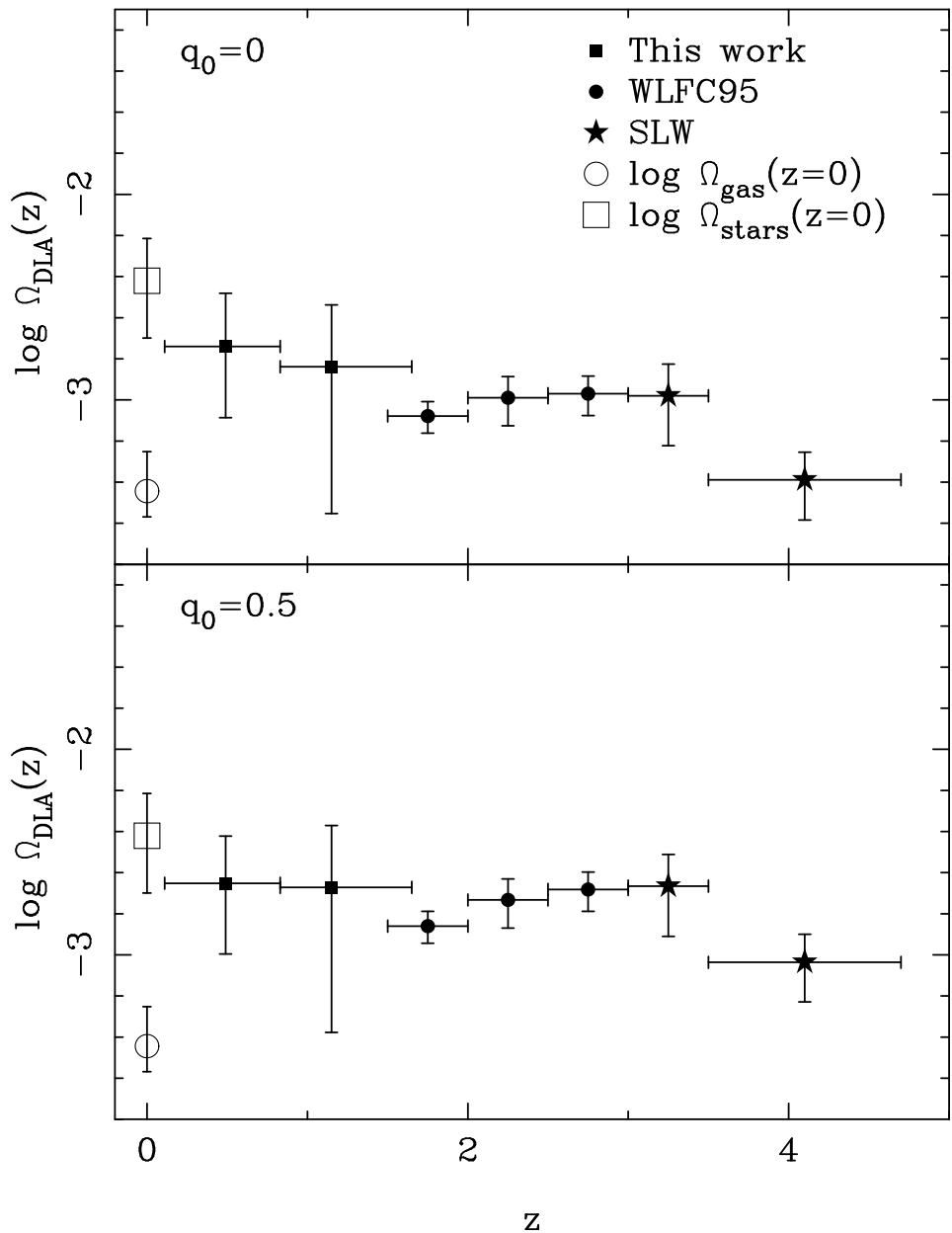,height=6.5in,width=5.0in}}
\caption{A plot of the logarithm of $\Omega_{DLA}(z)$ with 
$H_0=65$ km s$^{-1}$ Mpc$^{-1}$ for $q_0=0$ (top panel) and 
$q_0=0.5$ (bottom panel). The new low-redshift
results are shown as solid squares. The solid circles are from WLFC95 and
the solid stars are from Storrie-Lombardi \& Wolfe (in preparation). 
Vertical bars correspond to $1\sigma$ errors and horizontal bars indicate
bin sizes. The open circle at $z=0$ is the local 
neutral gas mass density as measured by Rao \& Briggs (1993) and the open 
square at $z=0$ is the local luminous mass density in stars 
(Fukugita, Hogan, \& Peebles 1998).}
\end{figure*}

The logarithm of $\Omega_{DLA}(z)$
for $q_0=0$ and $q_0=0.5$ is plotted in Figure 31. The high-redshift
results of WLFC95 and  Storrie-Lombardi \& Wolfe (2000, in preparation)
are shown as solid circles and stars, respectively.  In this case,
the completeness correction (see \S4.2) to the high-redshift results
on $\Omega_{DLA}(z)$ is small ($\approx 7$\%) relative to the
reported errors and has not been applied to the data. See \S6 for a
discussion of what are likely to be more significant errors in the
determination of $\Omega_{DLA}(z)$.  The new low-redshift points are
plotted as solid squares and are seen to be of comparable value to the
points at higher redshift.  The errors on the low-redshift points are
large owing to the small number of low-redshift DLA systems studied.
The open circle at $z=0$ is the local neutral gas mass density as
measured by Rao \& Briggs (1993). The open square at $z=0$ is the
local luminous mass density in stars as determined by Fukugita, Hogan, \&
Peebles (1998).

The results on $\Omega_{DLA}(z)$ are similar to the $n_{DLA}(z)$ results in that
they show strong evidence for evolution if the $z=0$ point is included.
If $q_0=0$, the comoving cosmological mass density of high-redshift
DLA absorbers is $\approx 4$ times larger than what is
inferred from the local 21 cm observations ($\approx 6.5$ times larger if
$q_0=0.5$).  Alternatively, if the $z=0$ point is excluded, there is no
convincing evidence for a decrease in $\Omega_{DLA}(z)$ with decreasing
redshift even at redshifts as low as $z \approx 0.5$.
In fact, the actual results reveal a tendency for $\Omega_{DLA}(z)$
to rise slightly with decreasing redshift in the interval $z \approx
2 - 0.5$.  As expected, this tendency is somewhat more pronounced in
a $q_0=0$ universe than in a $q_0=0.5$ universe.  However, the error
bars on the $\Omega_{DLA}(z)$ low-redshift points are fractionally
much larger than the $n_{DLA}(z)$ error bars and therefore more
problematic. This is because the determination of $\Omega_{DLA}(z)$ is
very sensitive to the small number of systems which have the highest
column densities. Thus, the current results do not provide {\it very
strong} constraints on the evolution of the amount of neutral gas in
the Universe in the redshift interval $0.1<z<1.6$.

Nevertheless, assuming that the $z=0$ point is correct (i.e. assuming we
have not missed a significant fraction of the $z=0$ neutral gas in local
21 cm emission surveys), the data would be consistent with the possibility
of a rapid decline in both $n_{DLA}(z)$ and $\Omega_{DLA}(z)$ from
$z \approx 0.5$ to $z=0$. If $q_0=0$, this would imply that
the DLA absorber population lost $\approx$ 75\% of its neutral gas mass over
the most recent $\approx$ 33\% of the age of the Universe (if $q_0=0.5$,
$\approx$ 85\% of the mass would be lost during the 
most recent $\approx$ 46\% of the age of the Universe).
We also note that, except for the $z=4.1$ data point,
all the $z>0$ data points in Figure 31 lie within $1\sigma$ of
constant $\log\Omega_{DLA}(z)=-3.0$ for $q_0=0$ and 
$\log\Omega_{DLA}(z)=-2.8$ for $q_0=0.5$. The $z=4.1$ data point is 
discrepant from a constant value at about the $2\sigma$ level.

{\it What do these results imply?} In the context of the
conversion of neutral gas into stars and global galaxy formation as
might be tracked by $\Omega_{DLA}(z)$, {\it the current data provide no
evidence which would support the view that the bulk of galaxy formation
happened, for example, at redshifts $z>1$}.  This is in contradiction
with earlier conclusions reached by   LWT95 on low-redshift DLA; their
results showed a steady decline in $\Omega_{DLA}(z)$ from $z\approx3.5$ to
$z\approx 0.008$. The constant
neutral gas mass density from $z\approx 4$ to $z\approx 0.5$ that our 
new results imply is also at odds with the evolution of
the star formation density in the Universe as derived from the UV
luminosity density of galaxies.  Steidel et al. (1999) show that after
correction for effects of dust obscuration, the star formation
density remains roughly constant from $z\approx 4$ to $z\approx 1$.
The data for $z<1$ from Lilly et al. (1996) indicate that the star formation
density falls by a factor of $\approx 5$ from $z\approx 1$ to $z\approx 0.2$.
The star formation density and the gas mass density cannot both be constant
over the same range of redshifts ($1<z<4$) if they track the same objects. 
This implies that DLA galaxies and star-forming galaxies trace different
populations (see also Pettini et al. 1999, Steidel et al. 1999 and references
therein). However, star-forming galaxies must have neutral gas from
which to make stars. There are possibly three reasons why searches for DLA
systems are not finding star-forming galaxies: (1) the intensive star-forming phase
of UV-luminous galaxies is short-lived, (2) background quasars are
obscured by dust in the UV-luminous galaxies, and/or (3) the UV-luminous galaxies 
are likely to be the brightest galaxies at the epochs they are detected (whereas,
random sight lines that pick up the DLA population are more likely 
to intersect \HI\ regions in the more common type of object, i.e., $L^*$
or fainter galaxies.  Additional evidence from metallicity studies 
and direct imaging also point to DLA galaxies as being a slowly evolving
population.  We will return to the discussion on the nature of
DLA galaxies in \S7. 

\section{The Column Density Distribution Function of DLA Systems}

The column density distribution function, $f(N_{HI},z)=d^2{\cal
N}(z)/dN_{HI}dX$, specifies the number of absorbers per unit
\HI\ column density per unit absorption distance, where the
absorption distance is $X(z)=\frac{1}{2}[(1+z)^2-1]$ if $q_0$=0 and
$X(z)=\frac{2}{3}[(1+z)^{1.5}-1]$ if $q_0$=0.5. Thus, $f(N_{HI},z)dN_{HI}dX$
is the number of absorbers with \HI\ column density between $N_{HI}$ and
$N_{HI} + dN_{HI}$ in absorption distance interval $dX$. 

There are several reasons why it is useful to study this function. For
example, its shape can reveal information on the geometry of the absorbing
regions (e.g. Milgrom 1988). More ideally, evolutionary models which
also take into account the global physical state of the gas (the H$^+$,
H$^0$, and H$_2$ fractions) and the conversion of gas into stars would
predict $f(N_{HI},z)$. This prediction could then be empirically tested.
Unfortunately, the statistics are still too poor to realize this
goal in detail. However, we can form a sample which will permit us to
derive the most accurate results currently possible on the \HI\ column
density distribution function of low-redshift DLA systems (\S5.1). We
accomplish this by adopting a rationale that allows us to form the
largest possible unbiased low-redshift sample. After correcting the
sample for Malmquist bias, the normalized cumulative column density
distribution can be derived.  We then use the KS
test to compare the low-redshift normalized cumulative column density
distribution function to ones derived from high-redshift DLA systems
(\S5.2) and local spirals (\S5.3).  Most notably, we show that the
comparisons reveal clear evidence for differences in the shapes of
these three column density distribution functions, which we interpret
as being indicative of a major evolutionary trend.  These results are
then converted to absolute DLA column density distribution functions to
show the redshift evolution (\S5.4).

\subsection{The Low-Redshift DLA Column Density Distribution}

\subsubsection{Rationale for Sample Selection}

The selection criteria for constructing a DLA column density 
distribution are different than what were used to define the
RT, and hence, the RT-DLA samples (\S3). We now explain how this sample
was constructed. 
To derive the most accurate results possible on the low-redshift DLA
column density distribution function we should use all available unbiased
information on all known low-redshift DLA systems.  Table 7 lists all of the 
confirmed low-redshift ($z<1.65$)
classical DLA ($N_{HI} \ge 2 \times 10^{20}$ atoms cm$^{-2}$) systems from
cosmologically intervening neutral gas that are currently known to us, 
without regard to possible bias. By confirmed systems we mean ones for 
which good UV data have been obtained on the Ly$\alpha$ absorption-line 
profile or for which 21 cm absorption lines have been detected.
For both of these cases, a reliable estimate of the \HI\
column density can be made. This is preferably done by fitting a Voigt
profile to the damping wings of the Ly$\alpha$ absorption line (\S2),
although use of the radio data and the assumption that the gas has 21
cm spin temperature of $T_s \approx 700$ K can also be used to produce a
reliable estimate (Lane et al. 2000, in preparation). Thus, the systems 
included in Table
7 form a biased sample of low-redshift classical DLA systems. Excluded
from Table 7 are any non-classical ``damped Ly$\alpha$ systems'' with
\HI\ column densities more than 2$\sigma$ less than $2 \times 10^{20}$
atoms cm$^{-2}$  (although damping wings on the Ly$\alpha$ absorption-line
profile may still be easily visible in such systems), and all potential
DLA systems which have been proposed on the basis of certain metal-line
ratios, x-ray absorption, or still even less-direct evidence. Also excluded
are 21 cm absorption systems that are not {\it cosmologically intervening}
along the sight-line to the quasar, but instead more likely arise in the 
quasar's host galaxy (e.g. Carilli et al. 1998).

\begin{deluxetable}{ccccclc}
\tablewidth{7.0in}
\tablenum{7}
\tablecaption{Confirmed Low-Redshift Classical DLA Systems}
 \tablehead{ 
\colhead{QSO} & 
\colhead{V} &
\colhead{$z_{em}$} &
\colhead{$z_{DLA}$} &
\colhead{$N_{HI}/10^{20}$} &
\colhead{Notes} &
\colhead{Reference}\\[.2ex]
\colhead{} &
\colhead{} &
\colhead{} &
\colhead{} &
\colhead{atoms cm$^{-2}$} &
\colhead{} &
\colhead{}
}
\startdata
0143$-$015 & 17.7 & 3.14 & 1.613  & 2.0  & \nodata & 1 \\
0218+357   & 20.0 & 0.94 & 0.687  & 13  & Biased (red quasar) & 2\\
0235$+$164 & 15.5 & 0.94 & 0.524  & 50  & Biased (21 cm) &  3\\
0248$+$430 & 17.6 & 1.31 & 0.394  & 36  & \nodata & 1 \\
0302$-$223 & 16.4 & 1.41 & 1.010  & 2.3 & \nodata  & 1 \\
0454$+$039 & 16.5 & 1.34 & 0.859  & 4.7 & \nodata & 1 \\
0500+019   & 21.2 & $>$0.58 & 0.585\tablenotemark{a} & 43\tablenotemark{b} & 
                       Biased (red quasar) &  4 \\
0738$+$313 & 16.1 & 0.63 & 0.091  & 15  & \nodata & 1, 5 \\
\nodata  & \nodata & \nodata  & 0.221  & 7.9 & \nodata & 1, 5 \\
0809+483   & 17.8 & 0.87 & 0.437  & 6.3 & Biased (21 cm) &  6\\
0827$+$243 & 17.3 & 0.94 & 0.518  & 2.0 & \nodata & 1 \\
0933$+$732 & 17.3 & 2.52 & 1.478  & 42  & \nodata & 1 \\
0935$+$417 & 16.3 & 1.98 & 1.372  & 3.3 & \nodata & 7 \\
0952$+$179 & 17.2 & 1.47 & 0.239  & 21  & \nodata & 1\\
0957$+$561 & 17.0 & 1.41 & 1.391  & 2.1 & \nodata & 1 \\
1122$-$168\tablenotemark{c} & 16.5 & 2.40 & 0.681  & 2.5 & \nodata & 8 \\
1127$-$145 & 16.9 & 1.19 & 0.313  & 51  & \nodata & 1 \\ 
1209$+$107 & 17.8 & 2.19 & 0.633  & 2.0 & \nodata  & 1 \\
1229$-$021 & 16.8 & 1.04 & 0.395  & 5.6 & Biased (21 cm) &  6\\
1328+307   & 17.3 & 0.85 & 0.692  & 15  & Biased (21 cm) &  6\\
1354$+$258 & 18.0 & 2.00 & 1.418  & 32  & \nodata & 1 \\
1622$+$239 & 17.5 & 0.93 & 0.656  & 2.3 & \nodata & 1 \\
1830$-$211 & 22.0 & 0.89 & 0.192  & 2.5 & Biased (red quasar) &  9\\
\enddata

\tablenotetext{a}{\ This system could be the result of absorption by the quasar
host galaxy.}

\tablenotetext{b}{\ The column density was determined assuming a spin temperature
of $T_s=700$K (Lane et al. 2000, in preparation).}

\tablenotetext{c}{\ de la Varga \& Reimers (1997) also detect a DLA candidate
towards Q0515$-$442 at $z=1.15$ using IUE spectra. In the past, IUE spectra have 
been found to be unreliable sources for DLA candidates and we have therefore 
not included this system in Table 7.}

\tablerefs{(1) This work, (2) Carilli, Rupen, \& Yanny 1993,  (3) Cohen et al. 1999,
(4) Carilli et al. 1998, (5) Rao \& Turnshek 1998, (6) Boiss\'e et al. 1998,
(7) Jannuzi et al. 1998, (8) de la Varga \& Reimers 1997, (9) Wiklind \& Combes 1998}
\end{deluxetable}

To derive the correct shape for the column density distribution function,
all systems identified on the basis of a bias in \HI\ column density must be
excluded.  In Table 7 we note which of the systems were selected on
the basis of such a bias. The remaining systems in Table 7 form our
``unbiased sample.''  As before (\S2.4), we exclude the four previously
known 21 cm absorbers. 
We also exclude cosmologically intervening 21 cm absorbers that have
been identified in red quasars (e.g. Carilli et al. 1998 and references
therein) since their selection is based on the bias of observing red
(i.e. dust reddened) quasars. Three systems fit into this category, and in
some cases these also involve gravitational lensing which might produce
an additional bias.  On the other hand, over the last several years four
low-redshift DLA systems were identified in the Ly$\alpha$
forest serendipitously, without regard to column density bias. The DLA
systems discovered in this way were not identified using a technique
which would bias the \HI\ column densities to higher values.  Therefore,
these serendipitously-discovered systems are included in our unbiased
sample. Specifically these are the $z=1.372$ DLA system in Q0935+417
which was found in the {\it HST} QSO Absorption Line Key Project survey
(Jannuzi et al. 1998), the $z=0.681$ DLA system in Q1122$-$168
(de la Varga \& Reimers 1997), and the $z=0.0912$ system in OI 363 and 
the $z=1.613$ system in UM 366 that we found in our {\it HST} spectra.
(Note that these four systems do
not have \MgII information and, therefore, are not part of the RT and
RT-DLA samples.) In summary, the present total low-redshift sample
consists of 23 low-redshift DLA systems (Table 7), seven of which are
excluded due to selection bias. Thus, before correction for Malmquist
bias, we have amassed an unbiased sample of 16 systems to
study the low-redshift DLA column density distribution function.

The reasons for making a correction for Malmquist bias were discussed in
\S4.1. To approximately correct our sample for this effect we have given
the five systems with $N_{HI} \approx 2\times10^{20}$ atoms cm$^{-2}$
a weight of 0.5 in the column density distribution analysis.  

\subsubsection{Normalized Cumulative Column Density Distribution Function}

Shown in Figure 32 as a solid curve with open circles is the normalized
low-redshift DLA cumulative column density distribution function. This
was derived using the Malmquist-bias-corrected sample discussed above which 
has mean redshift $<z> \approx 0.78$.
Since the curve is normalized, we can compare the shape of this curve
to the shapes of similarly derived column density distribution functions
for high-redshift DLA absorbers and local spirals. This will be done in
\S5.2 and \S5.3, respectively.

\subsection{Comparison with Results at High Redshift}

\begin{figure*}
\plotone{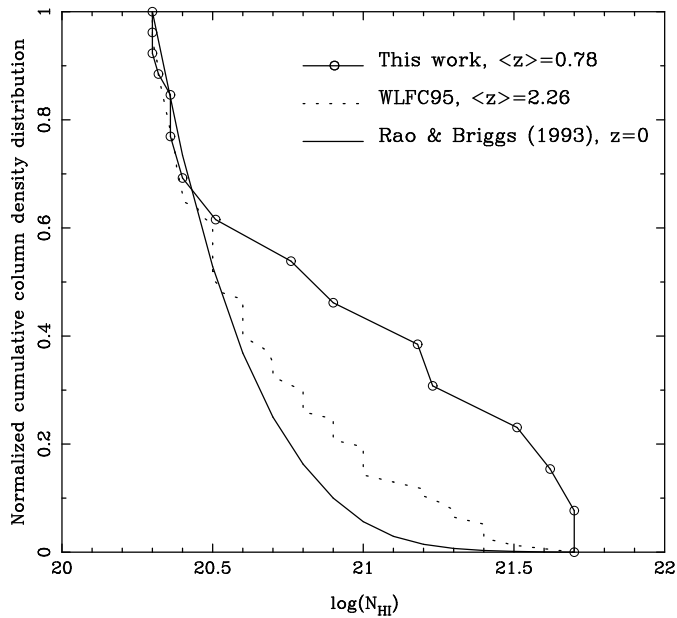}
\caption{The normalized cumulative column density distributions for  
the low-redshift DLA sample (solid curve with open circles),
the WLFC95 high-redshift DLA sample (dotted curve), and local
galaxies (solid curve). The WLFC95 and RT-DLA samples have been corrected 
for Malmquist bias. Data for local galaxies are from Rao \& Briggs (1993).}
\end{figure*}

  In order to compare our
low-redshift result with results obtained at higher redshift, we have
derived the high-redshift DLA normalized cumulative column density
distribution function using the data of WLFC95.  We did this by forming
an unbiased sample of high-redshift column densities as follows. First,
we included all 43 high-redshift ($z>1.65$) systems with higher-resolution
follow-up spectroscopy which have been used to confirm, with little doubt,
that these systems are damped.  In these cases \HI\ column densities were
derived by fitting Voigt damping profiles to the Ly$\alpha$ absorption
lines. We then included all 15 candidate DLA systems that have Ly$\alpha$
rest equivalent widths $>10$ \AA.  These systems do not have follow-up
spectroscopy, but their large rest equivalent widths formally indicate
that they have $N_{HI} > 2\times10^{20}$ atoms cm$^{-2}$, although a few
may not. We converted these rest equivalent widths to column densities
using the relation $N_{HI} = 2(W^{\lambda1216}_{0}/10 $\AA$)^{1/2}
\times 10^{20}$ atoms cm$^{-2}$. These 58 systems  comprise
the high-redshift ($z>1.65$) statistical sample of WLFC95. All
but one of these systems lies in the redshift interval $1.65<z<3.0$.
Finally, in keeping with the procedure adopted in \S4.2, we assumed that
18 of the remaining 89 ``untested'' candidate systems ($\approx$ 20\%) are
damped and, in reality, have $N_{HI} \ge 2\times10^{20}$ atoms cm$^{-2}$.
When adding these 18 systems to the unbiased high-redshift sample, we
have included them evenly over the column density interval $20.3 \le
\log(N_{HI}) \le 20.4$, consistent with likely measurement errors.
Recall that WLFC95 conclude that they expect these untested candidates to
be confirmed at a rate lower than one in four. However, evidence for an
effective success of at least 20\% is readily apparent in figure 2 of
Lanzetta et al. (1991).  We chose a success rate of 20\% since some of
the systems included in the first two steps may actually have $N_{HI}$
slightly smaller than $2\times10^{20}$ atoms cm$^{-2}$.  This makes an
approximate correction for the Malmquist bias in the high-redshift sample.
The resulting normalized cumulative column density distribution function
derived from this unbiased high-redshift sample is plotted on Figure 32
along with the low-redshift result. The high-redshift result is derived
from an effective sample size of 76 DLA systems with mean redshift $<z>
\approx 2.26$.

A KS test comparing the low- and high-redshift results indicates that there
is only a 2.8\% probability that the
two samples are derived from the same parent population.
This comparison refers only to the shapes of
the column density distribution functions and not their normalizations.
The low-redshift sample clearly contains a much larger fraction of
high column density systems.  
Thus, our analysis indicates that the
interpretation of the current data are inconsistent with an earlier
interpretation of the \HI\ column density distribution of low-redshift
DLA systems (LWT95). Further discussion of this result is deferred to 
\S5.5.  

\subsection{Comparison with Results Derived from Local Spiral Galaxies}

Rao \& Briggs (1993) used \HI\ emission observations of 27 local
spirals and information on the spiral galaxy luminosity function to
construct a local \HI\ column density distribution function. 
The more recent study of Zwaan et al. (1999) is consistent
with the Rao \& Briggs result.  
A comparison of the Rao \& Briggs (1993) result to the one derived from
high-redshift DLA statistics by Lanzetta et
al. (1991) showed clear evidence for evolution in the
number of systems with redshift, but the error bars on the high-redshift
points did not permit any meaningful conclusions to be drawn from a
comparison of the shapes of the local and high-redshift \HI\ column
density distributions.  Now, with improved statistics from high-redshift
DLA systems (WLFC95) and the results on low-redshift DLA systems
reported here, it is possible to perform a more meaningful comparison
of the shapes of the column density distribution functions versus
redshift. The one caveat which might influence comparisons with
the local results is the fact that the beam size used in \HI\ 21 cm
emission surveys might be too large to resolve out the local \HI\ cloud
structure. If this effect were severe enough, it would cause the highest
column densities to be under-represented in the local 21 cm emission
surveys. We believe the importance of this effect is minimal and
discuss it in more detail in \S6.5.
Of course, this is not a problem in the pencil-beam QSO absorption-line
surveys. The result on  the local \HI\
cumulative column density distribution derived by Rao \& Briggs
(1993) is included on Figure 32.  A KS test comparing the
low-redshift DLA result with the local H I result indicates that there
is only a 0.01\% chance that the two samples are derived from the same
 parent population\footnote{A KS test comparing the high-redshift
DLA result with the local \HI\ result indicates that there is only a 
2.0\% chance that these two samples are derived from the same parent 
population.}. Again, the comparison refers only to the shapes of
the column density distribution functions and not their normalizations.
The comparison shows that the local \HI\ sample contains a much smaller
fraction of very high column density gas than is found in the low-redshift
DLA sample.  In fact, the deficit of high column density gas in the
local \HI\ sample is even more pronounced than in the high-redshift DLA
sample (\S5.2).

\subsection{The DLA Absolute Column Density Distribution Function versus 
Redshift}

\begin{figure*}
\plotone{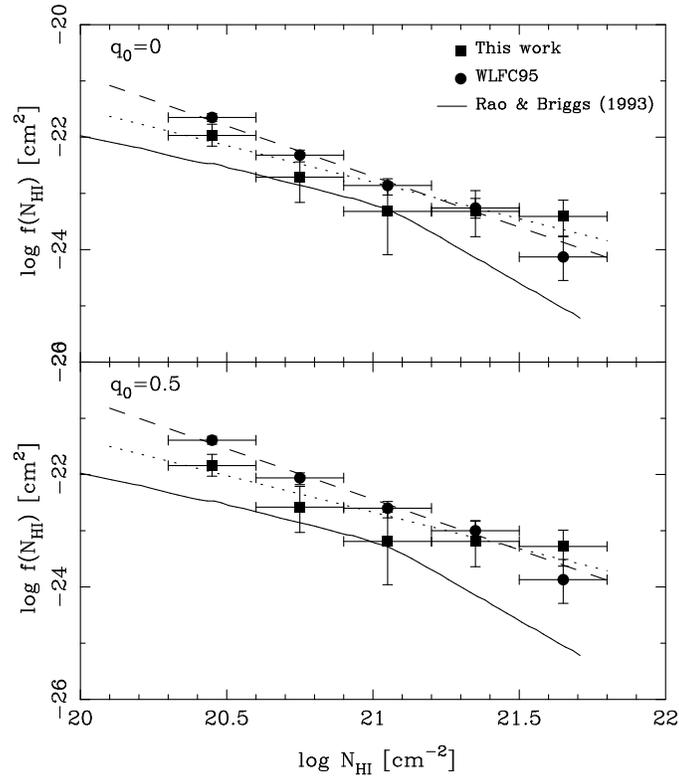}
\caption{The logarithm of the absolute column density distribution function, 
$f(N_{HI})$, for the low-redshift DLA sample (solid squares), 
the WLFC95 high-redshift DLA sample (solid circles), and local galaxies
(solid line) for $q_0=0$ and $q_0=0.5$.  Vertical bars correspond to $1\sigma$
errors and horizontal bars indicate bin sizes. The dashed line is a power-law
fit to the WLFC95 data with $\beta=1.8$ and the dotted line is a power-law
fit to the RT data with $\beta=1.3$.}
\end{figure*}

\begin{figure*}
\plotone{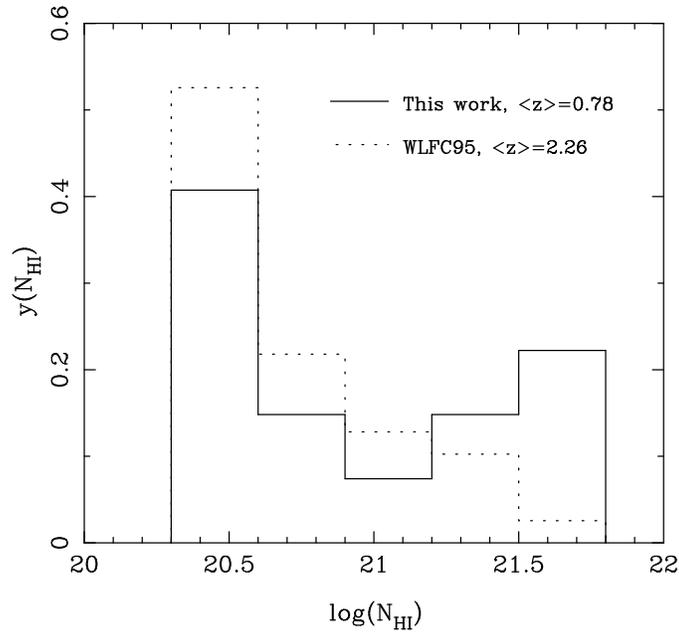}
\caption{The fraction of DLA systems detected as a function of column
density. The bin size is 0.3 in $\log N_{HI}$. Note the higher fraction
of the highest column density systems in the low-redshift sample.}
\end{figure*}

The absolute column density distribution functions at
$z=0$, low redshift ($<z>\approx 0.78$), and high redshift 
($<z> \approx 2.26$) are shown in Figure 33 assuming $q_0=0$ and
$q_0=0.5$. For $z=0$ this function is taken directly from Rao \& Briggs (1993). 
The data points at low and high redshift are determined by first estimating
the fraction of systems detected as a function of column density, $y(N_{HI},z)$,
for the low- and high-redshift samples, respectively. This is shown in Figure
34. Note the higher fraction of the highest column density systems in the low-redshift
sample.  Then, $f(N_{HI},z)$ is given by
\begin{equation}
f(N_{HI},z) = C(z) \frac{y(N_{HI},z)}{\Delta N_{HI}}
\end{equation}
where the constant, $C(z)$, is inversely proportional to the 
absorption distance interval of the survey (see \S5). In turn, $C(z)$
can be determined from $n_{DLA}(z)$ at $<z>=0.78$ and $<z>=2.26$ by using the 
expression relating  $n_{DLA}(z)$ and $f(N_{HI},z)$,
\begin{equation}
n_{DLA}(z)=(1+z)(1+2q_0z)^{-1/2} \sum f(N_{HI},z) \Delta N_{HI}.
\end{equation}
Thus,
\begin{equation}
C(z) = \frac{n_{DLA}(z)}{(1+z)(1+2q_0z)^{-1/2}} \times \frac{1}{\sum y(N_{HI},z)}.
\end{equation}
Note that $\sum y(N_{HI},z)$ is just the total number of systems found in the 
survey. The resulting values of $C(z)$ for the WLFC95 and RT samples are given in 
Table 8 for cosmologies with $q_0=0$ and $q_0=0.5$;
$f(N_{HI},z)$ then follows from equation (11) and Figure 34. 

\begin{deluxetable}{ccccc}
\tablenum{8}
\tablewidth{6.0in}
\tablecaption{Column Density Distribution Function Parameters }
\tablehead{ 
\colhead{Sample} &
\colhead{$<z>$} &
\multicolumn{2}{c}{$C(z)$} &   
\colhead{$\beta$} \\ [.2ex] \cline{3-4} 
\colhead{} &
\colhead{} &
\colhead{$q_0=0$} &
\colhead{$q_0=0.5$} &
\colhead{} 
}
 
\startdata
WLFC95 & 2.26 & $(1.1\pm0.2)\times10^{-3}$ & $(2.0\pm0.4)\times10^{-3}$ & 1.8 \\
RT   & 0.78 & $(3.9\pm2.4)\times10^{-3}$ & $(5.2\pm3.2)\times10^{-3}$ & 1.3 \\
Local galaxies & 0 & \nodata & \nodata & 1.2 (log $N_{HI}<21$) \\
               &   &         &         & 2.9 (log $N_{HI}>21$) \\
\enddata
\end{deluxetable}

In Table 8 we also present the results of parametric fits to the data on
the absolute column density distribution functions. These fits are
over-plotted on the binned data (Figure 33). Given the statistical
accuracy of the data, the results at both high and low redshifts are
fitted by a single power law. It has been argued that such power-law fits
are unphysical (e.g. WLFC95, Storrie-Lombardi, Irwin, \& McMahon 1996a, 
Pei, Fall, \&
Hauser 1999) because, for example, they cause the mass in DLA systems to
diverge when they take the form $f(N_{HI},z) \sim N_{HI}^{ -\beta}$
with $\beta \leq 2$.  While we agree that the power-law fits can be
unphysical, we have nevertheless opted to use them because the data at
high and low redshifts do not offer support for any particular model. 
For example, the exponential-disk model
fits to the column density distributions presented by WLFC95 are arbitrary
in that the data provide no evidence for the breaks that are shown in
their figure 7. In addition, at high redshift the data are in excellent
agreement with a single power-law fit, while too few systems are known at
low redshift to adopt a form other than a single power law.  Moreover,
as we discuss in \S5.5, the shapes of the column density distribution
functions at high and low redshifts appear to be inconsistent with a set
of randomly distributed gas-rich disks.  Therefore, we believe that it is
more appropriate to parameterize the column density distribution functions
in a model-independent way.  At both high and low redshifts, we adopt the
view that there is a cutoff in \HI\ column density at $N_{HI} \approx
5 \times 10^{21}$ atoms cm$^{-2}$. So far, this is the maximum \HI\
column density found for a DLA system at any redshift.

On the other hand, the absolute \HI\ column density distribution at
$z=0$ is clearly not well-represented by a single power law.  There is
clear evidence for a break in the column density distribution at $N_{HI}
\approx 10^{21}$ atoms cm$^{-2}$. We represent the data at $z=0$ using
a  broken power-law (Table 8), but again we assume that there is
a cutoff in the \HI\ column density distribution at $N_{HI} \approx 5
\times 10^{21}$ atoms cm$^{-2}$.

To summarize, our results suggest that at $N_{HI} \approx 2.2
\times 10^{21}$ atoms cm$^{-2}$, $f(N_{HI},<z>\approx2.2)
\approx f(N_{HI},<z>\approx0.8$); however $\beta \approx 1.8$
at $<z>\approx2.2$, while $\beta \approx 1.3$ at $<z>\approx0.8$.
The value of $f(N_{HI},z=0$) is correspondingly lower at all \HI\ column
densities, with $\beta=1.2$ for $N_{HI} < 10^{21}$ atoms cm$^{-2}$
and $\beta=2.9$ for $N_{HI} > 10^{21}$ atoms cm$^{-2}$.

\subsection{Discussion}

Despite our findings on the evolution of $n_{DLA}(z)$ and
$\Omega_{DLA}(z)$, our results on the normalized DLA column density
distribution function, $f(N_{HI},z)$, indicate that there is significant
evidence for real evolution in $f(N_{HI},z)$ even after excluding the
$z=0$ data points (Figures 32 and 33). When we compared these trends using the 
normalized shape of the column density distribution function we found 
them to be highly significant (\S5.2 and \S5.3)\footnote{Also, 
Storrie-Lombardi et al. (1996a) report some evidence for both a steepening 
in the column density distribution function and a decrease the DLA incidence
at the very highest redshifts ($z \approx 4$) that are reachable in surveys.}.
For example, a comparison of the low-redshift and high-redshift normalized
column density distribution functions indicates that there is only an
$\approx$ 2.8\% chance that they are drawn from a population of absorbers
with the same properties.

This result may seem surprising since there was only marginal evidence
for real evolution in $n_{DLA}(z)$ and $\Omega_{DLA}(z)$ per comoving
volume ($q_0=0$ universe) after the $z=0$ data are excluded.
However, a close examination of the data shows that the $n_{DLA}(z)$ and
$\Omega_{DLA}(z)$ results are consistent with the $f(N_{HI},z)$ results.
For a non-evolving $q_0=0$
universe our results show that $n_{DLA}(z)(1+z)^{-1}$ marginally
decreases with decreasing redshift (i.e. in Figure 30 a $\gamma = 1.5$
power law fits the DLA data better than a $\gamma = 1.0$ power law), while
$\Omega_{DLA}(z)$ marginally increases with decreasing redshift (Figure
31). These trends ultimately stem from the differences in $f(N_{HI},z)$
at high and low redshifts, and as discussed above,
our results (Figure 32 and 33) indicate that real evolution has occurred  
in $f(N_{HI},z)$.  The results on $n_{DLA}(z)$ and
$\Omega_{DLA}(z)$ do not explicitly contain any information as a function
of column density. Thus, $f(N_{HI},z)$ provides new information which
$n_{DLA}(z)$ and $\Omega_{DLA}(z)$ do not provide.  

{\it What are the possible origins for the observed evolutionary trend in
the column density distribution function?} We consider this question by
discussing two conceptual models for DLA absorbers.  The first model is
the Giant Hydrogen Cloud (GHC) model proposed by Khersonsky \& Turnshek
(1996). In this model DLA galaxies are simply presumed to contain
GHCs that can be approximated by spheroids. The second model is the
exponential-disk (ED) model which was recently  discussed by WLFC95.
In the ED model the \HI\ column density falls off exponentially with
increasing radius, in approximate agreement with 21 cm observations of
\HI\ in gas-rich spirals (e.g. van Gorkom 1993).  Moreover, at increasing
\HI\ column densities, the column density distribution function must
fall off as $\sim$ $N_{HI}^{ -3}$ for randomly inclined disks (Milgrom
1988; WLFC95).  This is consistent with what is observed locally (Rao \&
Briggs 1993; Zwaan et al. 1999), where the $f(N_{HI}) \sim N_{HI}^{
-3}$ behavior applies at $N_{HI} > 10^{21}$ atoms cm$^{-2}$. In the
ED model of WLFC95, it is supposed that the $f(N_{HI})\sim N_{HI}^{ -3}$ 
behavior sets in at increasingly larger \HI\ values
with increasing redshift.  We note that the GHC and ED models are not
necessarily inconsistent with each other since it is possible to envisage
roughly spherical GHCs arranged in a galactic disk.

We will consider the case where the $z=0$ data must fit into any
evolutionary trend, as this seems to be a reasonable constraint.

{\it What is the cause of the flattening of the $f(N_{HI},z)$ distribution
from high to low redshift?} First, recall that the observed flattening
from $<z> \approx 2.2$ to $<z> \approx 0.8$ is just the opposite of
the evolutionary trend reported by  LWT95.  We will consider the 
evolution in the context of both the GHC and ED models.
For the GHC model, one would expect $f(N_{HI},z)$ to flatten if accretion
processes were taking place that caused the relative build-up of more
massive \HI\ clouds in the central regions of the absorbing structure.
As an illustrative example, consider a case noted by Milgrom (1988).
Namely, he points out that for random sight-lines through identical
spherical systems where the column density at impact parameter $l$
is $N_{HI}(l) = N_0[1+(l/r_c)^2]^{-c}$, a power-law column density
distribution with $\beta = (c+1)/c$ would result for lines of sight that
are outside the core. Here, $N_0$ is the column density at $l=0$, $r_c$
is the core radius, and $c$ is the power-law index which specifies
the run of the neutral gas density distribution with cloud radius,
$n_0(r)\sim[1+(r/r_c)^2]^{-(c+1/2)}$.  In this illustrative example we
would expect $c \approx 1.3$ at $<z> \approx 2.2$ and $c \approx 3.3$
at $<z> \approx 0.8$.  This corresponds to much more centrally condensed
systems at lower redshift.  Thus, in the context of this example, such
processes might be associated with galaxy collapse or mergers of galaxies.
In particular, if this type of process corresponded to both a relative
increase in the masses of GHCs and an increase in their particle density
(i.e. somewhat smaller absorbing structures) with decreasing redshift,
but with a relatively small conversion rate of gas into stars, it might
explain: (1) the relative flattening of the shape of the \HI\ column
density distribution from $<z> \approx 2.2$ to $<z> \approx 0.8$, (2)
the small decrease in $n_{DLA}(z)$ per comoving volume with decreasing
redshift, and (3) $\Omega_{DLA}(z)$ showing no evidence for evolution
(or a small increase) with decreasing redshift.  This scenario
would also seem to be consistent with the measurements of DLA absorber
metallicities (\S7.1) and kinematics (\S7.2).  On the other hand,
in the context of the ED model it is predicted that the column density
distribution should eventually fall off as $\sim N_{HI}^{-3}$ at the
highest column densities (WLFC95; Milgrom 1988). Such a fall-off would
also be predicted if the \HI\ gas were in  randomly distributed thin layers
(Milgrom 1988). However, Figures 32 and 33 show that the high-redshift
sample does not follow such a trend, and that the discrepancy is even
worse for the low-redshift sample. As mentioned above,
WLFC95 have speculated that such a fall off would take place at larger
column densities ($N_{HI} > 10^{21}$ atoms cm$^{-2}$) with increasing
redshift.  But with the small existing statistical samples of very large
column density systems, it is clear that no such evidence for a $\sim
N_{HI}^{-3}$ fall-off exists.

{\it What is the cause of the steepening of the $f(N_{HI},z)$ distribution
for $N_{HI} > 10^{21}$ atoms cm$^{-2}$ from low redshift to z=0?} In
the context of the GHC model, Khersonsky \& Turnshek (1996) suggested that a 
high enough rate of conversion of gas into
stars would naturally lead to enough UV radiation to destroy the more
massive higher column density clouds, eventually causing $f(N_{HI},z)$
to steepen. This concept is qualitatively consistent with the current
data in that one would surmise that as $\Omega_{DLA}$ decreases by a
factor of $\approx 4-6.5$ from $z \approx 2.0 - 0.5$ to $z=0$ there must at
some point be an increase in UV radiation owing to the conversion of gas
into stars.  At the same time, the column density distribution would
be expected to fall off as $\sim N_{HI}^{-3}$ at the highest column
densities in an ED model, and this is observed for $N_{HI} > 10^{21}$
atoms cm$^{-2}$ at $z=0$.  Of course, since randomly distributed gaseous
disk galaxies were used to derive the $z=0$ column density distribution
function (Rao \& Briggs 1993; Zwaan et al. 1999), the result at $z=0$
is expected. But the details of how the destruction of massive GHCs would
finally lead to the column density distribution
observed at $z=0$ are unclear.  Nevertheless, {\it the lack of a high
column density fall-off following the $\sim N_{HI}^{ -3}$ functional
form in the low-redshift DLA sample provides independent evidence that
at $<z> \approx 0.8$ the DLA absorbers we have detected do not predominantly
arise in sight-lines through randomly oriented gaseous disks.}

The above discussion offers some interesting possibilities for
interpreting the evolution of the $f(N_{HI},z)$ distribution from
high redshift to $z=0$.  However, while it is clear that evolution
is being observed, we cannot as yet really hope to use the results
to provide too many details of this evolution due to the small number
statistics. Moreover, detailed modeling of the $f(N_{HI},z)$ distribution
would be complicated by the possibility that the observed $f(N_{HI},z)$
distribution arises from a population of DLA absorbers that has a range of
properties, as opposed to sight-lines through randomly oriented identical
absorbers.  Indeed, the evidence for evolution in the shape of the
$f(N_{HI},z)$ distribution suggests that the population of DLA absorbers
must have a variety of properties.  Only at $z=0$ do we observe clear
evidence for the $\sim N_{HI}^{-3}$ fall-off at high column densities,
consistent with the gaseous disks of spiral galaxies.  Owing to the lack
of evidence for an $\sim N_{HI}^{-3}$ fall-off at high column densities,
our results on $f(N_{HI},z)$ at high and low redshifts indicate that these
samples of DLA absorbers are not exclusively derived from a population of randomly
oriented gaseous disks, apparently ruling out the paradigm that DLA systems can be
used to trace the evolution of the gaseous disk components of galaxies
from high to low redshifts. 

\section{The Limitations of the Results}

While we believe that the conclusions drawn thus far have been relatively
conservative, there should be concerns about the reliability of
some of these results and the future application of this technique.
Some important issues are noted below. A clear understanding of these
issues will require larger sample sizes and further study.

\subsection{Concerns about Small Number Statistics}

Although the results presented here represent the best assessment so far
of DLA statistics at $z<1.65$ from a systematic survey, the numbers are still
very small.  It should be clear that a much larger low-redshift DLA
survey would have to be performed in order to accurately follow
the cosmological evolution of neutral hydrogen gas at low redshifts.

Our estimate of the error in the incidence of DLA in two redshift bins,
$n_{DLA}(z \approx 0.5,1.2)$, as reported in \S4 was relatively 
straightforward. We used the Poissonian error corresponding to the small
number of detected systems and combined this with the error on $n_{MgII}(z)$
estimated from the SS92 \MgII survey to estimate the final error.
However, an associated serious problem involves the determination of
$\Omega_{DLA}(z \approx 0.5,1.2)$ and the corresponding errors. The
reported $\Omega_{DLA}$ and errors at these redshifts (Figure 31) were
derived using equations (9) and (10) in \S4. However, for consistency
we must also have
\begin{equation}
\Omega_{DLA} =  \frac{\mu 8 \pi G m_{H}}{3cH_0}
\int\limits_{N_{HI}=2\times10^{20}}^{N_{HI}=N_{HI}^{max}} 
N_{HI} f(N_{HI},z) dN_{HI}.
\end{equation}
Thus, the value of $\Omega_{DLA}$ depends on $N_{HI}^{max}$.
When the column density distributions are parameterized by power laws,
$\Omega_{DLA}$ will diverge if the power-law exponent is $\beta \leq 2$
and $N_{HI}^{max}$ is permitted to be arbitrarily large. This is one
reason why it has been tempting to fit the column density distribution
function with an exponential-disk model in the past. However, by doing
so, the possible uncertainty in $\Omega_{DLA}(z)$ is minimized based
on the speculation that we understand the model for DLA absorbing
structures.  Even in a power-law model with a cutoff at $N_{HI}^{max}$ for
$f(N_{HI},z)$, it might be reasonable to expect that $N_{HI}^{max}$ depends
on redshift. Thus, depending on the chosen value of $N_{HI}^{max}$,
there could be a systematic error in the derived $\Omega_{DLA}(z)$
that is redshift dependent. As an example of the uncertainty, we point
out that for a power law with index $\beta \approx 1.7$, the value of
$\Omega_{DLA}$ increases by a factor of $\approx 2.5$ when $N_{HI}^{max}$
is increased from $5 \times 10^{21}$ atoms cm$^{-2}$ to $5 \times 10^{22}$
atoms cm$^{-2}$. In DLA surveys the largest \HI\ column densities are $\approx
5 \times 10^{21}$ atoms cm$^{-2}$, so this is the value we adopt for
the $N_{HI}^{max}$ at all redshifts. But, for example, Braun (1997) has
reported values as large as $\approx 4 \times 10^{22}$ atoms cm$^{-2}$
in local spirals.  Of course, the actual value of $N_{HI}^{max}$ in DLA
systems is very poorly determined because of small number statistics,
and this problem is amplified because the highest column density systems
are the most rare. Until sample sizes are significantly larger, we cannot 
rule out the possibility that this effect is causing us to underestimate 
$\Omega_{DLA}$ by a factor of as much as $\approx 2 - 3$ in a redshift 
dependent way.

\subsection{The Possibility of Gravitational Lensing Bias}

It has been argued that low-redshift surveys for DLA lines using
bright QSOs may suffer a gravitational lensing bias in the sense that
the incidence of DLA absorption will be systematically overestimated
(Smette et al. 1997). This would only happen for massive lensing
galaxies.  For determination of results on $\Omega_{DLA}(z<1.65)$
this effect is potentially important.  For example, if the two highest
column density DLA systems contributing to the lowest redshift bin
in Figure 31 were due to gravitational lensing bias, it would cause
$\Omega_{DLA}(z<0.8)$ to be severely overestimated. While the QSOs
exhibiting low-redshift DLA lines observed with {\it HST}-WFPC2 do not show any
evidence of being lensed (Le Brun et al. 1997), the ones that contributed 
the most to the high $\Omega_{DLA}(z<0.8)$ values have not been imaged 
with {\it HST}, and so the possibility of lensing by the very high column density
low-redshift DLA galaxies in our survey has not been studied with the 
greatest possible sensitivity.  On the other hand,
our ground-based studies suggest that the luminous counter-parts of
low-redshift DLA absorbers are often not highly-luminous, and therefore
massive, galaxies (e.g. Rao \& Turnshek 1998; \S7.3)

\subsection{Reliance on \MgII Statistics}

The derived statistical results on low-redshift DLA absorbers 
depend on our statistical understanding of the \MgII population of
absorbers. At the present time the statistical errors on the incidence
of \MgII absorption-line systems are $\approx$ 47\% 
 at $z \approx 0.5$ and $\approx$ 58\%  at $z
\approx 1.2$. In the future, if we are to expand and further develop
this technique, larger systematic surveys for QSO \MgII absorption-line
systems (especially with \w $> 0.5$ \AA) will have to be undertaken.

\subsection{Complications Due to the Presence of Dust}

Fall \& Pei (1989) originally suggested that dust in high column
density absorbers had some influence on the statistics of DLA
absorption, although at that time the observational evidence was 
not very convincing. More recently, Pei et al. (1999)
discussed the effects of dust on $\Omega_{DLA}(z)$, but used the 
less reliable low-redshift data point of   LWT95 to show that a significant 
amount  ($\approx$68\%) 
of neutral gas is being missed by DLA surveys. Since we derive a higher
value for $\Omega_{DLA}(z)$ at low redshift, this effect may not be 
as severe. A comparison with their models suggests that we might be
missing $\approx11$\% of the neutral gas mass at $z\approx0.5$.

Pettini et al. (1997a) have used depletions of Cr
relative to Zn in DLA absorbers and found evidence for dust, but after
taking into account the normally low metallicities of DLA systems
the inferred dust-to-gas ratios in the known population of
DLA absorbers are $\approx 1/30$ of the Galaxy value.  On the
other hand, Carilli et al. (1998) have identified some DLA absorbers by
finding 21 cm absorption in a sample of 15 red quasars that were 1 Jy
flat-spectrum radio sources. Based on their success, they conclude that
optically-selected samples may miss $\approx$ 1/2 of the high column
density QSO absorption-line systems due to obscuration by dust. Thus,
there is reason for concern, especially since the effect is likely to
be redshift dependent. The distribution of dust-to-gas ratios in high
column density systems must be investigated in greater detail before
we can hope to understand how dust affects our results on $n_{DLA}$
and $\Omega_{DLA}$.

\subsection{The \HI\ Column Density Distribution at $z=0$}

The \HI\ column density distribution at $z=0$, and hence $n_{DLA}(z=0)$ 
(see equation 12),
was determined using radial \HI\ column density profiles of a complete
sample of 27 large-diameter gas-rich galaxies (Rao \& Briggs 1993). 
At $z=0$, spirals afford
the largest cross-section to line-of-sight absorption at column densities
$N_{HI}>10^{19}$ atoms cm$^{-2}$ (see figure 5 in Rao \& Turnshek 1998)
and contribute $>$95\% to the local interception probability. Moreover,
$\approx$ 89\% of the \HI\ mass in the local Universe resides in spiral 
galaxies later than type S0 (Rao \& Briggs 1993). Thus, the $z=0$ ``DLA'' 
statistics should be accurately described
by the properties of local spiral galaxies. This is equivalent to saying that
 pencil-beam QSO surveys for DLA at $z=0$ will most likely intercept 
large diameter spirals. Thus, the statistics derived from such a survey 
should reproduce the properties of the local spiral galaxy population. 

The one caveat in using 21 cm emission measurements to describe column 
density distributions within galaxies is that spatial resolution is limited 
by the finite size of the radio beam. The technique measures the total 
\HI\ mass within the radio beam to a high degree of accuracy, but what one 
really derives is  an average \HI\ column density within the 
area sampled by the beam. Thus, information about the highest column 
densities residing in any knots or clumps smaller than the radio beam 
would be lost.  The extent to which this 
affects the $z=0$ statistics depends on the total cross-sectional area of any
high-column-density clumps and how much smaller they are than the beam size.
If such an effect is important,  the
most significant consequence might be an underestimation of $f(N_{HI})$ at 
column densities $N_{HI}>10^{21.5}$ atoms cm$^{-2}$ (Zwaan et al. 1999), but any 
effect on $n_{DLA}(z=0)$ 
is likely to be small since the highest column densities would still 
contribute only a fraction of the total cross-section over all 
$N_{HI}>2\times 10^{20}$ atoms cm$^{-2}$.
Note that the finite radio beam size does not affect the determination 
of $\Omega_{DLA}(z=0)$ since the integrated \HI\ mass within the beam is a
well determined quantity. The extent to which high column densities
are under-represented in 21 cm emission surveys can be estimated by
comparing the values of $\Omega_{DLA}(z=0)$ determined by 
(1) directly measuring
the \HI\ mass of galaxies as done by Rao \& Briggs (1993) and
(2) integrating the $f(N_{HI})$ distribution over all column densities
(see equation (9) in WLFC95 and equations (8), (9) and (10) in Rao \& Briggs 1993).
The first is a true measure of the total \HI\ mass including the
contribution from the highest column density clumps. If the exclusion of the
highest column density material in the $f(N_{HI})$ distribution were severe, 
then the second estimate of the total mass would be significantly lower.
However, this comparison was carried out in Rao \& Briggs (1993) and the 
two values were found to be equal within the errors. Therefore, we conclude
 that any high
column density gas that may have been missed in 21 cm emission surveys is
not significant, and that the $f(N_{HI})$ distribution at $z=0$ in Figure 33
is a fair representation of the local gas distribution.

A related point involves \HI\ self-absorption within a galactic disk. We note
that if the inclination of a disk is large, the measured \HI\ 
flux in a radio beam might be affected by \HI\ self-absorption within 
the disk.  This problem would 
underestimate both the mass and the mean column density measured within the 
beam by approximately 15\%. We first discussed this in RTB95. 
While we took this effect into account 
by increasing the upper error bar on $\Omega_{DLA}(z=0)$ by 15\%, we did not
include errors in the measured values of $N_{HI}$ versus radius for our
 sample of spiral galaxies in the determination of $f(N_{HI},z=0)$.

Finally , there is the issue of small \HI\ masses ($< 2\times 10^7 h^{-2}_{65}$
M$_{\odot}$) and the inability to detect them in local \HI\ surveys
(Zwaan et al. 1997). This is a 
limitation of our current knowledge of the distribution of \HI\ in the
local Universe. Observationally, the existence of absorbing objects with such
masses (e.g. gas-rich dwarfs) that also contain high column density regions
($N_{HI}>10^{21}$ atoms cm$^{-2}$) cannot be ruled out.

\section{The Nature of the DLA Absorber Population}

It is reasonable to assume that all DLA absorbers are
associated with some sort of galaxy since
there are no known DLA systems which lack metal-line absorption. Thus,
we can infer the presence of star formation, stellar nucleosynthesis,
and stellar mass loss.  In the GHC model for DLA systems (Khersonsky \&
Turnshek 1996)
noted earlier (\S5.5), it was  assumed that DLA systems
are associated with the neutral gas in GHCs that must naturally be
associated with various types of gas-rich galaxies and proto-galaxies.
Such galaxies would be the progenitors of different types of galaxies
(ellipticals, spirals, irregulars, dwarfs, etc.) that are observed
at the present epoch. This is consistent with the idea that the DLA
systems generally track neutral gas in, for example,
both spheroidal and disk components of galaxies and proto-galaxies (and
possibly other gaseous structures as well) from high redshift to the
present epoch. Alternatively, the ED model is much more specific in that
it relates DLA absorbers to galactic disk systems, which furthermore
are thought to be relatively luminous (WTSC86; Wolfe 1988; Lanzetta et al. 
1991; Steidel 1995).  Initially there were
many good indirect reasons to adopt the ED model (see Wolfe 1995 and
references therein). However, now there are some direct results which
contradict the single DLA-\HI\-disk paradigm. When evidence from DLA
metallicity measurements, kinematic measurements/simulations, and the
direct imaging of galaxies associated with DLA systems is considered,
the DLA-GHC model is seen to be a much more favorable framework for
interpretation of the DLA systems. Here we summarize this evidence.
The evidence supports the conclusion, drawn from analysis of DLA column
density distributions at low and high redshifts (\S5), 
that  DLA systems  do not exclusively trace the gaseous disk components 
of galaxies. Instead they trace a mixture of types.

\subsection{Evidence from Measurements of the Gas-Phase Metallicity}

Pettini and collaborators (e.g. Pettini et al. 1997b, Pettini et al. 1999,
Pettini et al. 2000) have measured metallicities in many of the
high-redshift DLA systems and some of the known low-redshift ones.
Some of their measurements pertain to systems which are not classically
damped, with $N_{HI} \ge 2 \times 10^{20}$ atoms cm$^{-2}$, but this does
not affect their broad conclusions.  The most extensive measurements are
for Zn which is a particularly good tracer of the gas phase metallicity since
it is not significantly depleted on to dust grains. 
While detailed interpretations of the metallicity measurements
have not reached a consensus (e.g. see Lu et al. 1996, Pettini et al. 1999,
Prochaska \& Wolfe 1999, and references therein), the broad results are clear. 
The typical column density weighted metallicity in the redshift interval 
$3<z<0.5$ is $\approx$ 1/12
of the solar value ($<$[Zn/H]$>\approx -1.1$], with no obvious tendency
for the metallicity to increase with decreasing redshift. However, there
is considerable scatter in the individual metallicities which is most
likely caused by the wide range of formation histories and galaxy types
responsible for the DLA systems. In fact, Pettini et al. (1999) have concluded
that the known DLA systems are unlikely to trace the galaxy population
responsible for the bulk of the star formation. Metallicity measurements
of the lowest-redshift systems in our low-redshift DLA sample will greatly
clarify the evolution of DLA metallicity to the lowest redshifts that
are currently attainable.

\subsection{Evidence from Measurements of the Kinematics}

High-resolution studies of the metal-line profiles associated with
DLA systems provide constraints on kinematic models of the absorbing
region. In general, a DLA absorbing region will include one or more very
high column density components as well as low column density
components that give rise to detectable metal lines. Using Keck HIRES
observations, Prochaska \& Wolfe (1997; 1998) have shown that the kinematics
of the metal-line profiles associated with high-redshift DLA absorbers
often have a ``leading-edge'' characteristic. Based on simulations, they
have argued that the kinematic profiles of the DLA absorbing regions as
a whole are more consistent with models of rotating \HI\ disks than any
other {\it single} type of model.  More recently, the kinematics
have also been shown to be consistent with gas in-fall due to merging
(Haehnelt, Steinmetz, \& Rauch 1998), randomly moving clouds in a spherical 
halo (McDonald \& Miralda-Escud\'e 1999), and multiple gas discs in a common 
halo (Maller et al. 2000).  Therefore, it 
seems most likely that, consistent with other current findings, the 
kinematics of DLA absorbing regions arise from a mix of kinematic 
structures, including some rotating gaseous disks.

\subsection{DLA Galaxy Imaging Studies}

There are some direct results from published {\it HST}-WFPC2 images of confirmed
($N_{HI}\ge2\times10^{20}$ cm$^{-2}$) low-redshift DLA
absorbers which suggest that the morphological types of DLA galaxies are
indeed mixed. Seven DLA absorbers in the redshift interval $0.395<z<1.010$
have so far been imaged with {\it HST}-WFPC2.  In one case (the $z=0.656$
system toward 3C 336 (1622+239)) there is no optical evidence for a
galaxy despite very deep observations (Steidel et al. 1997).  In the
other six cases the galaxies are likely to be spirals (the $z=0.437$
system toward 3C 196 (0809+483) and the $z=0.633$ system toward Q1209+107), 
two amorphous, LSB
galaxies (the $z=0.395$ system toward PKS 1229$-$021 and the $z=0.692$
system toward 3C 286 (1328+307)), and two compact objects (the $z=1.010$
system toward EX 0302$-$223 and the $z=0.860$ system toward PKS 0454+039)
(Le Brun et al. 1997).  One other higher-redshift DLA absorber imaged by
Le Brun et al. (1997) with {\it HST}-WFPC2 was also found to be a compact object
(the $z=1.776$ systems toward MC 1331+170).

In addition, ground-based observations of three of the lowest redshift
DLA absorbers in our low-redshift sample have been  published.
Initial WIYN imaging indicates that one is a
dwarf while the other is a dwarf and/or an object with very low
surface brightness (the $z=0.091$ and $0.221$ absorbers toward OI 363
(0738+313)) (Rao \& Turnshek 1998). In the third case, a luminous spiral
lies near the sight-line but several dwarfs are also present with even
smaller impact parameters (the $z=0.313$ system toward PKS 1127$-$14,
Lane et al. 1998a).  

Thus, the DLA galaxies associated with the DLA systems in our
low-redshift sample are not dominated by luminous gas-rich spirals.
Instead, the imaging results demonstrate that the morphological types of
DLA galaxies are  mixed and span a range in luminosities and surface
brightnesses. Ongoing imaging observations of other DLA absorbers in our 
low-redshift sample support this conclusion. 

\subsection{Discussion}

Several independent lines of evidence now lead us to conclude that
the DLA galaxy population represents a range of morphological types
and luminosities. It might then follow that the progenitors of spiral 
galaxies are not the main repositories of neutral gas at redshifts $z>0$. 
However, this is at odds with results near redshift $z=0$, where the bulk 
of the high-column-density neutral gas is found to reside in luminous gas-rich spirals.
Can these two apparently disparate observational results be reconciled?

When considered together, the following three results, (1) that $n_{DLA}(z)$ and 
$\Omega_{DLA}(z)$ show little evolution between redshifts $z\approx4$ and 
$z\approx0.5$, (2) that the metallicities of the DLA absorbers has remained 
roughly constant over this redshift interval, and (3) that many of the 
DLA galaxies at low redshift that have been identified through imaging 
are LSB or dwarf galaxies, might lead us to conclude that 
DLA absorbers are a slowly evolving population. In fact, 
 DLA galaxies probably do not trace star-forming galaxies
(see also Fynbo et al. 1999), and perhaps DLA absorbing regions are in a 
quiescent phase
of their evolution when they are not undergoing an episode of star formation. 
Thus, we might be seeing little evolution in the properties of DLA systems 
by virtue of an observational bias that selects similar types of absorbers at 
all redshifts, i.e., regions in intervening galaxies with high-column-density, 
low-metallicity (and hence, low-dust-content) gas.  If the models of Pei et 
al. (1999) are correct, then galaxies that are not tracked by DLA absorption
lines make up $\approx10$\% of the neutral gas at $z\approx0.5$ (using our new
low-redshift result) and  
$\approx60$\% at $z\approx 2.5$ where the effect of obscuration by dust is 
most severe (see their figure 7).  On the other hand, local \HI\ 21 cm 
emission-line surveys do not suffer from this bias. Thus, we are left with
some unanswered questions: (1) 
Are a majority of spiral-galaxy progenitors missing from DLA absorption-line
surveys and do they contain the missing gas fraction seen in the models of Pei 
et al. (1999)? (2) What fraction of the neutral gas mass is contained in 
star-forming galaxies at high redshift? (3) What are the present-day 
whereabouts of the neutral gas observed in high- and low-redshift DLA galaxies?

A self-consistent explanation of the observational data that has been 
obtained from QSO absorption lines and imaging studies of DLA galaxies,
UV luminous galaxies, faint blue dwarfs, etc. is still far from complete. 
Recall that the redshift interval $0<z<0.5$ includes the most recent
45\% of the age of the Universe for $q_0=0.5$ (33\% for $q_0=0$). Thus,
it is conceivable that significant changes in the properties of galaxies
have taken place over this time interval. Efforts on increasing the sample 
size of low-redshift DLA absorbers should be made. This will 
allow us to study the evolution in $n_{DLA}(z)$, $\Omega_{DLA}(z)$, and 
$f(N_{HI},z)$ in greater detail, and help assess the effects of biases
such as dust obscuration.

\section{Summary and Conclusions}  

We have used a novel technique, based on \MgII pre-selection, to
conduct a low-redshift ($z<1.65$) survey for DLA absorption systems
in QSO spectra. The survey was conducted in the UV with {\it HST}-FOS.
Our low-redshift results, when combined with results from DLA studies
at higher redshift, afford us the opportunity to study the distribution
of neutral gas in DLA systems back to a time when the Universe was
less than 10\% of its present age. Aside from DLA studies, no other
methods have revealed such large amounts of significantly redshifted
neutral gas. Thus, the study of DLA absorbers is currently the most
effective method for investigating the bulk of the known neutral gas mass of
the Universe. Our findings and conclusions are as follows:

(1) In our survey we uncovered 12 DLA lines in a sample of 
87 \MgII systems with \MgII rest equivalent width \w $\ge0.3$ \AA. 
Two more DLA systems with no  available \MgII absorption-line information
were discovered serendipitously. 

(2) The DLA absorbers are drawn almost exclusively from
the population of \MgII absorbers which have \w $\ge0.6$ \AA.
Approximately 50\% of the systems with \MgII$\lambda$2796 $\ge$0.5 \AA\
{\it and} \FeII$\lambda$2600 $\ge$0.5 \AA\ have DLA absorption lines which
meet the classical definition used in high-redshift surveys, $N_{HI}
\ge 2 \times 10^{20}$ atoms cm$^{-2}$.  We know of only one rare case
of a low-redshift DLA absorber that does not have \MgII$\lambda$2796
$\ge$0.5 \AA\  and \FeII$\lambda$2600 $\ge$0.5 \AA.

(3) The survey method resulted in a determination of the fraction of \MgII
systems with DLA, which in turn allowed us to determine the incidence
of low-redshift DLA absorption using earlier results on the incidence
of \MgII absorption systems. Neutral hydrogen column densities were determined
by fitting Voigt damping profiles to the \lya lines in the
{\it HST}-FOS UV survey spectra. This information was then used to deduce the
low-redshift cosmological neutral gas mass density of DLA absorbers and their
\HI\ column density distribution.

(4) In making these determinations, possible biases were carefully eliminated.
 For example, known 21 cm absorbers were excluded from the initial low-redshift
 sample and
the low-redshift sample (this study, $<z> \approx 0.8$) and high-redshift
sample (from the literature, $<z> \approx 2.2$) were corrected for
Malmquist bias before they were compared. A comparison with
inferred results on DLA at $z=0$ was also made; the $z=0$ results were 
derived from 21 cm emission observations of nearby gas-rich spirals,
since spirals are the dominant known reservoirs of large column density
neutral gas in the local Universe.

(5) The findings were discussed in the context of a $\Lambda = 0$ cosmology
with $q_0=0$ and $q_0=0.5$.
Assuming that the $z=0$ results are not missing any significant fraction
of the local neutral gas, the incidence of DLA systems per unit redshift,
$n_{DLA}$, is found to decrease with decreasing redshift by an amount
significantly larger than what is expected in a no-evolution universe.
Specifically, we found that the high-redshift incidence of DLA systems is
$\approx 6$ times larger than that found at $z=0$ in a $q_0=0$ no-evolution
universe ($\approx 10$ times larger if $q_0=0.5$). However, if the $z=0$ 
point is removed from the analysis, there is only mild evidence for 
a real decrease in incidence per comoving volume down to $z\approx
0.5$. Thus, it is possible that most of the real evolution occurs between
$z\approx 0.5$ and $z=0$.

(6) The DLA cosmological neutral gas mass density also shows strong
evidence for evolution when compared with the $z=0$ results. If
$q_0=0$, the comoving cosmological mass density of high-redshift DLA
absorbers is $\approx 4$ times larger than what is inferred from  local
21 cm observations ($\approx 6.5$ times larger if $q_0=0.5$).  However,
if the $z=0$ results are ignored, the cosmological mass density of neutral
gas in low-redshift DLA absorbers is observed to be comparable to that
observed at high redshift (and, in fact, formally somewhat higher).  Thus,
we cannot point to a trend in the DLA data themselves which indicates that
$\Omega_{DLA}$ at low redshift is approaching the value derived for the
cosmological mass density of neutral gas found in present-day spirals.
Again, it is possible that most of the real evolution occurs between
$z\approx 0.5$ and $z=0$.

(7) The \HI\ column density distribution of the low-redshift DLA
sample is found to differ in comparison to the high-redshift
DLA sample and that inferred
for local spirals. The low-redshift DLA absorbers exhibit a
larger fraction of very high column density systems in comparison to
determinations at both high redshift and locally. The observed trend
suggests that the column density distribution starts out relatively
steep at high-redshift ($f \sim N_{HI}^{-1.8}$), it then flattens at the
redshifts covered in our low-redshift survey ($f \sim N_{HI}^{-1.3}$),
and then locally it starts off relatively flat ($f \sim N_{HI}^{-1.2}$)
but then steepens at $N_{HI} > 10^{21}$ atoms cm$^{-2}$ ($f\sim
N_{HI}^{-2.9}$), becoming even steeper than observed at high redshift.  We
emphasize that the column density distributions of DLA absorbers at high
and low redshifts are not observed to fall off as $f\sim N_{HI}^{-3}$
with increasing column density. An $f \sim N_{HI}^{-3}$ fall-off is
theoretically predicted for disk-like systems and this is, in fact,
what is observed locally in spiral samples.

(8) While our understanding is that it is unlikely that the $z=0$
studies have missed any significant component of the neutral gas, we
also understand that existing \HI\ 21 cm emission surveys are only very
sensitive to neutral gas masses $M_{HI} > 2\times 10^7 h^{-2}_{65}$ M$_{\odot}$.
Thus, if the $z=0$ results are missing a significant component of the
neutral gas mass, it almost certainly lies in this regime. However
unlikely, if this should turn out to be the case, there would then be
evidence for only mild intrinsic evolution in the DLA population over
the redshift interval $3>z>0$.

(9) The decrease in the overall incidence of DLA absorbers with decreasing
redshift, accompanied by the increase in the relative number of high
column density systems with decreasing redshift, explains the lack of
evidence for evolution in the amount of neutral gas locked up in the DLA
absorbers from high redshift to $z \approx 0.5$. This lack of evidence
for significant conversion of gas into stars (or some other phase),
 suggests that the gas contained in  DLA systems has
not experienced much star formation, and is consistent with the lack
of evolution seen in their metallicities. Moreover, recent 
results from imaging studies show that DLA galaxies 
comprise a mix of morphological types with a large fraction of them being
low-surface-brightness or dwarf galaxies. Thus, DLA absorption lines seem to
be tracking a slowly evolving population of objects. Indeed,
there might be little evolution in the properties of DLA systems for the simple
reason that observational biases select similar types of absorbers at 
all redshifts, i.e., regions in intervening galaxies with high-column-density, 
low-metallicity (and hence, low-dust-content) gas. 

(10) DLA systems trace the bulk of the observable neutral gas content of the 
Universe at redshifts $z>0$, while at the present epoch, the neutral gas mass 
density and cross-section are dominated by luminous spirals. In order to 
understand the evolution of neutral gas in the Universe, we must address the
question of the  present-day whereabouts of the neutral gas mass
observed in high- and low-redshift DLA systems.
Can models predict the decline of a factor of $\sim$ 5 in the neutral gas
mass density from $z\approx0.5$ to $z=0$ especially if DLA galaxies are a 
non-star-forming and slowly-evolving population? 

(11) Finally, in order to improve our understanding of the neutral gas
content of the Universe to the point where we can accurately follow
its evolution to $z=0$, we will have to collect and analyze
large samples of low-redshift DLA systems. In addition to improving the
statistical uncertainties, this would permit a better determination
of the shape of the \HI\ column density distribution at large column
densities, effectively allowing a determination of the maximum \HI\
column density which might be a function of redshift. Understanding this
turn-down in column density is important for the determination of 
$\Omega_{DLA}$. A large optical
ground-based survey for large rest equivalent width QSO \MgII absorption
lines would be a reasonable initial step. Additionally, however, in order
to fully take advantage of the information that could be  provided by
a large DLA survey, we would have to do more than simply increase the
sample size. We would also have to design observational programs that
clarify the uncertainties or systematic effects caused by potential dust
obscuration and possible gravitational lensing bias. The sensitivity of 
the \HI\ 21 cm emission surveys should also be pushed to lower limits, 
to the point where we can definitively assess the possibility of very high
column density DLA absorption arising in very low mass local objects
($M_{HI} < 10^{7}$ M$_{\odot}$).

\acknowledgements
We would like to thank the staff of the Space Telescope Science
Institute for supporting this work. We would also like to thank our
colleagues who are collaborating with us in our efforts to identify the
visible counterparts of the DLA galaxies from our low-redshift sample:
Dr. Frank Briggs, Dr. Jacqueline Bergeron, Wendy Lane, Dr. Eric Monier,
Daniel Nestor, and Dr. Alain Smette.  We would specifically like to thank 
Wendy Lane and Dr. Lisa Storrie-Lombardi for sharing their data in 
advance of publication, and Dr. Anuradha Koratkar for providing the spectrum
of PG 0117+213. We have also benefitted from discussions with Jane Charlton,
Chris Churchill, and Max Pettini. DAT would like to thank Drs. Valery Khersonsky,
Ken Lanzetta, and Art Wolfe for past collaborations on this topic. This
work was supported in part by STScI grant GO-06577.01-95A and NASA LTSA 
grant NAG5-7930. 

\end{document}